%% file: Final.tex
\documentclass[fleqn,10pt]{wlscirep}
\usepackage[utf8]{inputenc}
\usepackage[T1]{fontenc}
\usepackage{bm}
\usepackage{comment}
\usepackage{caption}
\usepackage{amsmath}
\usepackage{color}
\usepackage{lineno}
\usepackage[textwidth=1.7cm]{todonotes}
\DeclareMathOperator*{\argmin}{arg\,min}
\DeclareMathOperator*{\argminNp}{arg\,min ^{N_p}}
\usepackage[acronym]{glossaries}
\usepackage{algorithm}
\usepackage{algpseudocode}
\usepackage{siunitx}
\PassOptionsToPackage{hyphens}{url}\usepackage{hyperref}
\title{Neuromorphic Overparameterisation and Few-Shot Learning in Multilayer Physical Neural Networks}

\author[1,2*]{Kilian D. Stenning}
\author[1,2$\dagger$]{Jack C. Gartside}
\author[3,$\dagger$]{Luca Manneschi}
\author[1]{Christopher T. S. Cheung}
\author[1]{Tony Chen}
\author[1]{Alex Vanstone}
\author[4]{Jake Love}
\author[1]{Holly Holder}
\author[5]{Francesco Caravelli}
\author[6,7,8]{Hidekazu Kurebayashi}
\author[4]{Karin Everschor-Sitte}
\author[3]{Eleni Vasilaki}
\author[1,2]{Will R. Branford}
\affil[1]{Blackett Laboratory, Imperial College London, London SW7 2AZ, United Kingdom}
\affil[2]{London Centre for Nanotechnology, Imperial College London, London SW7 2AZ, United Kingdom}

\affil[3]{University of Sheffield, Sheffield S10 2TN, United Kingdom}
\affil[4]{Faculty of Physics and Center for Nanointegration Duisburg-Essen (CENIDE), University of Duisburg-Essen, 47057 Duisburg, Germany}
\affil[5]{Theoretical Division (T4), Los Alamos National Laboratory, Los Alamos, New Mexico 87545, USA}
\affil[6]{London Centre for Nanotechnology, University College London, London WC1H 0AH, United Kingdom}
\affil[7]{Department of Electronic and Electrical Engineering, University College London, London WC1H 0AH, United Kingdom.}
\affil[8]{WPI Advanced Institute for
 Materials Research, Tohoku University, Sendai, Japan.}
\affil[$\dagger$]{These authors contributed equally}

\affil[*]{Corresponding author e-mail: k.stenning18@imperial.ac.uk}

\begin{abstract}

Physical neuromorphic computing, exploiting the complex dynamics of physical systems, has seen rapid advancements in sophistication and performance. Physical reservoir computing, a subset of neuromorphic computing, faces limitations due to its reliance on single systems. This constrains output dimensionality and dynamic range, limiting performance to a narrow range of tasks. Here, we engineer a suite of nanomagnetic array physical reservoirs and interconnect them in parallel and series to create a multilayer neural network architecture. The output of one reservoir is recorded, scaled and virtually fed as input to the next reservoir. This networked approach increases output dimensionality, internal dynamics and computational performance.  We demonstrate that a physical neuromorphic system can achieve an overparameterised state, facilitating meta-learning on small training sets and yielding strong performance across a wide range of tasks. Our approach's efficacy is further demonstrated through few-shot learning, where the system rapidly adapts to new tasks.

\end{abstract}
\begin{document}

\flushbottom
\maketitle
\thispagestyle{empty}
\section*{Introduction}
In artificial intelligence (AI) and machine learning, the performance of models often scales with their size and the number of trainable parameters. Empirical evidence supports the advantages of overparameterised regimes\cite{zou2020gradient,zou2019improved} , where models, despite having a large number of parameters, avoid overfitting, generalise effectively, and learn efficiently from a limited number of training examples. 

Physical neuromorphic computing aims to offload processing for machine learning problems to the complex dynamics of physical systems\cite{markovic2020physics,mizrahi2018neural,gartside2022reconfigurable,allwood2023perspective,schuman2022opportunities,tanaka2019recent,nakajima2020physical,milano2021materia,chumak2021roadmap,papp2021nanoscale,cucchi2022hands,vidamour2023reconfigurable,wright2022deep}. Physical computing architectures range from implementations of feed-forward neural networks\cite{wright2022deep,torrejon2017neuromorphic} with tuneable internal weights, to reservoir computing, where complex internal system dynamics are leveraged for computation\cite{tanaka2019recent}. Neuromorphic schemes stand to benefit from the advantages of operating in an overparameterised regime. A primary use case of neuromorphic systems is in edge-computing\cite{cao2020overview} where a remotely situated device locally performs AI-like tasks. For example, an exoplanet rover vehicle performing object classification can operate more efficiently by processing data on-site rather than transmitting large datasets to cloud-servers\cite{schuman2022opportunities}. Here, acquiring large datasets and transmitting them to cloud-servers for processing is often inefficient making the ability to compute well and adapt to new tasks with small training sets a desirable function. 

In physical reservoir computing, the internal dynamics of a system are not trained. Instead, only a set of weights to be applied to the readout layer are trained, reducing training costs when compared to neural network architectures. Typically, only a single physical system is employed with a fixed set of internal dynamics, resulting in a lack of versatility and overly specialised computation which is fixed at the fabrication stage. 
In contrast, the brain possesses a rich set of internal dynamics, incorporating multiple memory timescales to efficiently process temporal data\cite{hasson2015hierarchical}.
To mimic this, research on software-based reservoirs has shown that combining multiple reservoirs with differing internal dynamics in parallel and series network architectures significantly improves performance\cite{jaeger2007discovering,manneschi2021exploiting,moon2021hierarchical,gallicchio2017deep,gallicchio2017echo,gallicchio2018design,ma2020deepr,goldmann2020deep}. Parallel networks have been physically implemented\cite{van2017advances,liang2022rotating,tanaka2019recent}, however they lack inter-node connectivity for transferring information between physical systems - limiting performance. Translating series-connected networks (often termed hierarchical or deep) to physical systems is nontrivial and so-far unrealised due to the large number of possible inter-layer configurations and interconnect complexity.

Here, we present solutions to key problems in the physical reservoir computing field: We fabricate three physical nanomagnetic reservoirs with high output dimensionality and show how increasing the complexity of system dynamics can improve computational properties (Nanomagnetic reservoirs, Figure \ref{Samples}). We then develop and demonstrate a methodology to interconnect arbitrary reservoirs into networks. We demonstrate the computational benefits of the networked architecture and compare performance to software. Reservoir outputs are experimentally measured with network interconnections made virtually and outputs combined offline during training (Multilayer physical neural network, Figure \ref{RC_network}). We explore the overparameterised regime, made possible through the networking approach, where the number of network output channels far exceeds the size of the training set. The physical networks architectures do not overfit, show enhanced computational performance and are capable of fast learning with limited data. This approach is applicable to all physical systems and methods to achieve overparameterisation in an arbitrary system are discussed (Overparameterisation, Figure \ref{OP}). We demonstrate the power of operating in an overparameterised regime by implementing a few-shot learning task using model-agnostic meta-learning\cite{wang2020generalizing,vanschoren2019meta,finn2017model}. The physical network is able to rapidly adapt to new tasks with a small number of training data points (Learning in the overparameterised regime, Figure \ref{Metalearning}).

Whilst we use nanomagnetic reservoirs to demonstrate the benefits of networking and overparameterisation, the methodology can be applied to any physical system. Additionally, we discuss the scalability of our nanomagnetic computing scheme and calculate the theoretical power consumption of a device. The scheme described here lifts the limitations of low-dimensionality and single physical systems from neuromorphic computing, moving towards a next-generation of versatile computational networks that harness the synergistic strengths of multiple physical systems. All data and code is publicly available\cite{NeuroOverParam}.

\section*{Results}
\subsection*{Nanomagnetic reservoirs}

\begin{figure*}[htb!]
    \includegraphics[width=\textwidth]{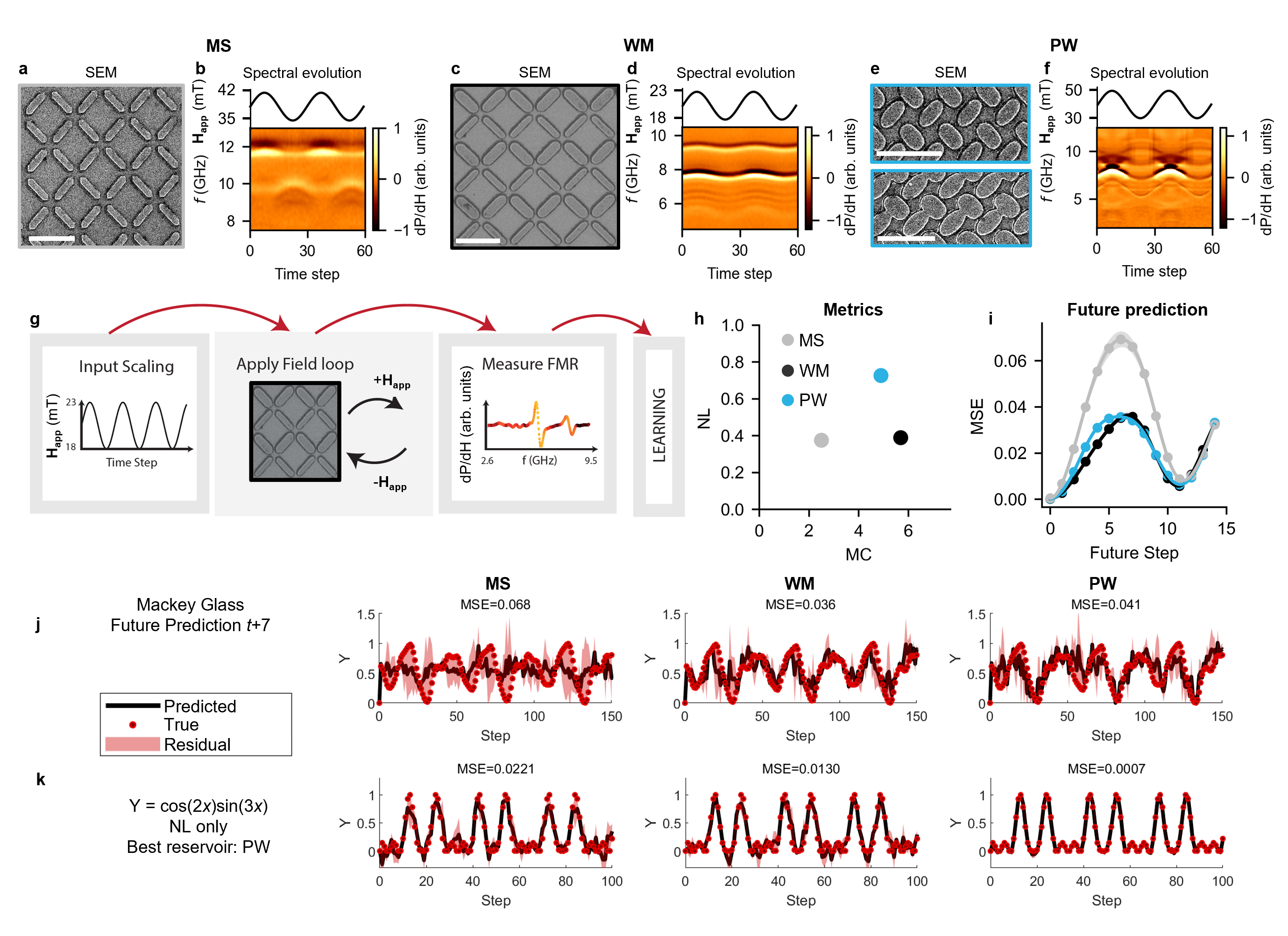}
    \caption{\textbf{Nanomagnetic reservoirs.}
    a,b) Square macrospin-only artificial spin ice (MS).
    c,d) Width-modified square artificial spin-vortex ice (WM). 
    e-f) Disorderded pinwheel artificial spin-vortex ice (PW). Bars vary from fully disconnected to partially connected.
    a,c,e) Scanning electron micrographs (SEM). All scale bars correspond to \SI{1}{\micro\meter}.
    b,d,f) Ferromagnetic resonance spectroscopy (FMR) spectral evolution from a sinusoidal field-series input. Scale bar represents amplitude of FMR signal dP/dH (arb. units).
    g) Reservoir computing schematic. Data is applied via magnetic field loops (+$\mathbf{H}_{\rm app}$ then -$\mathbf{H}_{\rm app}$) which leads of collective switching dynamics in the array. FMR output spectra measured at -$\mathbf{H}_{\rm app}$ is used as computational output. 
    h) Memory-capacity (MC) and nonlinearity (NL) of the reservoirs. Variations in sample design produce a diverse set of metrics. 
    i) Mean-squared error (MSE) when predicting future values of the Mackey-Glass time-series. High memory-capacity and low nonlinearity (WM) gives best performance. Shading is the standard deviation of the prediction of 10 feature selection trials.
    j) Attempted predictions of \textit{t}+7 of the Mackey-Glass equation. No single reservoir performs well.
    k) Transforming a sine-wave to cos(2\textit{x})sin(3\textit{x}). PW has 31.6 $\times$ lower MSE than MS. Shading represents the residual of the prediction.
    }
    \label{Samples}
\end{figure*}

The physical reservoirs used here are nanomagnetic arrays which have both nonlinearity and fading memory (i.e. a temporal response which depends on both current and previous inputs)\cite{gartside2022reconfigurable,jensen2020reservoir}. 
Each array comprises many individual nanomagnetic elements, each with its own magnetisation ‘state’. When applying an external input, the elements in an array may switch magnetisation state. The switching process depends on the external input, the current state of an element and the states of neighbouring elements through dipolar coupling, which gives rise to collective switching and high-frequency dynamics which we utilise for computation. Here, we drive switching with external magnetic fields. The array state is read by measuring the absorption of injected microwaves (i.e. measuring resonance) using ferromagnetic resonance spectroscopy (FMR), producing a spectra that is highly correlated with the collective state of the array\cite{vanstone2021spectral,gartside2022reconfigurable,jungfleisch2016dynamic,kaffash2021nanomagnonics}.

Figure \ref{Samples} shows scanning electron micrographs (Figures \ref{Samples} a,c,e) and FMR spectral evolution of three nanomagnetic reservoirs (labelled macrospin (MS), width-modifed (WM)\cite{gartside2022reconfigurable} and pinwheel (PW), see Supplementary Note 1 for details) when subject to a sinusoidal field-input (Figures \ref{Samples} b,d,f). A schematic of the computing scheme is shown in Figure \ref{Samples} g). The three arrays are designed to produce different and history-dependent responses to explore the effects of networking system with distinct dynamics. MS is a square lattice with bars only supporting macrospin magnetisation states. WM is a width-modified square lattice capable of hosting both macrospin and vortex spin textures. PW is a disordered pinwheel lattice with structural diversity throughout the sample. A detailed discussion of the design is provided in Supplementary Note 1.
The high readout dimensionality of FMR is key to achieving both strong performance and overparameterisation. Other techniques for readout exist, such as magnetoresistance \cite{hu2023distinguishing}, albeit at a lower output dimensionality. During training, weights are applied to recorded data offline via ridge regression which transforms the FMR spectra to a 1D time-series with the aim of closely matching the target waveform (see Methods for further details).

We assess reservoir performance via the mean-squared error (MSE) between the target and the reservoir prediction as well as two metrics:  memory-capacity and nonlinearity. Memory-capacity measures the ability of the current state to recall previous inputs\cite{love2023spatial}, which arises from history-dependent state evolution. Nonlinearity measures how well past inputs can be linearly mapped to the current state\cite{love2023spatial}, and can arise from a number of physical system dynamics. In this work, these dynamics include resonant frequencies shifts from changing microstates and input field and the shape of FMR peaks. Metric calculations are further described in the Methods section, Task Selection, and Supplementary Note 2. These metrics allow a coarse mapping between physical and computational properties, enabling comparison of different systems. Figure \ref{Samples} h) shows the nonlinearity (NL) and memory-capacity (MC) metrics. Our array designs produce a diverse set of metrics, ideal for exploring the computational benefits of networking later in this work.

Throughout this work, we focus on two input time-series: a sine-wave and the chaotic Mackey-Glass time-series\cite{mackey1977oscillation} (described in the Methods section, Task Selection). These datasets are chosen as they are compatible with 1D global field input and can be used to devise a number of computational tasks with varying requirements. We employ short training datasets (200 data points) to reflect real-world applications with strict limitations on data collection time and energy.

We wish to evaluate the best possible computation from our readout. To do so, we employ a feature algorithm selecting the best performing combination of readout features (i.e. frequency channels) and discarding noisy or highly correlated features which do not improve computation (see Methods for details).

Figure \ref{Samples} i) shows the mean-squared error (MSE) of each reservoir when predicting various future values (x-axis) of the Mackey-Glass time-series. These tasks require high memory-capacity, as attaining a good prediction requires knowledge of previous inputs, with WM and PW outperforming MS due to their richer internal switching dynamics (Supplementary Note 1). All arrays exhibit performance breakdown at longer future steps, evidenced by the periodic MSE profiles in Figure \ref{Samples} i). This is clear for the \textit{t}+7 task taken from the high-MSE region (Figure \ref{Samples} j), where predictions do not resemble the target. At later future steps, prediction quality improves due to the quasi-periodicity of the Mackey-Glass equation (\textit{t}+11 is similar to t). 

This breakdown in performance is common in single physical systems and software reservoirs which often do not possess the range of dynamic timescales (e.g. retaining enough information about previous inputs) to accurately predict future steps and provide a true prediction where performance gradually decreases when predicting further into the future\cite{manneschi2021exploiting,gallicchio2018layering,goldmann2020deep}. 

Figure \ref{Samples} k) shows the performance when transforming a sine-wave to cos(2x)sin(3x). The disordered PW array achieves up to 31.6 $\times$ improvement for transformation tasks with no additional energy cost, highlighting the significance of system geometry and design. Performance for a history-dependent nonlinear transform task (nonlinear autoregressive moving average, NARMA, transform\cite{jaeger2002adaptive}) and further sine-transformations are shown in Supplementary Note 3. NARMA transforms display similar periodic profiles. For sine-transformations, no one array performs the best at all tasks, further highlighting the lack of versatility of single physical systems.

\subsection*{Multilayer physical neural network}

\begin{figure*}[htb!]
    \centering
    \includegraphics[width=\textwidth]{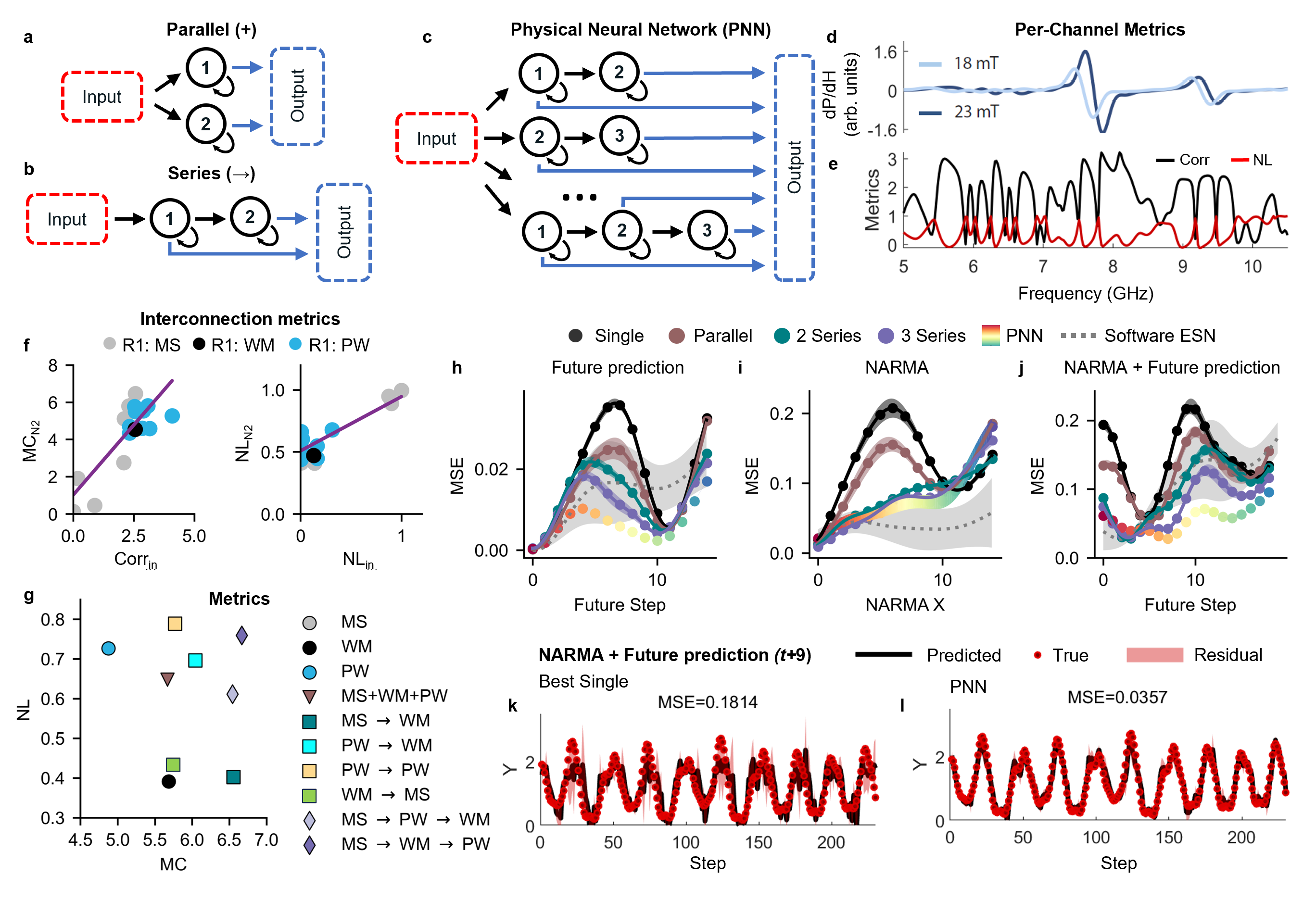}
    \caption{\textbf{Multilayer physical neural networks with complex nodes.} Schematics of a) parallel (+) networks, b) series ($\rightarrow$ networks and c) physical neural networks (PNN). Network nodes are recurrent, non-linear nanomagnetic reservoirs with high output dimensionality. In parallel networks, input is fed to multiple nodes independently. In series, the output of one node is virtually fed as input to the next. PNN networks combine series networks in parallel. All interconnections are made virtually, as opposed to physical interconnection. The response of every node is combined offline to create the output of a network.
    d) WM FMR amplitudes at maximum (dark blue) and minimum (light blue) input fields. 
    e) Frequency-channel signal correlation (Corr) to previous time steps and nonlinearity (NL). These metrics are used to guide which frequency channel is used as input for the next node in a series network.
    f) Relationship between first reservoir node output metrics, correlation (Corr\textsubscript{in}) and nonlinearity (NL\textsubscript{in}), used for interconnections and measured second reservoir metrics, memory-capacity (MC\textsubscript{R2}) and nonlinearity (NL\textsubscript{R2}), when the second reservoir (R2) is WM. Lines represent linear fits. Higher metric scores are correlated with higher-scoring reservoir output channels.
    g) Memory-capacity (MC) and nonlinearity (NL) of selected single (circles), parallel (triangles) and series networks (squares for series length 2, diamonds for series length 3). Networks have a broad enhancement of metrics versus single arrays. PNN's can take any metric combination.
    MSE profiles for h) Mackey-Glass future prediction, i) NARMA transformation and j) future prediction of NARMA-7 processed Mackey-Glass for the best single (WM), parallel (MS+WM+PW), 2 series (MS$\rightarrow$WM), 3 series (MS$\rightarrow$WM$\rightarrow$PW), and PNN. Also shown is the performance of a software echo-state network with 100 nodes. MSE profiles are significantly flattened for the PNN. Shading is the standard deviation of the prediction of 10 feature selection trials or 10 echo-state networks. k,l) Example predictions for \textit{t}+9 of the NARMA7 processed Mackey-Glass signal. Shading represents the residual of the prediction
    }
    \label{RC_network}
\end{figure*}
To overcome the limitations of single physical systems, we now interconnect individual reservoirs to form networks, where each reservoir acts as a complex node with memory, nonlinearity and high output-dimensionality (as opposed to traditional neural network nodes which have no memory and only 1 output dimension). 

We begin by making parallel and 1D series networks. In parallel networks, inputs and readout for each array are performed independently and combined offline (Figure \ref{RC_network} a). In series networks, the output from one node is used as input to the next (Figure \ref{RC_network} b). In this work, interconnections are made virtually, as opposed to real-time interconnection. We experimentally record all data from the first node and select one output channel to be used as input to the next node. After scaling to an appropriate field range, this time-series is then passed to the next node and its response is recorded. The full readout from both layers is then combined offline during training and prediction. We trial 49 total architectures (4 parallel and 36 2-series and 9 3-series networks with different configurations of arrays interconnections). Full details of each architecture can be found in the supplementary information. We combine the outputs of every series / parallel architecture offline to create a network that is analogous to a physical neural network (PNN) (Figure \ref{RC_network} c). 

As each node has $\sim$200 readout channels and the optimisation time required to evaluate all interconnections is not practically feasible. Seeking a more efficient solution, we explore how the output-channel characteristics from one node affect the memory-capacity and nonlinearity of the of the next node in a series network. Figures \ref{RC_network} d,e) show the per-channel nonlinearity (NL) and correlation (Corr) to previous inputs values for WM (see Methods for calculation details). Certain frequency channels are highly correlated with previous inputs (e.g. 7.9 - 8.1 GHz) whereas others are highly nonlinear (e.g. 7 GHz). By feeding the output of one channel as input to the next node we can evaluate how output characteristics affect the next node's memory-capacity and nonlinearity. Figure \ref{RC_network} f) shows this relationship when the second node is WM (other architectures shown in Supplementary Note 4). Corr\textsubscript{in} and NL\textsubscript{in} refer to the correlation and nonlinearity of the first node output channel. MC\textsubscript{N2} and NL\textsubscript{N2} are the memory-capacity and nonlinearity of the second node. Points are colour-coded depending on which array acts as the first node (consistent with previous Figures).  Both MC\textsubscript{N2} and NL\textsubscript{N2} follow an approximately linear relationship. A linear relationship is also observed when comparing correlations to specific previous inputs (Supplementary Note 5). As such, one may tailor the overall network metrics by selecting output-channels with certain characteristics. This interconnection control goes beyond conventional reservoir computing where interconnections are made at random\cite{manneschi2021exploiting}, allowing controlled design of network properties.

Figure \ref{RC_network} g) shows the memory-capacity and nonlinearity of selected single, parallel and series sub-networks. Parallel network nonlinearity and memory-capacity (triangular markers) lie between the nonlinearity and memory-capacity from the constituent reservoirs. Memory-capacity does not increase in parallel as no information is directly transferred between network nodes. Series networks (square and diamond symbols for series network depth of 2 and 3 respectively) show memory and nonlinearity improvements above any single reservoir. The MS$\rightarrow$PW$\rightarrow$WM network (purple diamond) has both high memory and nonlinearity. PNN's can be devised to possess any metric combination as it comprises all sub networks. The ordering of arrays in series networks is important, with memory enhancement only observed when networks are sequenced from low (first) to high (last) memory (e.g. MS$\rightarrow$WM has a larger memory capacity than WM$\rightarrow$MS), a phenomenon also seen in software reservoirs\cite{manneschi2021exploiting} and human brains\cite{hasson2015hierarchical}. To date, there has been an open question as to whether physical reservoirs are analogous to software echo-state networks. These results suggest that the two are comparable, and that methods used to improve software echo-state networks can be transferred to physical reservoir computing.

We now evaluate the performance of these networks using the feature selection methodology\cite{manneschi2021exploiting,manneschi2021sparce} described in the Methods section, Learning Algorithms. 
Figures \ref{RC_network} h-j) compares the MSE for the best configuration of each network architecture when predicting future values of the Mackey-Glass equation (h), performing a NARMA transform (i) and combining NARMA transform and prediction (j) (additional architectures and tasks are shown in Supplementary Note 6). Parallel arrays (brown lines) do not show significant MSE reductions for these tasks as there is no information transfer and memory increase in this architecture. 2 and 3 series networks show significant decreases in MSE for all tasks, which improve as the series network is extended. The PNN outperforms other architectures across all future time step prediction tasks (Figures \ref{RC_network} h,j) with significant MSE vs $t$ flattening demonstrating higher-quality prediction. This is particularly evident when predicting \textit{t}+9 of the NARMA transformed Mackey-Glass equation (Figures \ref{RC_network} k,l) and when reconstructing the Mackey-Glass attractor (Supplementary Note 7). For NARMA transform (Figure \ref{RC_network} h), both the PNN and three-series network show a linear MSE vs $t$ profile, indicative of reaching optimum performance. Additionally, the PNN performs well across a host of nonlinear signal-transformation tasks (Supplementary Note 6), outperforming other network architectures at 9/20 tasks.

A strength of the PNN is the increased readout dimensionality. The results in Figure \ref{RC_network} h-j) use 40 - 100 outputs for the single, parallel and series networks and $\sim$13,500 outputs for the PNN. This number of outputs is far greater than the size of the training set, placing the PNN in the overparameterised regime (discussed later). In Supplementary Note 8, we constrain the PNN to have a similar number of outputs to the other architectures and find performance is similar to that of a three-series network. As a comparison, the grey dashed line in h-j) represents the average performance of a 50 randomly initialised 100-node echo state networks (ESN) (further software comparison provided in Supplementary Note 9). We find that single arrays are well matched to ESNs with $\sim$ 20 nodes and PNNs are well matched to ESNs with 500 nodes. Additionally, we initialise three ESNs with similar characteristics to nanomagnetic arrays and interconnect them. Interconnected ESNs, whilst functionally different, also display improvements during interconnection, highlighting both the general nature of the technique, and strengthening the analogy between physical reservoirs and software echo-state networks.

\subsection*{Overparameterisation}
\begin{figure*}[htb!]
    \includegraphics[width=\textwidth]{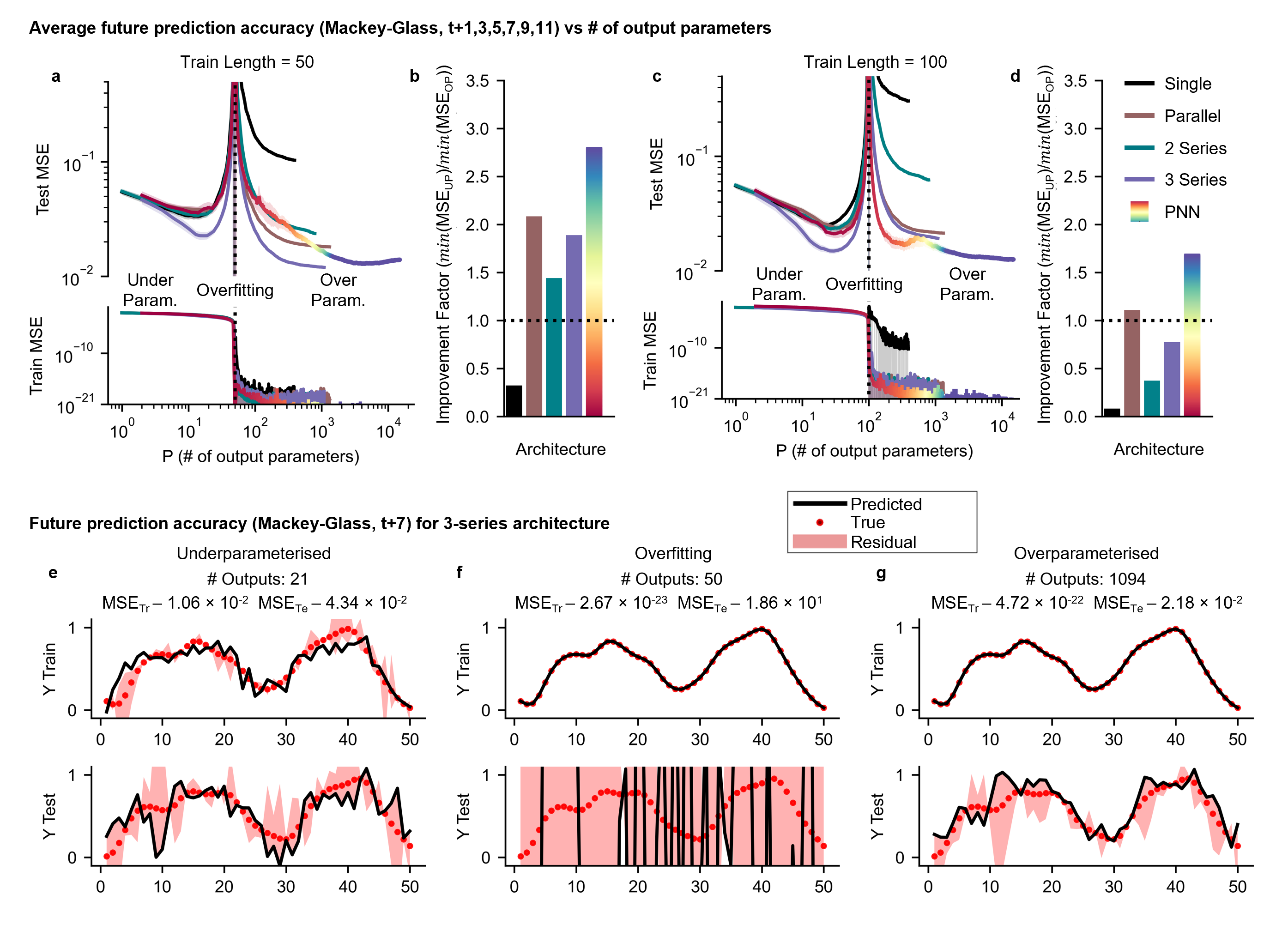}
    \caption{\textbf{Neuromorphic overparameterisation.} Train and test MSE for selected architectures when predicting future values of the Mackey-Glass equation when varying the number of output parameters P (i.e. FMR output channels) for training set lengths of a) 50 and c) 100. MSE is the mean of 50 random combinations of outputs across 6 different prediction tasks (Mackey-Glass, \textit{t}+1,3,5,7,9,11). Shading is the standard error of the MSE over 50 trials of randomly selecting features. b,d) Improvement factors (\(min\)(MSE\textsubscript{UP}) / \(min\)(MSE\textsubscript{OP})) when moving from underparameterised (UP) regime to overparameterised (OP) regime. Each network shows three regimes: an underparameterised regime when the number of parameters is small, an overfitting regime where the number of parameters is near the size of the training set and an overparameterised regime when the number of parameters exceeds the size of the training set. Example train and test predictions for 3 series architecture in the underparameterised regime (e), overfitting regime (f) and overparameterised regime (g).
    }
    \label{OP}
\end{figure*}
We now explore the computational benefits of reaching an overparameterised regime, where the number of network outputs exceeds the size of the training dataset. To reach overparameterisation there are three key parameters:  the size of the length of the training dataset, the number of output parameters, which here refers the number of FMR channels, used during the training phase, and the effective dimensionality of the readout. The last point is critical, as the number of parameters can be arbitrarily increased by increasing the measurement resolution or repeating measurements but doing so results in increasingly correlated outputs which have no impact during training. Here, we show how engineering a rich readout response can be used to reach overparameterisation in physical systems.

Figures \ref{OP} a,c) shows the train and test MSE for the best networks from each architecture (equivalent to Figure \ref{RC_network}) when varying the number of output parameters for training set lengths of a) 50 and c) 100. Figures \ref{OP} b,d) show the improvement factors (min(MSE\textsubscript{UP})/min(MSE\textsubscript{OP})) when comparing the MSEs between the underparameterised (UP) and overparameterised (OP) regime. All architectures beyond single systems show improvements when operating in the overparameterised regime. Here, MSE is an average over 6 prediction tasks (Mackey-Glass, \textit{t}+1,3,5,7,9,11) and 50 random trials of adding parameters (see Methods for details).

For each network and training length, we see three regimes with example predictions for each regime shown in Figures \ref{OP} e-g) for the 3 series architecture: When the number of output parameters is less than the training length, the system is underparameterised and train and test MSE decrease as the number of parameters increases (Figure \ref{OP} e)). This is where the majority of neuromorphic schemes operate. When the number of output parameters is close to the length of the training set, the networks enter an overfitting regime where MSE follows a ‘U-trend’ and performance deteriorates with increasing number of outputs. Here, there are multiple solutions during training and the network is able to perfectly fit the target, leading to vanishingly small training errors. However, the fitted weights are arbitrary and when presented with unseen data, produce predictions with no resemblance to the target (Figure \ref{OP} f). Surprisingly, this trend does not continue as the the number of outputs increases. Instead, the test performance substantially improves. Here, the network enters an overparameterised regime  and overcomes overfitting\cite{belkin2019reconciling,nakkiran2021deep,zou2020gradient,zou2019improved}. Instead of memorising the training data, the network can generalise and learn the underlying behaviour of the task, resulting in improved test performance (Figure \ref{OP} g). Crucially, for some networks, the overparameterised MSE is lower than the underparameterised MSE. This phenomenon has been observed in software deep learning (sometimes referred to as a double-descent phenomena\cite{belkin2019reconciling,nakkiran2021deep}), but not physical computing systems. 

The extent to which MSE recovers depends on the effective dimensionality of the readout and train length. For single reservoirs, the limited internal dynamics produce a highly correlated set of outputs. Whilst MSE reduces in the overparameterised regime, the network is unable to fully overcome the effects of overfitting resulting in high overparameterised MSE and improvement factors below 1. The networking approach increases the effective dimensionality of the readout. In parallel networks this is achieved by having distinct nodes with different dynamic responses. In series networks, nodes receive unique inputs which produce different dynamic regimes. For short training lengths (Figures \ref{OP} a,b), the enhanced output dimensionality produces a beneficial overparameterised regime with improved MSE. Improvements increase as the network size increases and even series networks with one node type can achieve a beneficial overparameterised regime (Supplementary Note 10). For longer train lengths (Figures \ref{OP} c,d), only the PNN reaches a beneficial overparameterised regime due to the increased effective readout dimensionality required to overcome overfitting. Task also plays a role, with more challenging prediction tasks demanding higher effective readout dimensionality to reach a beneficial overparameterised regime (Supplementary Note 10).
Interestingly, the PNN architecture shows signatures of triple descent (\textit{N}\textsubscript{train} = 50, P $\sim$ 150 and \textit{N}\textsubscript{train} = 100, P $\sim$ 600) where a peak in MSE is observed in the overparameterised regime, a sign of superabundant overparameterisation\cite{adlam2020neural}. Here we use linear regression to train networks weights. When using gradient-descent, a similar trend is observed showing the results are robust to different training methods (Supplementary Note 11) 

These results can be applied to any arbitrary system to improve computation and reach an overparameterised state. To produce a diverse set of outputs, both physical system and readout should be designed in tandem. Readout improvements can be obtained by increasing measurement resolution, fabricating multiple electrodes or reading at different external biases (e.g. here we measure FMR at the input field inducing field dependent resonant shifts). Moving away from homogeneous systems will produce a greater breadth of internal dynamics, improving performance and enabling a useful overparameterised regime. Examples include introducing structural variation throughout the system (as with the PW array here) or biasing physically separated regions. 

The networking approach is an effective method of improving readout dimensionality. Combining distinct systems is beneficial as the breadth of dynamics is likely to be higher, but networking the same physical system can produce overparameterisation. Enriched dynamics and increased effective readout dimensionality can be achieved by operating the physical system in different dynamic regimes, for example, by changing external conditions to access different dynamics\cite{lee2023task}. Alternatively, by varying the input sequence, either through a random mask (a common technique known as virtual nodes) or by feeding the output of one node as input to another, a diverse parallel / series network can be created. Even in the limit of a single physical system with one readout per time-step, overparameterisation can be achieved through a combination of parallel and series networking, provided that the internal dynamics are rich enough.

\subsection*{Learning in the overparameterised regime} 

\begin{figure*}[htb!]
    \centering
    \includegraphics[width=\textwidth]{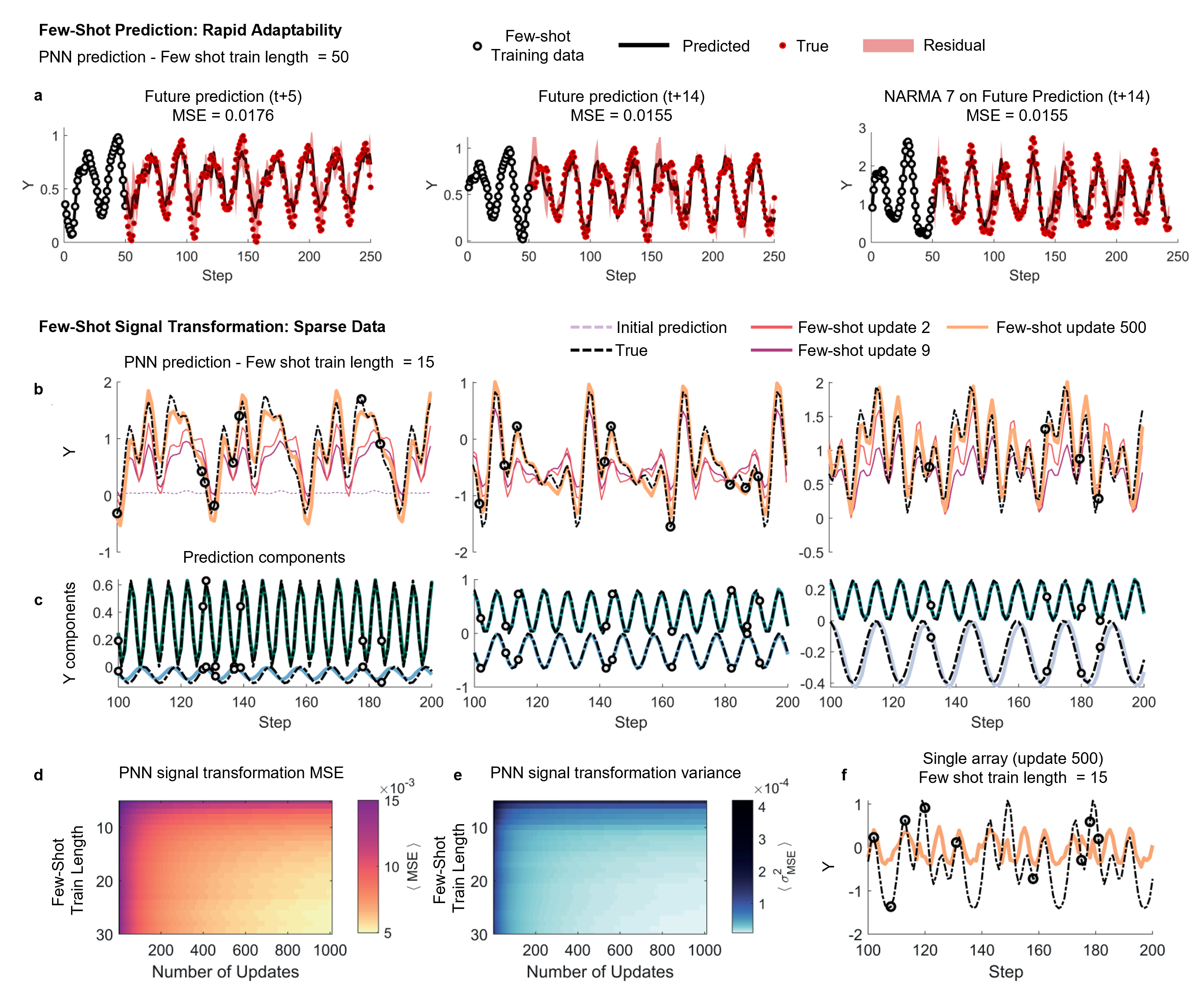}
    \caption{\textbf{Few-shot learning.} a) System predictions (black lines) in the overparameterised regime when rapidly adapting to 50 training data-points (grey circles) at the beginning of the series. The PNN is able to rapidly learn the task and shows strong performance on the test data set (target points in red). Shading is the residual between true and predicted values.  b) Sine transformations with sparse training data.  The system is asked to learn five different frequency components of a time-varying function (dashed black line). Here, the PNN is shown just 15 sparsely separated training data points (grey circles). As the system is updated (coloured plots), the PNN prediction steadily improves. c) Two example frequency components of the targets in b) which the PNN learns. d) and e) Average error, < MSE >, and variance of the error, < $\sigma$\textsuperscript{2} MSE >, computed over 500 signal transformation tasks as the train length and updates vary. The system shows good generalisation across all tasks as shown by the low variance. f) Few-shot signal transformation using a single array after 500 updates. The single array fails completely. 
    }
    \label{Metalearning}
\end{figure*}

We now showcase the computational advantages of physical neural networks operating in the overparameterised regime. The high dimensionality and complexity of the network readout permits rapid learning with a limited number of data-points\cite{belkin2019reconciling,nakkiran2021deep,zou2020gradient,zou2019improved,finn2017model}.
This characteristic is a particularly desirable feature for any neuromorphic computing system as it allows rapid adaptation to changing tasks/environments in remote applications where collecting long training datasets carries a high cost. To demonstrate this, we show a challenging fast few-shot learning adaptation for previously unseen tasks using a model-agnostic meta learning approach\cite{wang2020generalizing,finn2017model}.

Figure \ref{Metalearning} a) shows the system prediction when predicting (left-to-right) the $t+5$, $t+14$ and NARMA-processed $t+14$ Mackey-Glass signals. The PNN is trained on just the first 50 data points of the signal, highlighted by black circles. The PNN is able to learn the underlying system dynamics and provide good predictions demonstrating the power and adaptability of the overparameterised regime. We note that the MSE's achieved are are comparable to those in Figures \ref{RC_network} e,g) when using feature selection with the PNN. As such, this meta-learning can achieve strong results with a 75$\%$ reduction in training set size (50 training data points here vs 200 previously used).

To further showcase the computational capabilities of the PNN, we now demonstrate few-shot learning where the seen training data are sparsely distributed throughout the target dataset, representing for instance a very low sampling rate of a physical input sensor in an edge-computing use-case. In Figure \ref{Metalearning} b), the system is driven by a sinusoidal input and asked to predict a target of the form \textbf{\~{y}}(t)=$\sum_{n}^{N\textsuperscript{$\omega$}}$ $a_{n}\sin$($nt$ + $\theta_{n}$) i.e. simultaneously predict amplitude, frequency and phase changes - a task that requires a range of temporal dynamics in the network, often used as a meta-learning benchmark task\cite{finn2017model}. The values $a_n$ and $\theta_n$ are sampled randomly at the beginning of each task from continuous uniform distributions (details in the Methods section, Task Selection). The goal is to train the system to generalise to a set of $a_n$ and $\theta_n$ values, and then rapidly adapt to a new task with a limited set of sparse training points. 

In all previous tasks, the network is trained to produce a single output. Here, we simultaneously adapt to five distinct functions and sum the predictions to produce the final waveform. This increases the task difficulty as the network must be generalised to all possible amplitude, frequency and phase shifts, and any errors will be amplified in the final output. To achieve this, we use a variation of the MAML meta-learning algorithm\cite{finn2017model} applied to the frequency-channel outputs of the network, leaving all history-dependent and non-linear computation to the intrinsic dynamics of the physical network. 

Figures \ref{Metalearning} b) and c) show predictions of three example tasks for the overall target and two example sub-components from each task respectively. The network sees just 15 data points (highlighted by grey circles in panel b)) throughout the entire process. In Figure \ref{Metalearning} b),  the target (dashed black line) and predicted values after updating the generalised matrix a number of times are shown. At update 0, the predicted response does not match the target waveform as expected. As the network updates, the error between the prediction and target reduce. Further example tasks are provided in Supplementary Note 12.

Despite the limited information and the high variability of tasks, the network learns the underlying sinusoidal components and can adapt to different targets with the partial information available. To support this claim of generalisation, Figures \ref{Metalearning} d) and e) show the average error < MSE >, and variance of the error, < $\sigma$\textsuperscript{2} MSE >, respectively, calculated over 500 different tasks for various few-shot train lengths and number of updates. The error decreases as the number of updates and available data points is increased as expected, with strong performance observed for as little as 10 training data points. Crucially, the variance of the error across all tasks is low, demonstrating strong generalisation. Finally, Figure \ref{Metalearning} f) reports an example of prediction obtained when meta-learning is applied on a single reservoir, which fails completely at the task.

The meta-learning approach showcases the richness of the high-dimensional output space of the PNN architecture. The network is able to learn the general behaviour and dynamics of the input and a set of tasks enabling rapid adaptability. This removes the requirement for complete retraining, making essential progress toward on-the-fly system reconfiguration. 

\section*{Discussion}
We have demonstrated how interconnecting physical systems into larger networks can improve computational performance across a broad task-set as well as enhance task-agnostic computational metrics. We now discuss the challenges which must be addressed before the PNN architecture can be realised at a future device level, as well as the prospects for scaling nanomagnetic hardware.

Currently, the PNN is created offline, with the entire dataset for each node recorded sequentially, as opposed to each data point being passed through the entire network during each time step i.e. node 1 receives the entire dataset, then node 2 receives the output of node 1. Furthermore, a 48 node PNN is created with three unique physical systems, where each physical systems state is reset once the data has been passed through. In a future device, each data point must travel through the PNN in real-time, requiring 48 unique physical systems and measurement set-ups. Therefore, as PNN size increases, so does device complexity. Additionally, outputs from all nodes in the network are recorded and used during the learning phase which, in practice, requires surplus measurement time and memory. Our technique has been tested on systems which take 1D input. For physical systems with >1D inputs, we expect the technique to be equally, if not more, powerful.

Realising a device level PNN requires a reduction in the number of nodes and the number of measured outputs without sacrificing computational performance. This can be achieved by increasing the number of unique physical reservoirs thereby increasing the range of beneficial reservoir dynamics that each node has. Furthermore, the number of measured outputs can be optimised to only record significantly uncorrelated outputs for each node. Optimisation of PNN architecture, and inter-node connectivity (i.e. combining multiple output channels as input to a particular node) will allow further improvements.

We now assess the scalability of the nanomagnetic array-based computing scheme presented in this work. The computing architecture can be separated into three components: input, readout and weight multiplication. At present we use global magnetic field inputs which are unsuitable for scaling due to the large power, long rise-time and large spatial footprint. Ferromagnetic resonance-based readout, whilst theoretically fast, is currently achieved with a laboratory scale RF-source and a lock-in amplifier, resulting in slow data throughout ($\sim$ 20 s for one spectra) and high energy cost (Supplementary Note 13 calculates the power, energy and time for our current scheme compared to conventional processors). To improve the technological prospects of our scheme, input must be delivered on-chip, and readout powers and speeds must be reduced to be competitive with alternative approaches.

We now propose and benchmark potential ways to translate nanomagnetic arrays to device level (extended analysis and discussion is provided in Supplementary Note 13). We assume a nanomagnetic array with dimensions of \SI{5}{\micro\meter}m  $\times$ \SI{5}{\micro\meter} ($\sim$100 elements). By patterning the array on top of a current-carrying microstrip, array-specific Oersted fields can be generated. For a \SI{5}{\micro\meter}  $\times$ \SI{5}{\micro\meter} array patterned on top of an Cu microstrip, with width $w$ = \SI{5}{\micro\meter} and thickness $t$ = 50 nm in height. The current required to produce a 25 mT Oersted field is $\sim$ 200 mA \cite{kiermaier2012electrical} (see Supplementary Note 13 for calculation details). For Cu resistivity of 16.8 n$\Omega$ m\cite{matula1979electrical}, the resistance is R = 0.336 $\Omega$ consuming a minimum power P = 13.4 mW. For a pulse time of 1 ns, this equates to an energy of 13.4 pJ per input. Nanomagnet coercivity and therefore power consumption can be reduced by fabricating thinner nanomagnets. 

Instead, we can fabricate arrays in contact with a high spin-Hall angle material to switch macrospins via spin-orbit torque\cite{fukami2016spin}. This can be achieved with Ta thickness and magnetic layer thicknesses of around 5 nm and 1-2 nm respectively\cite{fukami2016spin} and current densities on the order of 10$^{11}$ Am\textsuperscript{-2}. Whilst reducing thickness reduces dipolar coupling, arrays of nanomagnets remain highly correlated at these thicknesses\cite{farhan2013direct,kapaklis2014thermal} and will retain the collective processing that thicker nanomagnets exhibit. For a \SI{5}{\micro\meter}  $\times$ \SI{5}{\micro\meter} Ta strip with 5 nm thickness, the corresponds to a current of 2.5 mA. For Ta resistivity of 131 n$\Omega$m\cite{milovsevic1999thermal}, R\textsubscript{Ta} = 26.2 $\Omega$ and P = 0.164 mW. Here, switching times can be as low as 250 ps \cite{bhowmik2014spin} giving input energies of 41 fJ. Spin-orbit torques decrease linearly with increasing ferromagnet thickness \cite{manchon2019current}. For 20 nm thick nanomagnets used in this work, the current density required to switch an element will be on the order of 10 - 40 $\times$ (based on micromagnetic simulations) due to thicker elements and higher coercive fields, giving currents in the range of 25 - 100 mA and powers in the range of 16.8 - 269 mW. Limited spin-diffusion in ferromagnets may further increase or even prevent switching in thicker elements. As such, for thicker nanomagnets, Oersted field switching may be preferable. 

Microwave readout can be implemented either sequentially (frequency-swept RF source) or in parallel (multiple channels simultaneously) depending on the technique.  For sequential readout out (as implemented in this work), we can utilise spin-torque FMR which converts magnetisation resonance to DC voltage. Assuming microwave powers in the range of 1 - 10 mW, the generated voltage at resonance is on the order 1 - 10 µV which can be amplified with a CMOS low noise high gain amplifier. Readout time per channel is theoretically limited to the time taken to reach steady precession ($\sim$ 14 ns\cite{ross2023multilayer}). As the generated voltage is DC, spectral information is lost, hence each channel must be recorded sequentially, giving a measurement time of $\sim$2.6 µs and energy of 2.8 nJ (Supplementary Note 13), dominating over the input speeds. Alternatively, we can pass mixed frequency RF-signals through a device patterned on a co-planar waveguide and detect the microwave absorption, for example via an RF noise source. Here, the readout power is higher (150 mW) but speed and energy are reduced (14 ns and 2.1 nJ respectively, Supplementary Note 13). The output signal can, for example, be sent to multiple spin-torque oscillators tuned to different frequencies serving as RF-diodes\cite{leroux2022convolutional,ross2023multilayer} allowing parallel detection of multiple channels at $\sim$ ns timescales.
Weight multiplication can be achieved by passing output signals to memristor cross-bar arrays which routinely operate with nW - µW powers\cite{wang2020pure}. As such, the power consumption, energy and time required to perform an operation of our proposed device is on the order of 1 - 160 mW, 1.5 - 2,800 ns and 2.1 - 2.9 nJ depending on the input and readout method used. At the lower end, this is within the power range for battery-less operation using energy harvesting technology\cite{shaikh2016energy, jebali2024powering}. For PNNs, nodes are measured sequentially hence the total operating power remains unchanged, and time and energy scale linearly with the number of nodes. Simultaneous measurement of physical nodes would increase power and reduce time. Supplementary Note 13 provides further comparison to CMOS hardware by calculating the number of FLOPs to update an echo-state network (the closest software analogue to nanomagnetic hardware). Our calculations indicate that, for large echo-state networks, the proposed devices have potential to operate at lower energy costs that conventional hardware.

To conclude, here we mitigate the limitations of physical reservoir computing by engineering networks of physical reservoirs with distinct properties.
We have engineered multiple nanomagnetic reservoirs with varying internal dynamics, evaluating their computational metrics and performance across a broad benchmark taskset. Our results highlight the computational performance gained from enriched state spaces, applicable across a broad range of neuromorphic systems. PW outperforms MS by up to 31.6$\times$, demonstrating careful design of system geometry and dynamics is critical, with computational benefits available via physical system design optimisation.

We then constructed a physical neural network from a suite of distinct physical systems where interconnections between reservoirs are made virtually and outputs are combined offline, overcoming the fundamental limitation of the memory/nonlinearity tradeoff that roadblocks neuromorphic progress.  We demonstrate that methods used to improve computational performance in software echo-state networks can be transferred to physical reservoir computing systems. The modular, reconfigurable physical neural network architecture enables strong performance at a broader range of tasks and allows the implementation of modern machine learning approaches such as meta-learning.
The high dimensionality enabled by the physical neural network architecture allows us to demonstrate the benefits of operating physical neuromorphic computing in an overparameterised regime to accomplish few-shot learning tasks with just a handful of distinct physical systems. Currently, network interconnections are made virtually. We expect the same performance improvements to be achieved with the next-generation of our scheme with real-time physical interconnections. Networks of complex nodes described here are largely unexplored, as such many open questions remain. Exploration of different network architectures from both a computational and device architecture perspective is crucial for optimising performance and for device fabrication where increasing the number of nodes and interconnections comes at a cost.

Our method of interconnecting network layers via assessing output-channel/feature metrics allows tailoring of the network metrics, bypassing costly iterative approaches.  The approach is broadly-applicable across physical neuromorphic schemes. If the required memory-capacity and nonlinearity for a given task are known, metric programming allows rapid configuration of an appropriate network. If the required memory-capacity and nonlinearity are not known or the task depends on more than these metrics, metric programming can be used to search the memory-capacity, nonlinearity phase space for pockets of high performance. The introduction of a trainable inter-layer parameter opens vast possibilities in implementing hardware neural networks with reservoirs serving as nodes and inter-layer connections serving as weights\cite{manneschi2024optimising}.

\newpage

\section*{Methods}
The Methods section is organised as follows: Experimental Methods includes the fabrication of samples, measurement of FMR response and the details of implementing reservoir computing and interconnecting arrays. Following this we discuss the tasks chosen and how they are evaluated in Task selection. Learning algorithms then provides a detailed description of the learning algorithms used in this work.  

\subsection*{Experimental Methods}
\subsubsection*{Nanofabrication}

Artificial spin reservoirs are fabricated via electron-beam lithography liftoff method on a Raith eLine system with PMMA resist. 25 nm Ni$_{81}$Fe$_{19}$ (permalloy) is thermally evaporated and capped with 5 nm Al$_2$O$_3$. For WM, a staircase subset of bars are increased in width to reduce its coercive field relative to the thin subset, allowing independent subset reversal via global field. For PW, a variation in widths are fabricated across the sample by varying the electron beam lithography dose. Within a \SI{100}{\micro\meter} $\times$ \SI{100}{\micro\meter} write-field, the bar dimensions remain constant.
The flip-chip FMR measurements require mm-scale nanostructure arrays. Each sample has dimensions of roughly $\sim$ 3 $\times$ 2 mm\textsuperscript{2}. As such, the distribution of nanofabrication imperfections termed quenched disorder is of greater magnitude here than typically observed in studies on smaller artificial spin systems, typically employing 10-100 micron-scale arrays. The chief consequence of this is that the Gaussian spread of coercive fields is over a few mT for each bar subset. Smaller artificial spin reservoir arrays have narrower coercive field distributions, with the only consequence being that optimal applied field ranges for reservoir computation input will be scaled across a corresponding narrower field range, not an issue for typical 0.1 mT or better field resolution of modern magnet systems. 
\subsubsection*{Magnetic Force Microscopy Measurement}

Magnetic force micrographs are produced on a Dimension 3100 using commercially available normal-moment MFM tips.
\subsubsection*{Ferromagnetic Resonance Measurement}

Ferromagnetic resonance spectra are measured using a NanOsc Instruments cryoFMR in a Quantum Design Physical Properties Measurement System. Broadband FMR measurements are carried out on large area samples $(\sim 3 \times 2~ \text{ mm}^2)$ mounted flip-chip style on a coplanar waveguide. The waveguide is connected to a microwave generator, coupling RF magnetic fields to the sample. The output from waveguide is rectified using an RF-diode detector. Measurements are done in fixed in-plane field while the RF frequency is swept in 20 MHz steps. The DC field are then modulated at 490 Hz with a 0.48 mT RMS field and the diode voltage response measured via lock-in. The experimental spectra show the derivative output of the microwave signal as a function of field and frequency. The normalised differential spectra are displayed as false-colour images with symmetric log colour scale.

\subsubsection*{Data Input and Readout}
Reservoir computing schemes consist of three layers: an input layer, a hidden reservoir layer, and an output layer corresponding to globally-applied fields, the nanomagnetic reservoir and the FMR response respectively. 
For all tasks, the inputs are linearly mapped to a field range spanning 35 - 42 mT for MS, 18 - 23.5 mT for WM and 30-50 mT for PW, with the mapped field value corresponding to the maximum field of a minor loop applied to the system. In other words, for a single data point, we apply a field at +$\mathbf{H}_{\rm app}$ then -$\mathbf{H}_{\rm app}$. 
After each minor loop, the FMR response is measured at the applied field -$\mathbf{H}_{\rm app}$ between 8 - 12.5 GHz, 5 - 10.5 GHz and 5 - 10.5 GHz in 20 MHz steps for MS, WM and PW respectively. The FMR output is smoothed in frequency by applying a low-pass filter to reduce noise. Eliminating noise improves computational performance\cite{gartside2022reconfigurable}. For each input data-point of the external signal $s(t)$, we measure $\approx 300$ distinct frequency channels and take each channel as an output. This process is repeated for the entire dataset with training and prediction performed offline. 
\subsubsection*{Interconnecting Arrays}
When interconnecting arrays, we first input the original Mackey-Glass or sinusoidal input into the first array via the input and readout method previously described. We then analyse the memory and nonlinearity of each individual frequency output channel (described later). A particular frequency channel of interest is converted to an appropriate field range. The resulting field sequence then applied to the next array via the computing scheme previously described. This process is then repeated for the next array in the network. The outputs from every network layer are concatenated for learning.

\subsection*{Task Selection}
Throughout this manuscript, we focus on temporally-driven regression tasks that require memory and nonlinearity. 
Considering a sequence of T inputs $s(t)$ = $\big [s(1),s(2),...,s(\rm{T}) \big]$, the physical system response is a series of observations $o(t)$ = $\big [o(1),o(2),...,o(\rm{T}) \big]$ across time. These observations can be gathered from a single reservoir configuration as in Figure \ref{Samples}, or can be a collection of activities from multiple reservoirs, in parallel or inter-connected as in Figure \ref{RC_network}.
In other words, the response of the system $o(t)$ at time t is the concatenation of the outputs of the different reservoirs used in the architecture considered.
The tasks faced can be divided into five categories:

\textbf{Sine Transformation Tasks.} The system is driven by a sinusoidal periodic input $o(t)=\sin(t)$ and asked to predict different transformations, such as $\tilde{y}(t)=\Big(|\sin(x/2)|, \sin(2x), \sin(3x), \sin^{2}(x),\sin^{3}(x), \cos(x), \cos(2x), \cos(3x), \rm{saw}(x), \rm{saw}(2x),... \Big)$. 
The inputs $[s(t), s(t+\delta t), ...]$ are chosen to have 30 data points per period of the sinusoidal wave, thus with $\delta t=2\pi/30$. The total dataset size is 250 data points. If the target is symmetric with respect to the input, the task only requires nonlinearity. If the target is asymmetric, then both nonlinearity and memory are required. 

\textbf{Mackey-Glass Forecasting.} The Mackey-Glass time-delay differential equation takes the form $\dfrac{ds}{dt} = \beta \dfrac{s_{\tau}}{1+s_{\tau}^{n}} - \lambda s$ and is evaluated numerically with $\beta$ = 0.2, n = 10 and $\tau$ = 17. Given $s(t)$ as external varying input, the desired outputs are $\tilde{y}(t)=\Big(s(t+\delta t), s(t+2\delta t),..., s(t+\rm{M}\delta t) \Big)$, corresponding to the future of the driving signal at different times. We use 22 data points per period of the external signal for a total of 250 data points. This task predominantly requires memory as constructing future steps requires knowledge of the previous behaviour of the input signal.

\textbf{Non-Linear Auto-Regressive Moving Average Tasks.} Non-linear auto-regressive moving average (NARMA) is a typical benchmark used by the reservoir computing community. The definition of the x-th desired output is $\tilde{y}_x(t) = \rm{NARMA}\Big\{s(t')|x\Big\}=As(t'-1) + Bs(t'-1) \sum_{n=1}^{x}s(t'-n) + Cs(t'-1)s(t'-x) + D$, where the constants are set to A = 0.3, B =  0.01, C = 2, D = 0.1. The input signal $s(t')$ is the Mackey-Glass signal, where the variable $t'$ is introduced to account for a possible temporal shift of the input. For $t'=t$,  $\tilde{y}_x(t)$ is the application of NARMA on the Mackey-Glass signal at the current time t, while for $t'=t+10\delta t$,  $\tilde{y}_x(t)$ is the result of NARMA on the input signal delayed by ten time steps in the future. The index $x$ can instead vary between one, defining a task with a single temporal dependency, and fifteen, for a problem that requires memory of fifteen inputs. Varying $x$ and $t'$, we can define a rich variety of tasks with different computational complexity. 

\textbf{Evaluations of Memory-Capacity and Nonlinearity.} Memory capacity and nonlinearity are metrics frequently used for the characterization of the properties of a physical device. While these metrics do not constitute tasks in the common terminology, we include them in this section for simplicity of explanation. Indeed, we use the same training methodology to measure MC and NL as in the other tasks faced. We evaluate these metrics with the Mackey-Glass time-delay differential equation as input. This gives results that are correlated to conventional memory-capacity and nonlinearity scores, with some small convolution of the input signal - negligible for our purposes of relatively assessing artificial spin reservoirs and designing network interconnections. 

For memory-capacity the desired outputs are $\tilde{y}(t)=\Big(s(t-\delta t), s(t-2\delta t),..., s(t-k\delta t) \Big)$, corresponding to the previous inputs of the driving signal at different times. To avoid effects from the periodicity of the input signal, we set $k$=8. The R$^{2}$ value of the predicted and target values is evaluated for each value of $\delta t$ where a high R$^{2}$ value means a good linear fit and high memory and a low R$^{2}$ value means a poor fit a low memory. The final memory-capacity value is the sum of R$^{2}$ from $k$ = 0 to $k$ = 8.

For nonlinearity, $x(t)$ is the input signal from t = 0 to t = -7 and  $\tilde{y}(t)$ is a certain reservoir output. For each output, the R$^{2}$ value of the predicted and target values is evaluated. Nonlinearity for a single output is given by 1 - R$^{2}$ i.e. a good linear fit gives a high R$^{2}$ and low nonlinearity and a bad linear fit gives low R$^{2}$ and high nonlinearity. Nonlinearity is averaged over all selected features.
    
Memory capacity and nonlinearity can be calculated using a single multiple frequency channels. For the single channel analysis, we perform the same calculations, but using just a single FMR channel.

\textbf{Frequency Decomposition, a Few-Shot Learning Task.} The network is driven by a sinusoidal input $s(t)$ and needs to reconstruct a decomposition of a temporal varying signal in the form of $\tilde{y}(t)$ = $\sum_{n}^{N_{\omega}} a_n\sin(nt+\theta_n)$, where $N_{\omega}=5$. The values of $a_n$ and $\theta_n$ are randomly sampled at the beginning of each task from uniform distributions. In particular, $a_n \in [-1.2 \ \ 1.2]$ and $\theta_n \in [0 \ \ \pi/n]$.  The output layer is composed of $N_{\omega}$ nodes, and the system is asked to predict a target $\tilde{y}(t)=\Big( a_1 sin(t+\theta_1), a_2 sin(2t+\theta_2),...,a_5 sin(5t+\theta_5) \Big)$ after observing the values of $\tilde{y}(t)$ over K data points i.e. time steps. The value of K adopted for the examples of Figures \ref{Metalearning} b) and c) is fifteen, but the network reports good performance even with $\rm{K}=10$ (Figures \ref{Metalearning} d) and e). To face this challenging task, we use the PNN in the overparameterised regime and a meta-learning algorithm to quickly adapt the read-out connectivity. Details of the meta-learning approach are given below in the Meta-learning section of the Methods.

\subsection*{Learning Algorithms}
The type of learning algorithm we use to select features and train the networks varies throughout the manuscript.
For comparisons between different systems (Figures \ref{Samples} \ref{RC_network}), training is accomplished through a features selection algorithm (discussed below) and optimisation of the read-out weights $W_{o}$. 
When exploring the effects of overparameterisation, we randomly select features and then optimise $W_{o}$ for those features.
We call $x(t)$ the representation at which the read-out weights operate, and we define the output of the system as $y(t)=W_{o}x(t)$. The vector $x(t)$ simply contains a subset of features of $o(t)$ defined via the feature selection algorithm or randomly picked as in the dimensional study on overparameterisation.  
In this setting, optimisation of $W_{o}$ is achieved with linear or ridge regression, which minimises the error function $\rm{E}=\sum_t||\tilde y(t)-W_{o} x(t)||^2+\lambda ||W_{o}||^2$.  
\newline 
For the simulations where we show the adaptability of the system with a limited amount of data (results of Figure \ref{Metalearning} and of Section Learning in the overparameterised regime), we used gradient-descent optimisation techniques, particularly Adam \cite{kingma2014adam}, to minimise the mean-squared error between prediction and target. All codes and data for the learning algorithms are available online (see code availability statement for details).

\subsubsection*{Feature Selection}
\begin{figure}[ht!]
    \centering
    \includegraphics[width=0.9\textwidth]{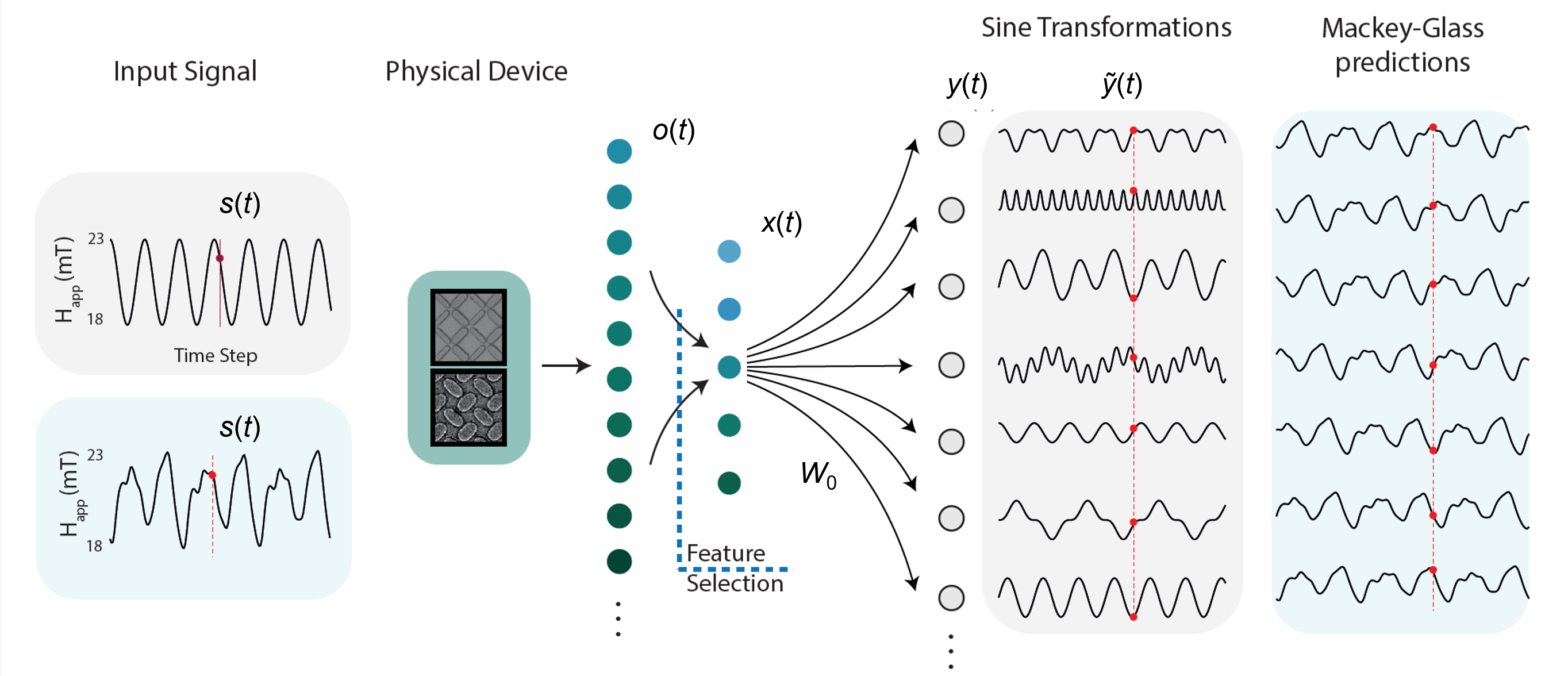}
    \caption{Schematic of offline learning. Input data is applied to the physical system producing a set of $\mathbf{o}(t)$ features. Feature selection reduces the size of $\mathbf{o}(t)$ to a new set of $\mathbf{x}(t)$ features. We then apply ridge regression to obtain a set of weights $W_{o}$ to obtain the transformation / prediction $\mathbf{y}(t)$ = $W_{o}$$\mathbf{x}(t)$}
    \label{Learning_scheme1}
\end{figure}

\begin{figure}[ht!]
    \centering
    \includegraphics[width=0.8\textwidth]{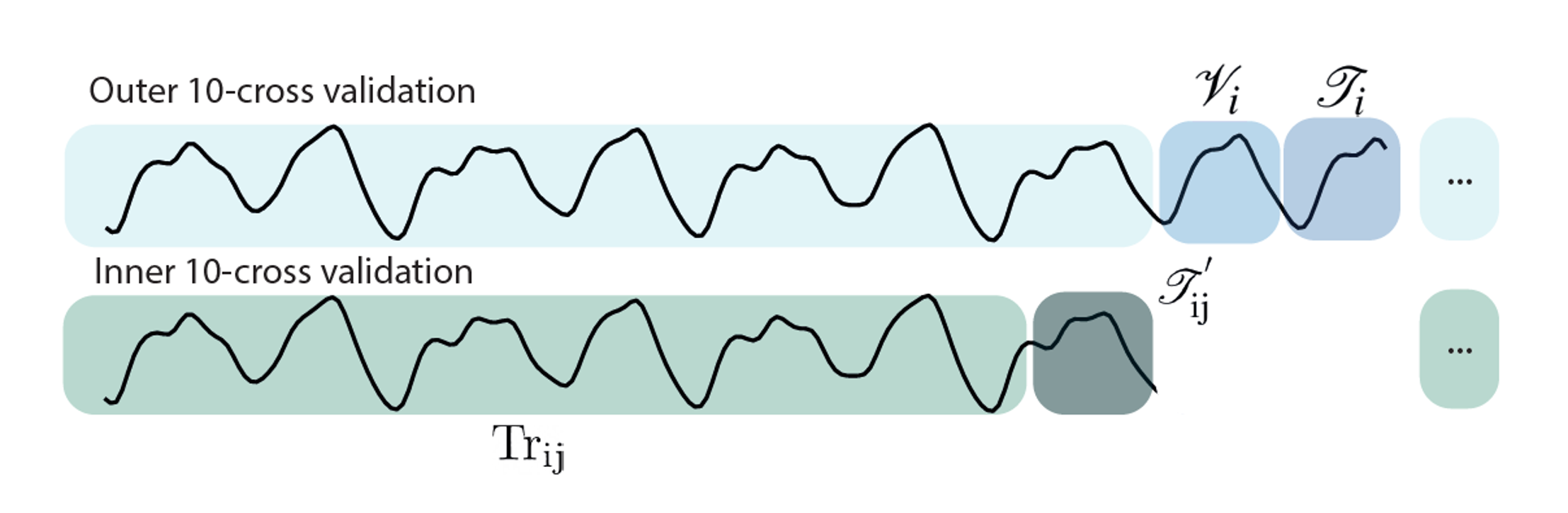}
    \caption{Data splitting during training. Schematic of the data splits for the inner and outer validation loops.}
    \label{Learning_scheme}
\end{figure}

The dimensionality of an observation $o(t)$ can vary depending on the architecture considered, spanning from $\approx 250$ dimensions when using a single arrays to $\approx$ 14,000 for the PNN of Figure \ref{RC_network}. The high readout dimensionality allows better separability of input data, however, high dimensional spaces constitute a challenge due to overfitting issues. As such, learning over a high-dimensional features' space with few data points constitutes a challenge and opportunity for physically defined reservoirs.
For this reason, we design a feature-selection methodology to avoid overfitting and to exploit the computational abilities of architectures with varying complexity (Figure \ref{Learning_scheme1}, Algorithms \ref{alg:hyp},\ref{alg:evol}). The methodology adopted can be at first described as a $10$ cross-validation (inner validation loop) of a $10$ cross-validation approach (outer validation loop), where the outer cross-validation is used to accurately evaluate the performance and the inner loop is used to perform feature-selection (Figure \ref{Learning_scheme}). For each split, feature selection is accomplished by discarding highly correlated features and through an evolutionary algorithm. The independent parts of this methodology are known, but the overall procedure is unique and can give accurate performance measurements for our situation, where we have a suite of systems with varying dimensionality to compare over limited data.   

\begin{algorithm}[h!]
\caption{Hyperparameter selection}\label{alg:hyp}
\begin{algorithmic}
\For 
{for each split in outer loop $i$}
\For {each split in inner loop $j$}
\For {each $\theta\in\Theta, \lambda\in\Lambda$}

\State Find optimal weights: 
\(\mathbf{W}^{*}_{o}=\argmin \limits_{\mathbf{W}_o \in \mathbb{R}^m} \rm{E}\Big\{ \rm{Tr}_{ij}|\mathbf{W}_{o}, \theta, \lambda\Big\}\)
\State Compute error on the test set 
\(\rm{E}\Big \{ \mathcal{T}^{'}_{ij}|\mathbf{W}^{*}_{o}, \theta, \lambda \Big \}\)

\EndFor
\State Select optimal hyperparameters as 
\(\theta^{*}_{i}, \lambda^{*}_{i}=\argmin \limits_{\theta \in \Theta, \lambda \in \Lambda} \sum_{j=1}^{10} \rm{E}\Big \{ \mathcal{T}^{'}_{ij}|\mathbf{W}^{*}_{o}, \theta, \lambda \Big \}\)
\State Find the corresponding boolean vector $\theta^{*}_{i} \rightarrow \textbf{f}^{(i)}$
\EndFor
\EndFor

\end{algorithmic}
\end{algorithm}

\begin{algorithm}[h!]
\caption{Evolutionary Algorithm}\label{alg:evol}
\begin{algorithmic}
\For 
{for each split in outer loop $i$}

\State Initialise parents $\textbf{F}_{p} = \big \{ \mathbf{f}(n) \big \} = \textbf{f}^{(i)}$

\Repeat 
\State Apply crossover and mutations to generate children $\textbf{F}_c$

\For{each split in inner loop j}
\For{each $\textbf{f} \in \textbf{F}_c$}
\State Find optimal weights: \(\mathbf{W}^{*}_{o}=\argmin \limits_{\mathbf{W}_o \in \mathbb{R}^m} \rm{E}\Big\{ \rm{Tr}_{ij}|\mathbf{W}_{o}, \textbf{f},\lambda^{*}_{i}\Big\}\)
\State Select new $\textbf{F}_p$ as the $\textbf{f}'s$ with minimal \( \sum_{j=1}^{10} \rm{E}\Big \{ \mathcal{T}^{'}_{ij}|\mathbf{W}^{*}_{o}, \textbf{f}, \lambda_{i}^{*} \Big \}\)
\EndFor
\EndFor

\Until{best performance on $\mathcal{V}_i$}

\EndFor

\end{algorithmic}
\end{algorithm}
 
We now describe the feature selection methodology in detail. Considering the response of a system across time $\mathbf{o}(t)$, the feature selection algorithm aims to select a subset of features $\mathbf{x}(t)$ that will be used for training and evaluation. We can consider feature selection as a boolean operation over the $\mathbf{o}(t)$ feature space, where a value of one (zero) corresponds to the considered feature being used (neglected). If $m$ is the dimensionality of $\mathbf{o}(t)$, the number of possible ways to define $\mathbf{x}(t)$ is $2^m-1$. As a consequence, the feature selection algorithm can lead also to overfitting. Therefore, we must implement an additional cross-validation step to ensure the selected features performance is general across the entire dataset.

Considering a specific split of the outer validation loop (Figure \ref{Learning_scheme}), where we select a validation set $\mathcal{V}_i$ and a test set $\mathcal{T}_i$ (comprising of the $10\%$ of the data each, i.e. 25 data points), we perform $10$ cross-validations on the remaining data to optimise hyperparameter values through grid-search. In this inner validation loop, each split corresponds to a test set $\mathcal{T}^{'}_{ij}$ (comprising again of the $10\%$ of the remaining data, without $\mathcal{V}_i$ and $\mathcal{T}_i$), where changing $j$ means to select a different test-split in the inner loop based on the $i$-th original split of the outer validation (Figure \ref{Learning_scheme}). 
The remaining data, highlighted in green in Figure \ref{Learning_scheme}, are used for training to optimise the read-out weights and minimise the error function $\rm{E}$ through ridge-regression previously described.

At this stage, we perform a grid-search methodology on hyperparameters $\theta$ and $\lambda$ which control directly and indirectly the number of features being adopted for training (Algorithm \ref{alg:hyp}).
The hyperparameter $\theta$ acts as a threshold on the correlation matrix of the features. Simply, if the correlation among two features exceeds the specific value of $\theta$ considered, one of these two features is removed for training (and testing). The idea behind this method is to discard features that are highly correlated, since they would contribute in a similar way during training. This emphasises diversity in the reservoir response. 
The hyperparameter $\lambda$ is the penalty term in ridge-regression. Higher values of $\lambda$ lead to a stronger penalisation on the magnitude of the read-out weights. As such, $\lambda$ can help prevent overfitting and controls indirectly the number of features being adopted. We should use a high value of $\lambda$ if the model is more prone to overfitting the training dataset, a case that occurs when the number of features adopted is high. Calling  $\rm{E}\Big \{ \mathcal{T}^{'}_{ij}|\mathbf{W}^{*}_{o}, \theta, \lambda \Big \}$ the error computed on the test set $\mathcal{T}^{'}_{ij}$ with weights $\mathbf{W}^{*}_o$ optimised on the corresponding training data ($\mathbf{W}^{*}_{o}=\argmin \limits_{\mathbf{W}_o \in \mathbb{R}^m} \rm{E}\Big\{ \rm{Tr}_{ij}|\mathbf{W}_{o}, \theta, \lambda\Big\}$ in Algorithm \ref{alg:hyp} and with hyperparameter values $\theta$ and $\lambda$ respectively, we select the values of the hyperparameters that correspond to the minimum average error over the test sets in the inner validation loop. Otherwise stated, we select the optimal  $\theta^{*}_{i}$ and $\lambda^{*}_{i}$ for the $i$-th split in the outer loop from the test average error in the inner $10$ cross-validation as $\theta^{*}_{i}, \lambda^{*}_{i}=\argmin \limits_{\theta \in \Theta, \lambda \in \Lambda} \sum_{j=1}^{10} \rm{E}\Big \{ \mathcal{T}^{'}_{ij}|\mathbf{W}^{*}_{o}, \theta, \lambda \Big \}$. This methodology permits to find $\theta^{*}_{i}$ and $\lambda^{*}_{i}$ that are not strongly dependent on the split considered, while maintaining the parts of the dataset $\mathcal{V}_i$ and $\mathcal{T}_i$ unused during training and hyperparameter selection. The sets $\Theta$ and $\Lambda$ correspond to the values explored in the grid-search. In our case, $\Theta=\{1, 0.999, 0.99, 0.98, 0.97,... \}$ and $\Lambda=\{ 1e-4,1e-3,1e-2,5e-2,1e-1\}$. Repeating this procedure for each split of the outer loop, we found the optimal $\theta^{*}_{i}$ and $\lambda^{*}_{i}$, for $i=1,...,10$. This concludes the hyperparameter selection algorithm described in Algorithm \ref{alg:hyp}. Selection of the hyperparameters $\theta^{*}_i$ permit to find subsets of features based on correlation measures. However, promoting diversity of reservoir measures does not necessarily correspond to the highest performance achievable. Thus, we adopted an evolutionary algorithm to better explore the space of possible combination of measurements (Algorithm \ref{alg:evol}). 

It is necessary now to notice how a value of $\theta^{*}_i$ corresponds to a $m$-dimensional boolean vector $\mathbf{f}^{(i)}$, whose $j$-th dimension is zero if its $j$-th feature $f^{(i)}_j$ is correlated more than $\theta^{*}_i$ with at least one other output. For each split $i$ in the outer loop, we adopted an evolutionary algorithm that operates over the $m$-dimensional boolean space of feature-selection, where each individual corresponds to a specific vector $\mathbf{f}$. 
At each evolutionary step, we perform operations of crossover and mutation over a set of $N_p$ parents $\mathcal{\mathbf{F}}_p=\big \{ \mathbf{f}(n) \big \}_{n=1,...,\rm{N}_p}$. For each split $i$ of the outer loop and at the first evolutionary step, we initialised the parents of the algorithm to $\mathbf{f}^{(i)}$.  We defined a crossover operation among two individuals $\mathbf{f}(i)$ and $\mathbf{f}(j)$ as  $\mathbf{f}=\rm{CrossOver}\big(\mathbf{f}(i), \mathbf{f}(j) \big)$ where the $k$-th dimension of the new vector $\mathbf{f}$ is randomly equal to $f(i)_k$ or $f(j)_k$ with the same probability. A mutation operation of a specific $\mathbf{f}(i)$ is defined as $\mathbf{f}=\rm{Mutation}\big( \mathbf{f}(i) \big)$ by simply applying the operator $not$ to each dimension of $\mathbf{f}(i)$ with a predefined probability $p_m$. The application of crossovers and mutations permits the definition of a set of $\rm{N}_c$ children $\mathbf{F}_c=\big \{ \mathbf{f}(n) \big \}_{n=1,...,\rm{N}_c}$ from which we select the $\rm{N}_p$ models with the highest performance over the test sets of the inner loop as parents for the next iteration. Otherwise stated, we selected the $\mathbf{F}_p$ vectors corresponding to the lowest values of the average error as $\mathbf{F}_p=\argminNp \limits_{\mathbf{f} \in \mathbf{F}_c} \sum_{j=1}^{10} \rm{E}\Big \{ \mathcal{T}^{'}_{ij}|\mathbf{W}^{*}_{o}, \mathbf{f}, \lambda^{*}_i \Big \}$ where $\argminNp$ selects the $\rm{N}_p$ arguments of the corresponding function with minimal values. We notice how a step of the evolutionary approach aims to minimise an error estimated in the same fashion as in the algorithm of Algorithm \ref{alg:hyp}, but this time searching for the best performing set $\mathbf{F}_p$, rather then the best performing couple of hyperparameter values $\lambda^{*}_i$ and $\theta^{*}_i$.

 Finally, we stop the evolutionary algorithm at the iteration instance where the average performance of $\mathbf{F}_p$ over $\mathcal{V}_i$ is at minimum and selected the model $\mathbf{f}^{*}_i$ with the lowest error on $\mathcal{V}_i$. The utilisation of a separate set $\mathcal{V}_i$ for the stop-learning condition is necessary to avoid overfitting of the training data. Indeed, it is possible to notice how the performance on $\mathcal{V}_i$ would improve for the first iterations of the evolutionary algorithm and then become worse. This concludes the evolutionary algorithm in Algorithm \ref{alg:evol}. At last, the overall performance of the model is computed as the sum of the mean-squared errors over the outer validation loop as $\mathcal{E}=\sum_{i=1}^{10} \rm{E}\Big \{ \mathcal{T}_{i}|\mathbf{W}^{*}_{o}, \mathbf{f}^{*}_i, \lambda^{*}_i \Big \}$. 
 Summarising, we can think the overall methodology as an optimisation of relevant hyperparameters followed by a fine-tuning of the set of features used through an evolutionary algorithm. The final performance and its measure of variation reported in the paper are computed as average and standard deviation over ten repetitions of the evolutionary algorithm respectively. 
  
\subsubsection*{Evaluating Overparameterisation}
To explore the effects of overparameterisation it is necessary to vary the number of network parameters (i.e. number of FMR channels). This is achieved as follows: first, a set of tasks is prepared for which the MSE will be evaluated. Here, we analyse performance when predicting future values of the Mackey-Glass equation. We average MSE when predicting \textit{t}+1, \textit{t}+3, \textit{t}+5, \textit{t}+7, \textit{t}+9 and \textit{t}+11. Next, we vary the number of parameters. To do so, we randomly shuffle a sequence of integers from 0 to \textit{N}, where \textit{N} is the total number of output channels for a given network e.g. \textit{N} = (328, 34, 273 ...). From this, we select the first P points of the sequence, giving a list of indexes referring to which channels to include in that sequence. For those channels, we perform a 5 cross-validation of different train and test splits and use linear regression to evaluate the train and test MSE for a given task and set of outputs. P is increased in steps of \textit{N}/500 for single, parallel and series networks and steps of \textit{N}/5000 for the PNN. For a given task, we repeat this process for 50 random shuffles to ensure we are sampling a broad range of output combinations for each network. The displayed MSEs are the average and standard error of these 50 trials over all tasks.

\subsubsection*{Meta-Learning}\label{Metalearning_methods}

\begin{figure}[ht!]
    \centering
    \includegraphics[width=1\textwidth]{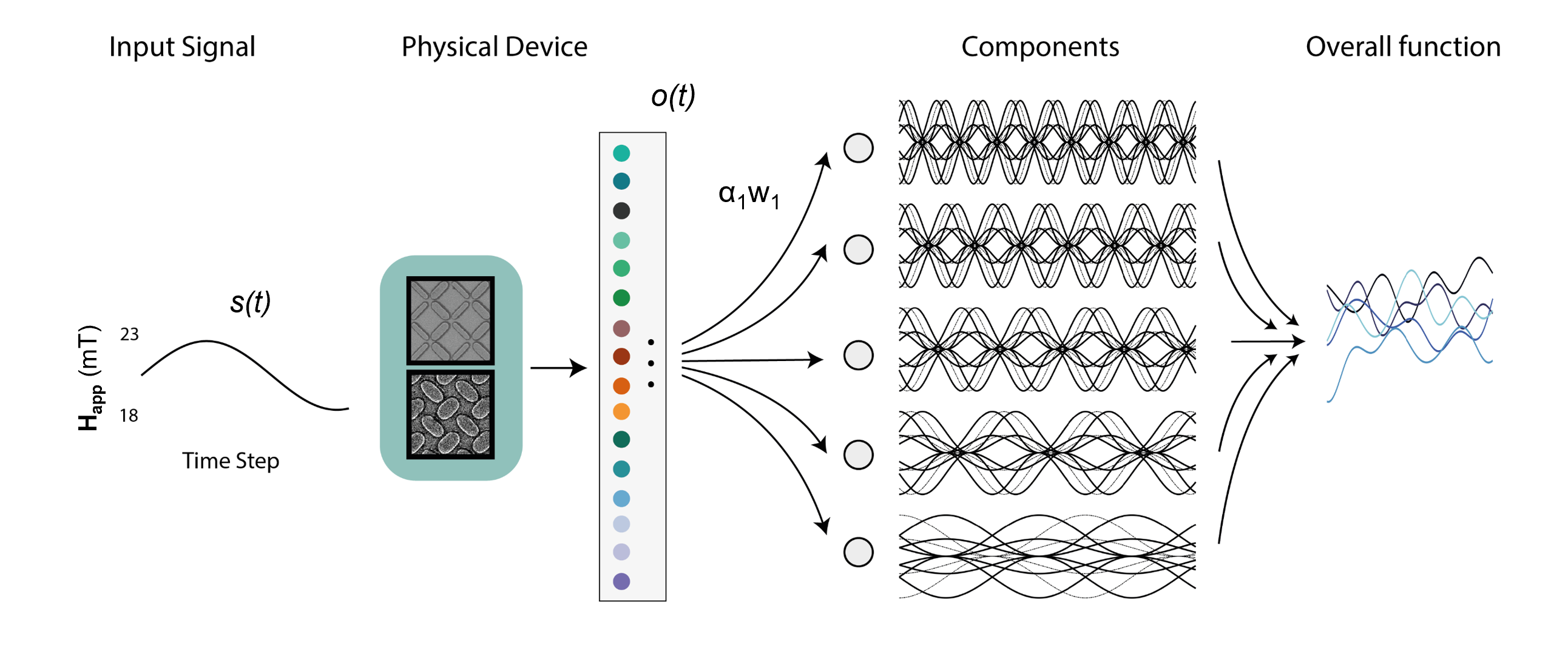}
    \caption{Schematic of the few-shot learning task. The input signal is applied to the multilayer neural network and the observations are recorded. The system is then trained to learn a subset of frequency decomposition with minimal training data.}
    \label{Meta_Learning_scheme}
\end{figure}

\begin{algorithm}[h!]
\caption{Meta-learning through MAML}\label{alg:maml}
\begin{algorithmic}
\State \textbf{Sample} a batch of tasks from $p(\mathcal{T})$
\For{each task $i$}
\State \textbf{Sample} K datapoints of device responses $\mathcal{D}_i=\Big \{...,(\tilde{y}(t_k),o(t_k)),... \Big\} $, $k = 1,...,K$

\For{number of inner loop steps $n = 1,...,\tilde{n}$}
\State $\mathbf{y}(t_k)=f(\mathbf{o}(t_k)|\boldsymbol{\alpha}_{i}(n),\mathbf{W}_o)$
\State $\boldsymbol{\alpha}_{i}(n+1) = \boldsymbol{\alpha} - \eta_1\nabla_{\boldsymbol{\alpha}_i(n)}\rm{E}_{\mathcal{T}_i}\big(\mathcal{D}_i|\boldsymbol{\alpha}_i(n), \textbf{W}_o\big)$

\State \textbf{Sample} Q datapoints of devices response $\mathcal{D}'_i=\Big \{...,(\tilde{y}(t_q),o(t_q)),... \Big\} $, $q = 1,...,Q$ for the meta-update
\EndFor
\State \textbf{Perform} the meta-update
\State $\mathbf{W}_o\leftarrow \mathbf{W}_o-\eta_2\nabla_{\mathbf{W}_o} \sum_i \rm{E}_{\mathcal{T}_i}\Big( \mathcal{D}'_i|\boldsymbol{\alpha}_i(\tilde{n}), \textbf{W}_o\Big)$
\State $\boldsymbol{\alpha}(0)\leftarrow \boldsymbol{\alpha}(0)-\eta_2\nabla_{\boldsymbol{\alpha}(0)} \sum_i \rm{E}_{\mathcal{T}_i}\Big( \mathcal{D}'_i|\boldsymbol{\alpha}_i(\tilde{n}), \textbf{W}_o\Big)$
\EndFor

\end{algorithmic}
\end{algorithm}

The goal of meta-learning is optimise an initial state for the network such that when a new task with limited data points is presented, the network can be quickly updated to give strong performance. A schematic of the meta-learning algorithm is presented in Figure \ref{Meta_Learning_scheme} and pseudo-code shown in Algorithm \ref{alg:maml}.

Let us consider a family of M tasks $\mathcal{T}=\{\mathcal{T}_1,\mathcal{T}_2,...,\mathcal{T}_{\rm{M}}\}$. Each task is composed of a dataset and a cost function $\rm{E}_{\mathcal{T}_i}$. A meta-learning algorithm is trained on a subset of $\mathcal{T}$ and asked to quickly adapt and generalise on a new test subset of $\mathcal{T}$. Otherwise stated, the aim of meta-learning is to find an initial set of parameters $\mathbf{W}(0)$ that permits learning of an unseen task $\mathcal{T}_i$ by updating the model over a small number of data points from $\mathcal{T}_i$. To achieve this, we use a variation of the MAML algorithm, which is now summarised. For a given task $\mathcal{T}_{i}$, the initial set of trainable parameters $\mathbf{W}(0)$ are updated via gradient descent over a batch of data points $\mathcal{D}_i=\Big \{...,(\mathbf{o}_{j},\tilde{\mathbf{y}}_{j} ),... \Big\} $, where $\mathbf{o}_j$ and $\tilde{\mathbf{y}}_{j}$ are the inputs and targets for the j-th data point respectively 
\begin{equation}
    \mathbf{W}_{i}(n)=\mathbf{W}_{i}(n-1)-\eta_1 \nabla_{\mathbf{W}_{i}(n-1)} \rm{E}_{\mathcal{T}_i} \big(\mathcal{D}_i| \mathbf{W}_i(n-1) \big) \label{Eq.grad_in}
\end{equation}
where $\eta_1$ is the learning rate adopted. 
Eq.\ref{Eq.grad_in} is repeated iteratively for $n=0,...,\tilde{n}$, where $\tilde{n}$ are the number of updates performed in each task. We notice how the subscripts $i$ on the parameters $\mathbf{W}$ are introduced because the latter become task-specific after updating, while they all start from the same values $\mathbf{W}(0)$ at the beginning of a task.
MAML optimizes the parameters $\mathbf{W}(0)$ (i.e. the parameters used at the start of a task, before any gradient descent) through the minimisation of cost functions sampled from $\mathcal{T}$ and computed over the updated parameters $\mathbf{W}(\tilde{n})$ i.e. the performance of a set of $\mathbf{W}(0)$ is evaluated based on the resulting $\mathbf{W}(\tilde{n})$ after gradient descent. Mathematically, the aim is to find the optimal $\mathbf{W}(0)$ that minimises the meta-learning objective $\mathcal{E}$
\begin{align}
    \mathcal{E}=\sum_{\mathcal{T}} \rm{E}_{\mathcal{T}_i}(\mathcal{D}'_i|\mathbf{W}_i(\tilde{n})) \label{Eq.E_out} \\
   \mathbf{W}(0)\leftarrow \mathbf{W}(0) -\eta_{2} \boldsymbol{\nabla}_{\mathbf{W}(0)}\sum_i \rm{E}_{\mathcal{T}_i}\Big( \mathcal{D}'_i |\mathbf{W}_i(\tilde{n})\Big) 
\end{align}
where the apex $'$ is adopted to differentiate the data points used for the meta-update from the inner loop of Eq.\ref{Eq.grad_in}, and $\eta_2$ is the learning rate for the meta-update.
Gradients of $\mathcal{E}$ need to be computed with respect to $\mathbf{W}(0)$, and this results in the optimisation of the recursive Eq.\ref{Eq.grad_in} and the computation of higher-order derivatives. In our case, we use the first-order approximation of the algorithm\cite{finn2017model}.
After the meta-learning process, the system is asked to learn an unseen task $\mathcal{T}_j$ updating $\mathbf{W}(0)$ through the iteration of Eq.\ref{Eq.grad_in} computed on a small subset of data of $\mathcal{D}_j$. 

In contrast to previous works, we adopted this learning framework on the response of a physical system. The optimisation occurs exclusively at the read-out level, leaving the computation of non-linear transformations and temporal dependencies to the physical network. The outcome of our application of meta-learning, depends on the richness of the dynamics of the nanomagnetic arrays.
In this case, the read-out connectivity matrix is not task-specific as in previous sections, we use the same $\mathbf{W}(0)$ for different tasks. 

We decompose the parameters of the system $\mathbf{W}$ into two sets of parameters $\{\mathbf{W}_o,\boldsymbol{\alpha}\}$ to exploit few-shot learning on the readout from the PNN. The output of the system is defined as $\mathbf{y}(t)=f(\mathbf{o}(t)|\boldsymbol{\alpha},\mathbf{W}_o)$, where the function f is linear and follows $f(\mathbf{o}(t)|\boldsymbol{\alpha},\mathbf{W}_o)=\boldsymbol{\alpha} \circ \big(\mathbf{W}_o  \mathbf{o}(t)\big)$, where $\circ$ stands for element by element multiplication. The inclusion of parameters $\boldsymbol{\alpha}$ as scaling factors for the weights has previously been used in \cite{salimans2016weight}, albeit in the context of weight normalisation.  
Here, we train the parameters $\boldsymbol{\alpha}$ and $\mathbf{W}_o$ with different timescales of the learning process, disentangling their contribution in the inner (task-specific updates, Eq.\ref{Eq.grad_in}) and outer loops (computation of the meta-learning objective of Eq.\ref{Eq.E_out}) of the MAML algorithm. Specifically, the $\boldsymbol{\alpha}$ parameters are updated for each task following
\begin{align}
    \alpha_i(n)=\alpha_i(n-1)-\eta_{\alpha} \dfrac{\partial \rm{E}_{\mathcal{T}_i}}{\partial \alpha_i(n-1)} \ \ \rm{for} \ n=1,...,\tilde{n} 
\end{align}
while the parameters $\mathbf{W}_o$ are optimised through the meta-learning objective via
\begin{align}
    \mathbf{W}_o\leftarrow \mathbf{W}_o-\eta_2{\mathbf{w}} \sum_{\mathcal{T}} \rm{E}_{\mathcal{T}_j}(\mathcal{D}_j|\alpha_i(\tilde{n})) \label{Eq.meta_w}
\end{align}
In this way, training of the parameters $\mathbf{W}_o$ is accomplished after appropriate, task-dependent scaling of the output activities. A pseudo-code of the algorithm is reported in Fig.\ref{Meta_Learning_scheme}. 

\subsubsection*{Echo State Network Comparison}

The Echo-state network of Figures \ref{RC_network} g-i) is a software model defined through 
\begin{align}
    \mathbf{x}(t+\delta t)=(1-\alpha)\mathbf{x}(t)+\alpha f(\mathbf{W}_{in}\mathbf{s}(t)+\mathbf{W}_{esn}\mathbf{x}(t))
\end{align}

where $\mathbf{x}$ is the reservoir state, $\mathbf{s}$ is the input, $\mathbf{W}_{esn}$ and $\mathbf{W}_{in}$ are fixed and random connectivity matrices defined following standard methodologies \cite{lukovsevivcius2012practical} and $\alpha$ is a scaling factor.
In particular, the eigenvalues of the associated, linearised dynamical system are rescaled to be inside the unit circle of the imaginary plane. Training occurs on the read-out level of the system. 
Echo-state networks and their spiking analogous liquid-state machines \cite{maass2011liquid} are the theoretical prototypes of the reservoir computing paradigm. Their performance can thus constitute an informative reference for the physically defined networks. We highlight two differences when making this comparison: first, the ESN is defined in simulations and it is consequently not affected by noise; second, the physically defined network has a feedforward topology, where the memory of the system lies in the intrinsic dynamics of each complex node and the connectivity is not random but tuned thanks to the designed methodology. \newline
For the results of Figure \ref{RC_network}, we evaluate the performance of a 100-node ESN. For each value of the dimensionality explored, we repeated the optimization process ten times resampling the random $\mathbf{W}_{in}$ and $\mathbf{W}_{esn}$. The black dots in Figure \ref{RC_network} report the average performance across these repetitions as the number of nodes varies. The black line reflects a polynomial fit of such results for illustrative purposes, while the grey area reflects the dispersion of the distributions of the results.  

\subsubsection*{Multilayer Perceptron Comparison}

The multilayer perceptron (MLP) comparison in the supplementary information is defined as follows. We have an MLP with four layers: an input layer, two hidden layers interconnected with a ReLu activation function and an output layer. We vary the size of the hidden layers from 1 - 500. Standard MLP's do not have any recurrent connections and therefore do not hold information about previous states, hence predictive performance is poor. We add false memory by providing the network with the previous T\textsubscript{seq} data points as input i.e. $\tilde{\mathbf{s}}(t)=\Big(s(t), s(t-1),..., s(t-(T_{seq}-1)) \Big)$. We vary T\textsubscript{seq} from 1 to 10 where T\textsubscript{seq} = 1 corresponds to a standard MLP only seeing the current input data point. Learning of MLP weights is performed using gradient descent, specifically Adam\cite{kingma2014adam}. 

\section*{Data availability statement}

The experimental data used in this study is available at Github and Zenodo\cite{NeuroOverParam} under accession codes https://github.com/StenningK/NeuroOverParam.git and https://doi.org/10.5281/zenodo.12721639.

\section*{Code availability statement}
The code developed in this study is available at Github and Zenodo\cite{NeuroOverParam} under accession codes https://github.com/StenningK/NeuroOverParam.git and https://doi.org/10.5281/zenodo.12721639.

\label{Bibliography}
\bibliography{bib.bib}

\section*{Acknowledgements}
KDS was supported by The Eric and Wendy Schmidt Fellowship Program and the Engineering and Physical Sciences Research Council (Grant No. EP/W524335/1).
This work was supported by the EPRSC grant EP/X015661/1 to WRB and HK\\
JCG was supported by the Royal Academy of Engineering under the Research Fellowship programme and the EPSRC ECR International Collaboration Grant ‘Three-Dimensional Multilayer Nanomagnetic Arrays for Neuromorphic Low-Energy Magnonic Processing’ EP/Y003276/1.
AV was supported by the EPSRC Centre for Doctoral Training in Advanced Characterisation of Materials (Grant No. EP/L015277/1) and EPSRC grant EP/X015661/1.\\
K.E.S. acknowledges funding from the German Research Foundation (Project No. 320163632) and from the Emergent AI Center funded by the Carl-Zeiss-Stiftung.\\
CC and TC performed the work as part of their Physics MSci project at Imperial College London.\\
Analysis was performed on the Imperial College London Research Computing Service\cite{hpc}.\\
The authors would like to thank David Mack for excellent laboratory management.

\section*{Author contributions}
KDS and JCG conceived the work and directed the project throughout.\\
KDS drafted the manuscript with contributions from all authors in editing and revision stages.\\
KDS and LM implemented the computation schemes.\\
LM developed the cross-validation training approach for reducing overfitting on shorter training datasets.\\
LM developed the feature selection methodology for selecting optimal features from many reservoirs in the multilayer physical neural network architecture.\\
LM designed and implemented the meta-learning scheme.\\
KDS designed and implemented the method of interconnecting networks.\\
CC and TC aided in analysis of reservoir metrics.\\
KDS, JCG and AV performed FMR measurements.\\
JCG and HH performed MFM measurements.\\
JCG and KDS fabricated the samples. JCG and AV performed CAD design of the structures.\\
JCG performed scanning electron microscopy measurements.\\
JL contributed task-agnostic metric analysis code.\\
FC helped with conceiving the work, analysing computing results and providing critical feedback.\\
EV provided oversight on computational architecture design.\\
KES, EV and HK provided critical feedback.\\
WRB oversaw the project and provided critical feedback. \\

\section*{Competing interests}
Authors patent applicant. Inventors (in no specific order): Kilian D. Stenning, Jack C. Gartside, Alex Vanstone, Holly H. Holder, Will R. Branford. Application number: PCT/GB2022/052501. Application filed. Patent filed in UK through Imperial College London. Patent covers the method of programming deep networks. The remaining authors declare no competing interests.

\newpage
\include{Supp}

\end{document}

%% file: Supp.tex
\renewcommand\refname{Supplementary References}
\renewcommand\figurename{Supplementary Figure} 
\makeatletter
\setcounter{figure}{0}
\renewcommand{\fnum@figure}{Supplementary Figure \thefigure}
\renewcommand{\fnum@table}{Supplementary Table \thetable}
\makeatother

\section*{Supplementary Information}

\subsection*{Supplementary note 1 - Ferromagnetic resonance spectroscopy and magnetic force microscopy images of the artificial spin systems}
\begin{figure}[h!]
    \centering
    \includegraphics[width=\textwidth]{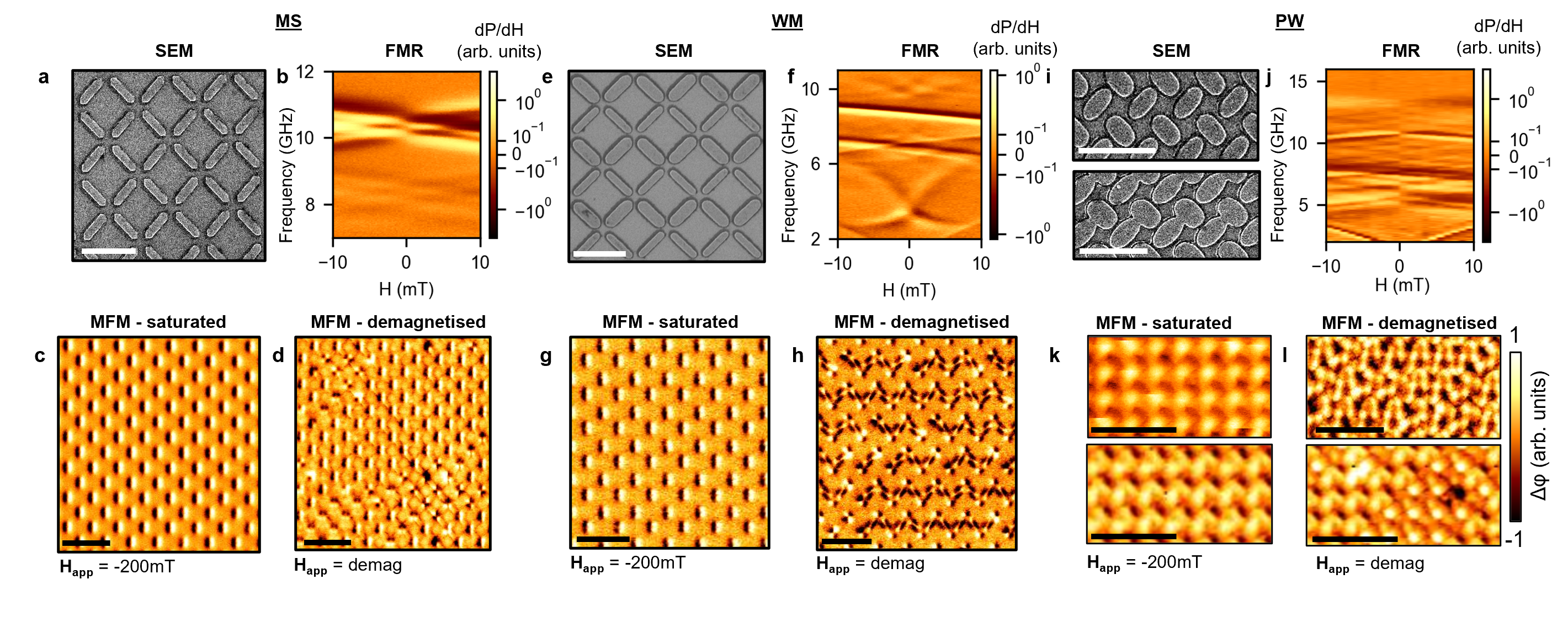}
    \caption{\textbf{Artificial spin system FMR response and bar states.} MS is shown in a-d), WM in e-h), PW in i-l). a,e,i) Scanning electron micrographs (SEM) of each aarray. b,h,n) Ferromagnetic resonance (FMR) heatmaps measured after AC-demagnetisation. MS (b) shows linear macrospin modes. WM (f) shows linear macrospin modes (6-7 GHz wide-bar, 8-9 GHz thin bar) and nonlinear vortex modes (2-6 GHz).  PW (j) exhibits rich linear and nonlinear modes. Magnetic force microscopy (MFM) images taken at remanence after field-saturation (c,g,k) and AC-demagnetisation (d,h,l). When saturated, all samples contain only  macrospins. When demagnetised, WM and PW (h,l) show vortices.}
    \label{FMR_MFM}
\end{figure}

The nanomagnetic arrays in this work are based on square and pinwheel artificial spin ice\cite{skjaervo2020advances}. Supplementary Figure \ref{FMR_MFM} shows scanning electron micrographs (SEM) (a,c,e) and ferromagnetic resonance (FMR) spectra after AC-demagnetisation (b,d,f). Three arrays were fabricated: 

MS is a square artificial spin ice (Supplementary Figure \ref{FMR_MFM} a,b). Bars are high aspect-ratio (530 nm $\times$ 120 nm) and only support macrospin states\cite{gartside2022reconfigurable}. WM is a width-modified artificial spin-vortex ice with a subset of wider, lower-coercivity bars (Supplementary Figure \ref{FMR_MFM} c,d). Bars are 600 nm $\times$ 200 nm (wide-bar)/125 nm (thin-bar). Wide bars host both macrospin and vortex states\cite{gartside2022reconfigurable} whereas thin bars host just macrospins. PW is a pinwheel-lattice artificial spin-vortex ice (Supplementary Figure \ref{FMR_MFM} e,f)\cite{macedo2018apparent,li2018superferromagnetism} with higher density and inter-island coupling. A gradient of bar dimensions are patterned across the sample, ranging from fully-disconnected (e, top) to partially-connected islands (e, bottom) giving a complex range of spectral-responses (Supplementary Figure \ref{FMR_MFM} f). Bar dimensions are constant across 100 $\times$ \SI{100}{\micro\meter}$^2$ (length 450 nm, width 240 nm / 265 nm (lower / upper panel) in Supplementary Figure 1 e). Islands support macrospins and vortices. 

MS FMR spectra comprise two dominant modes at 9.5 and 12 GHz with opposite linear gradients corresponding to macrospins aligned parallel (positive gradient) and anti-parallel (negative) to  \textbf{H}\textsubscript{app} (Supplementary Figure \ref{FMR_MFM} b). In both the field-saturated (Supplementary Figure \ref{FMR_MFM} c) and AC-demganetised state (Supplementary Figure \ref{FMR_MFM} d), only macrospin states are observed due to the high aspect-ratio of these bars. 

The WM FMR spectra comprises four dominant modes with rich responses: wide and thin bar linear macrospin modes (7 GHz and 9 GHz respectively), $\chi$-shaped vortex mode (2-6 GHz) and `whispering-gallery'-like high-frequency vortex mode\cite{schultheiss2019excitation} (10 GHz) (Supplementary Figure \ref{FMR_MFM} f). Vortices are observed in the wide-bars when the sample is in an AC-demagnetised state. 

PW displays a highly complex and non-linear FMR spectra (Supplementary Figure \ref{FMR_MFM} j) due to the broad range of bar dimensions and state variations in this sample. Bars host both macrospins and vortices (Supplementary Figure \ref{FMR_MFM} k,l)

\begin{figure}[h!]
    \centering
    \includegraphics[width=\textwidth]{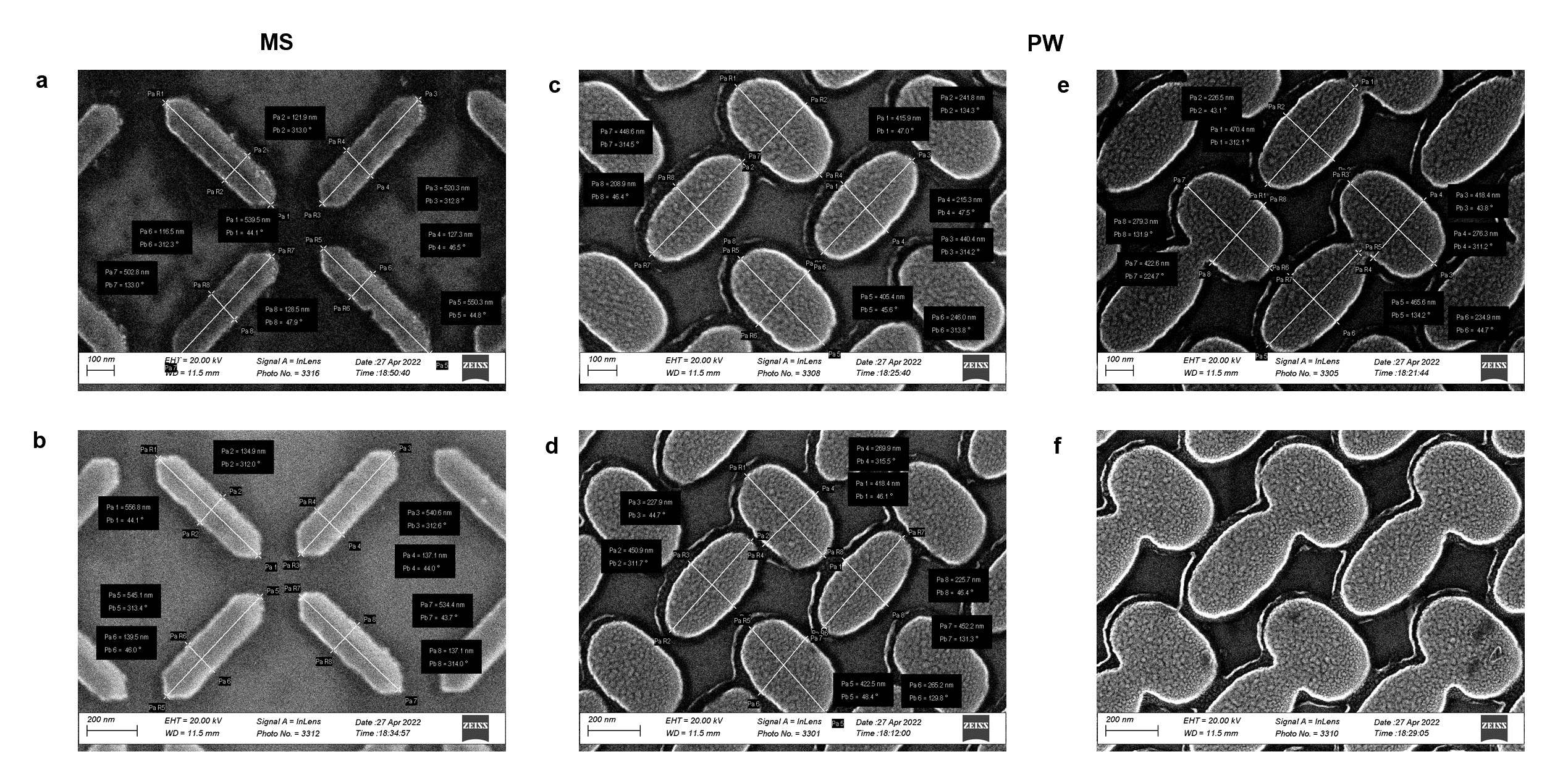}
    \caption{\textbf{Sample dimension variation.} Variation of dimensions across the MS (a,b) and PW (c-f) samples.}
    \label{Dimension_variation}
\end{figure}
 Supplementary Figure \ref{Dimension_variation} shows scanning electron micrographs of variation of dimensions across the MS (a,b) and PW (c-f) samples. In MS, two separate regions with widths of 124 nm and 137 nm are present. This manifests as slightly different coercive fields and resonant frequencies. In PW, a gradient of sample dimensions are fabricated across the surface of the chip. Bar subsets have slightly different widths to further enhance sample complexity. Widths range from 212 - 226 nm (thin) and 243 - 267 nm (wide) giving a broad distribution of resonant frequencies and coercive fields, enhancing output nonlinearity. Furthermore, some bars are connected (panel f) giving rise to complex magnetisation profiles and FMR spectral evolution.

\subsection*{Supplementary note 2 - k\textsubscript{max} selection for metric calculation.}
\begin{figure}[h!]
    \centering
    \includegraphics[width=\textwidth]{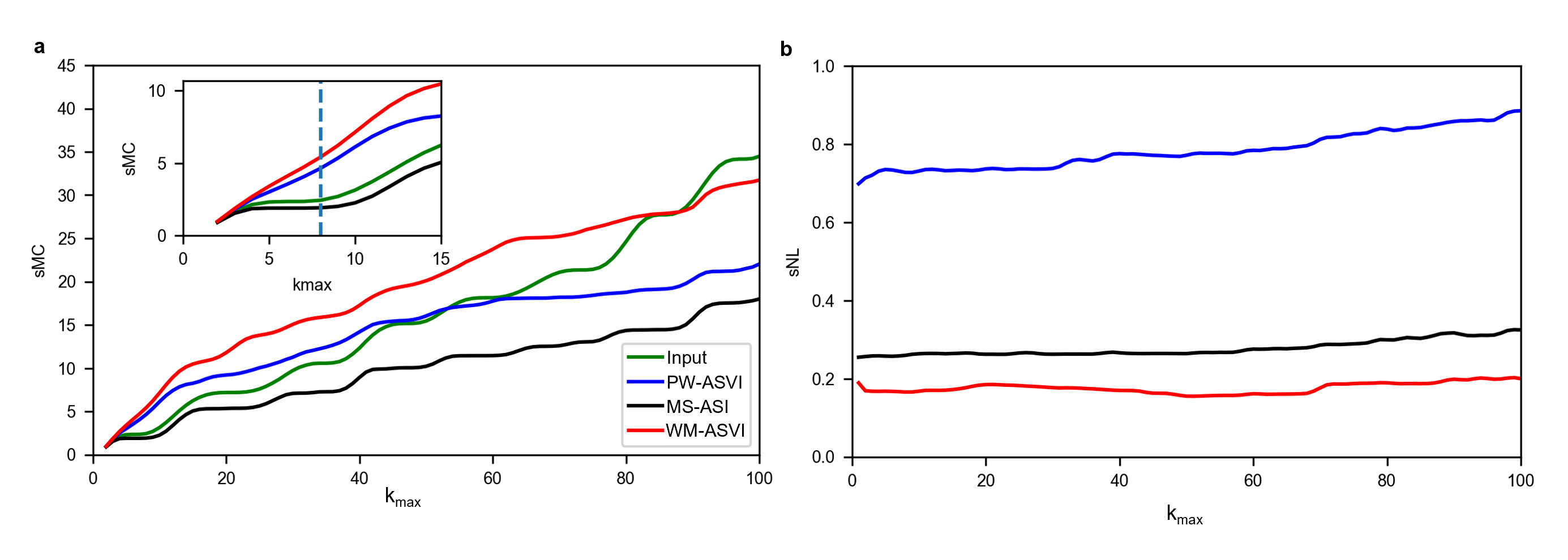}
    \caption{\textbf{Effects of k\textsubscript{max} when calculating memory-capacity (MC) and nonlinearity (NL).} a) memory-capacity (sMC) and b) nonlinearity (sNL) when varying how many previous inputs are used in the metric calculation (k\textsubscript{max}). Memory-capacity profiles show characteristic humps and continual rising indicative of correlation between current and past input profiles from the periodic nature of the Mackey-Glass input equation.}
    \label{metrics_k}
\end{figure}
Memory-capacity and nonlinearity  are typically obtained by inputting a random input signal, with no correlation between consecutive inputs, to the reservoir. When calculating memory-capacity, this ensures that any observed memory effects arise from reservoir states alone. However, our measurement scheme precludes this method. Data is encoded in magnetic field amplitude and readout in-field resulting in field-dependent frequency shifts in the FMR response. For random inputs, the sharp jumps between consecutive inputs causes information to shift between outputs (e.g. shifts of 0.5 GHz spanning 25 output frequencies are observed over a 5 mT input range). Linear regression is unable to process this type of shifting leading to misleadingly low calculated memory-capacity.

As such, we use the smoothly varying Mackey-Glass equation as an input signal to calculate the metrics. This ensures that any field shifts between consecutive input are minimised. Our FMR peaks are broad with microstate information held across multiple neighbouring outputs (e.g. 7 GHz and 7.02 GHz will be collinear). A disadvantage of this approach is that the Mackey-Glass equation is quasi-periodic. In addition to memory from the artificial spin reservoir states, this can lead to a memory arising from the similarity between current and previous inputs. As such, the value of k\textsubscript{max} with which the metrics are calculated must be chosen carefully to minimise the effects of the periodic input. 

  Supplementary Figure \ref{metrics_k} shows the memory-capacity and nonlinearity of the three artificial spin reservoir samples. Memory-capacity of the input signal with itself is also shown. All curves show characteristic humps and continuous rising in the memory-capacity profile due to the periodic nature of the MG input. Nonlinearity shows no variation with k\textsubscript{max}. We chose k\textsubscript{max} = 8 as this is at the end of the flat region on the input memory-capacity curve i.e. this is the maximum value before self-correlation from the next period starts.

\subsection*{Supplementary note 3 - Sine and NARMA transforms for each artificial spin reservoir}
\begin{figure}[h!]
    \centering
    \includegraphics[width=\textwidth]{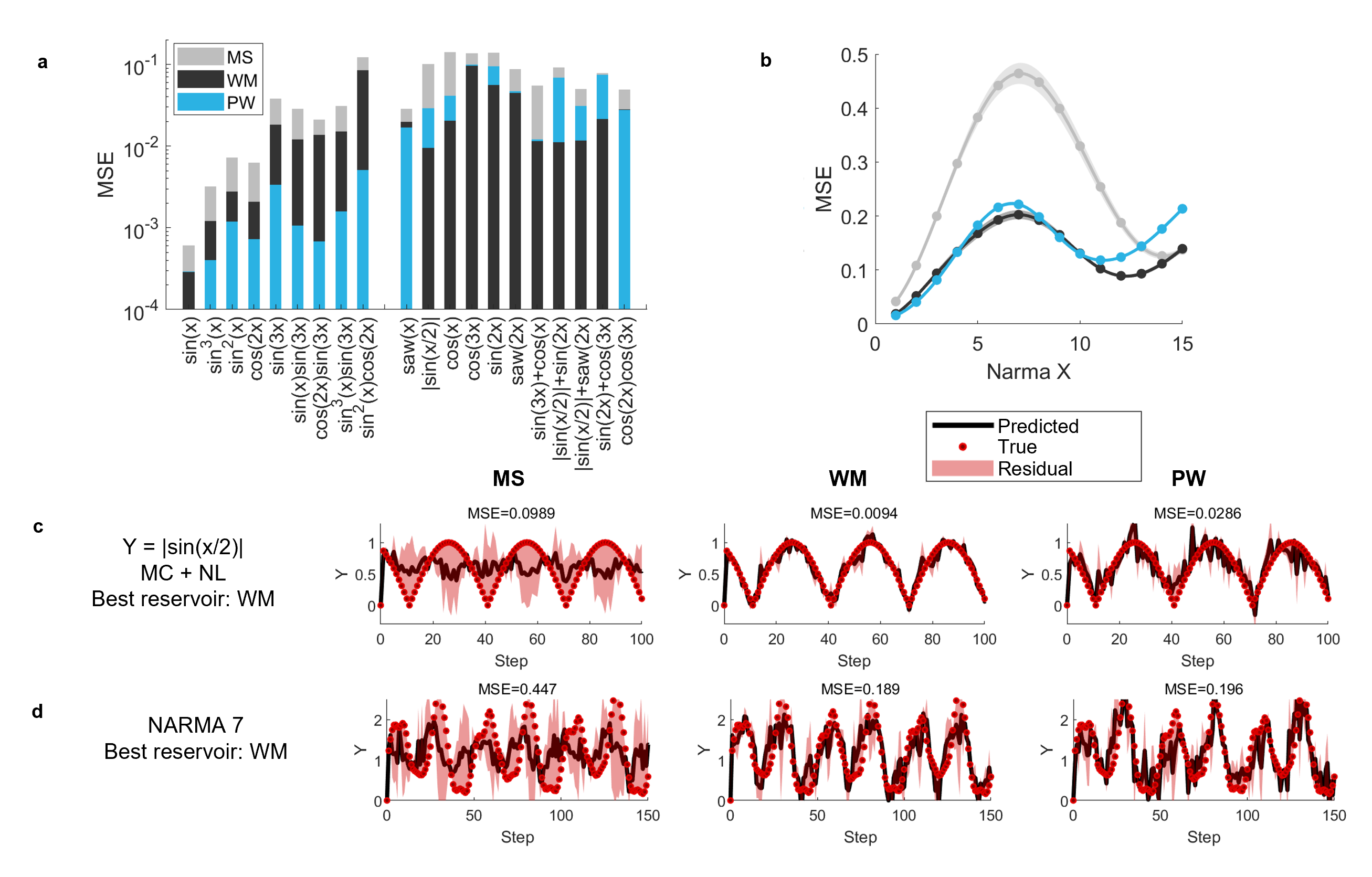}
    \caption{\textbf{Single reservoir sine and NARMA transforms.} a) MSE when transforming a sine input to a variety of targets. Tasks are chosen to be symmetric (nonlinearity only) or asymmetric (memory-capacity + nonlinearity) w.r.t the input. Vortices enhance performance up to 31.6$\times$ for symmetric tasks. Higher memory-capacity enhances asymmetric transformations.
    b) MSE for NARMA-transformation on the Mackey-Glass input signal. X corresponds to how far back the NARMA model is evaluated on. Samples with high memory-capacity perform well.
    Example predictions for |sin(x/2)| and NARMA7 are shown in panels c) and d) respectively.}
    \label{Individual_res_tranasforms}
\end{figure}
 Supplementary Figure \ref{Individual_res_tranasforms} a) shows the MSE for each reservoir when transforming a sinusoidal input to a variety of targets. Signal-transformations require varying memory-capacity and nonlinearity depending on the target waveform. Targets which are symmetric w.r.t. the input (eg. sin$^2$(x), sin(3x)sin(x)) require nonlinearity only, asymmetric waveforms w.r.t the input (e.g. saw(x), sin(x/2)) require both nonlinearity and memory-capacity as equivalent input values must be transformed to different output values across the wave-cycle. Highest nonlinearity PW dominates for symmetric transforms. Highest memory-capacity WM dominates for asymmetric transforms. Example plots for asymmetric |sin(x/2)| transform are shown in   Supplementary Figure \ref{Individual_res_tranasforms} c).

  Supplementary Figure \ref{Individual_res_tranasforms}b) shows performance when performing a NARMA-transform\cite{jaeger2002adaptive} on the Mackey-Glass input, evaluated as \(y(t) = Ay(t-1) + By(t-1) \sum_{n=1}^{X}y(t-n) + Cu(t-1)u(t-X) + D\) with A = 0.3, B =  0.01, C = 2, D = 0.1. X varies between 1-15, defined as `NARMAX'. This task adds additional nonlinear complexity vs. normal Mackey-Glass prediction, favouring both memory-capacity and nonlinearity with WM and PW performing similarly across all tasks. Example plots for the most-challenging NARMA7 are shown in  Supplementary Figure \ref{Individual_res_tranasforms} d).



\subsection*{Supplementary note 4 - Per-channel metric analysis for series connections}
\begin{figure*}[htb!]
    \includegraphics[width=\textwidth]{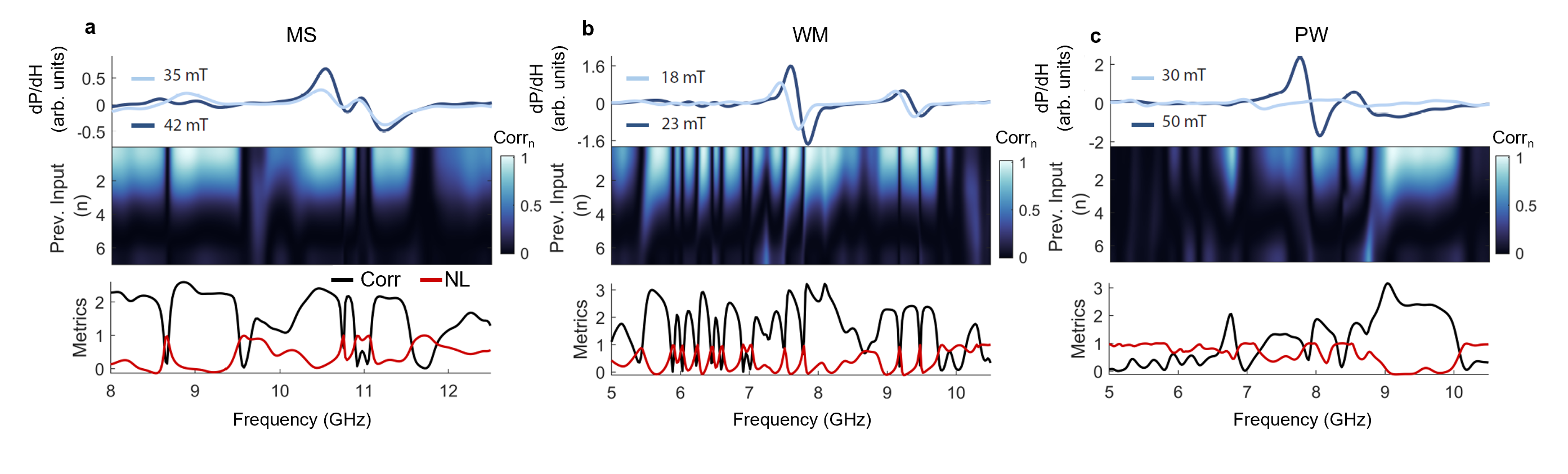}
    \captionof{figure}{\textbf{Per-channel metric analysis.}
    MS (a), WM (b) and PW (c) FMR amplitudes at maximum (dark blue) and minimum (light blue) input fields. Frequency-channel signal correlation (Corr\textsubscript{n}) for previous time-steps (n) ranging from 0-7. Memory-capacity (black) and nonlinearity (red) across all output frequencies is shown. Corr is a sum of Corr\textsubscript{n} from n = 0 to 7. 
    }
   \label{perchannel}
\end{figure*}

We assess metrics on a per-output channel basis, highlighting that memory and nonlinearity are provided by distinct spectral channels. Figures \ref{perchannel} a-c) (top) show FMR spectra at max (dark blue) and min (light blue) input field-amplitude, correlation to previous time steps for 0-7 previous time-steps (n) where Corr\textsubscript{n} is the correlation to the nth previous input, total correlation and nonlinearity of each output frequency-channel (bottom). Here, the correlation calculation is identical to the memory-capacity calculation, except that only one channel is used rather than multiple channels. As such, we call it correlation to avoid confusion with other discussions of memory capacity. For MS, high Corr\textsubscript{n} is limited to n $<$ 3. In contrast, WM and PW have some outputs which are correlated to 0-3 prior time-steps and others correlated to 4-7 steps (e.g. WM, 7.2 - 7.5 GHz). The presence of multiple correlation timescales is provided by vortex dynamics and is key to the strong prediction performance observed later.
In PW, the main FMR mode has high nonlinearity due to complex disordered microstate dynamics. The gradient of physical structures in the array provide more non-degenerate nonlinear responses and hence the highest nonlinearity score.

\subsection*{Supplementary note 5 - Vortex induced memory amplification.}

\begin{figure}[tb]
    \centering
    \includegraphics[width=\textwidth]{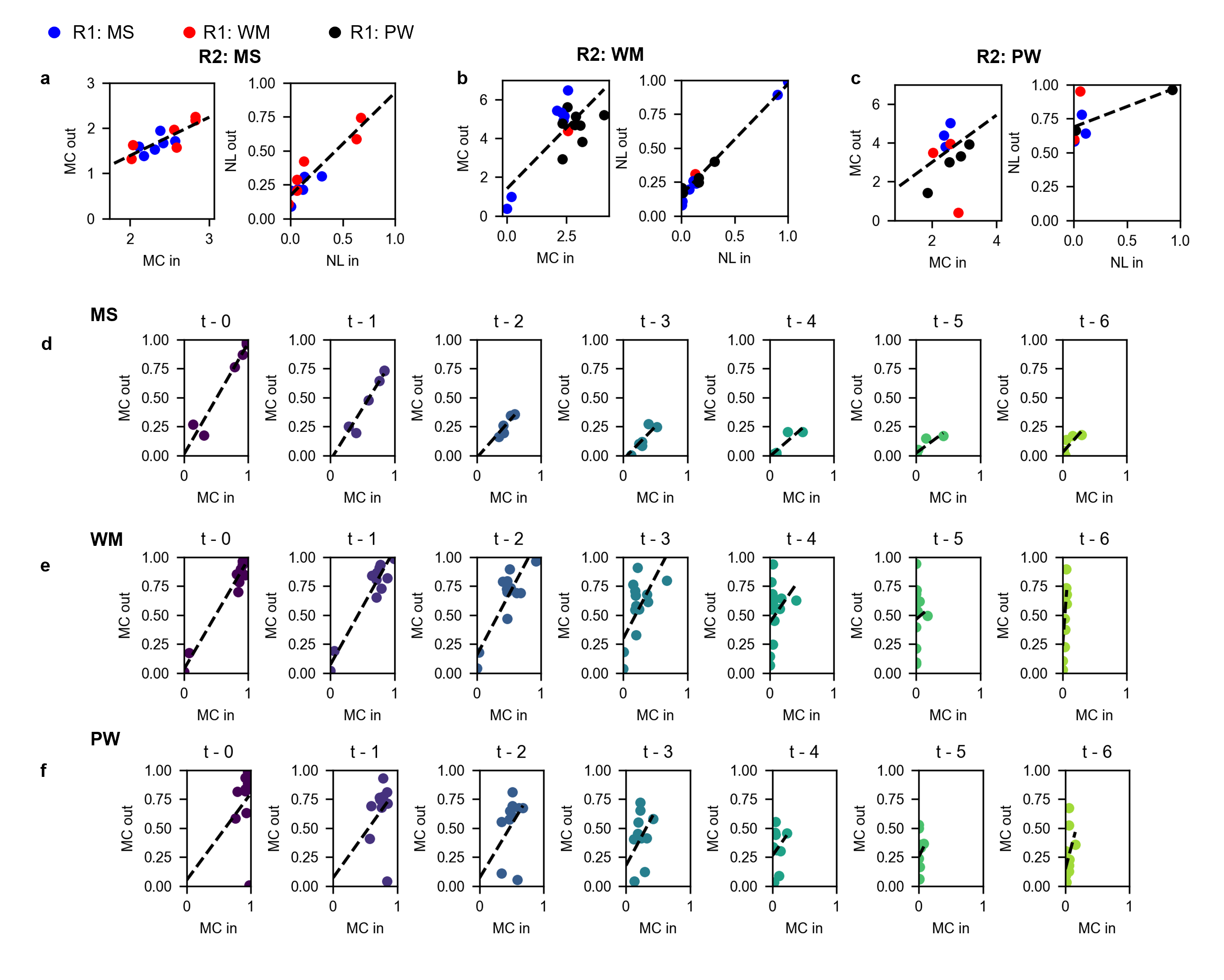}
    \caption{\textbf{Interconnection metrics and memory amplification.} Memory-capacity of reservoir 1 (MC\textsubscript{in}) vs Memory-capacity of reservoir 2 (MC\textsubscript{out})=when R2 is a) MS, b) WM and c) PW. High memory-capacity and nonlinearity are achieved when the interconnection memory-capacity and nonlinearity are high. d-f) MC\textsubscript{in} vs MC\textsubscript{out} when evaluating memory-capacity on specific previous inputs from t-0 (current input) to t-6. Dashed line represents a linear fit. 
    }
    \label{Mem_amplification}
\end{figure}
Supplementary Figures \ref{Mem_amplification} a-c) show the memory-capacity of reservoir 1 (MC\textsubscript{in}) vs memory-capacity of reservoir 2 (MC\textsubscript{out})=when R2 is a) MS, b) WM and c) PW.
Supplementary Figures \ref{Mem_amplification} d-f) shows the relationship for MC\textsubscript{in} vs MC\textsubscript{out} when calculating memory-capacity for specific previous inputs. In all cases, a linear trend is observed. For MS, the gradient of MC\textsubscript{in} vs MC\textsubscript{out} stays approximately the same as memory-capacity is evaluated to further previous inputs. For WM and PW, strong MC\textsubscript{out} is observed throughout. The gradient between input and output memory-capacity increases for further previous inputs. As such, the memory of the input signal is effectively amplified. The effect is strongest for WM. Small hints of long-term memory in the input signal translate to large contributions in the second reservoir. This provides additional information into why the ordering of reservoirs is crucial, as memory amplification at the end of the network is key for improved performance.

\subsection*{Supplementary note 6 - Interconnecting high-dimensional physical systems}
\begin{figure*}[htb!]
    \includegraphics[width=\textwidth]{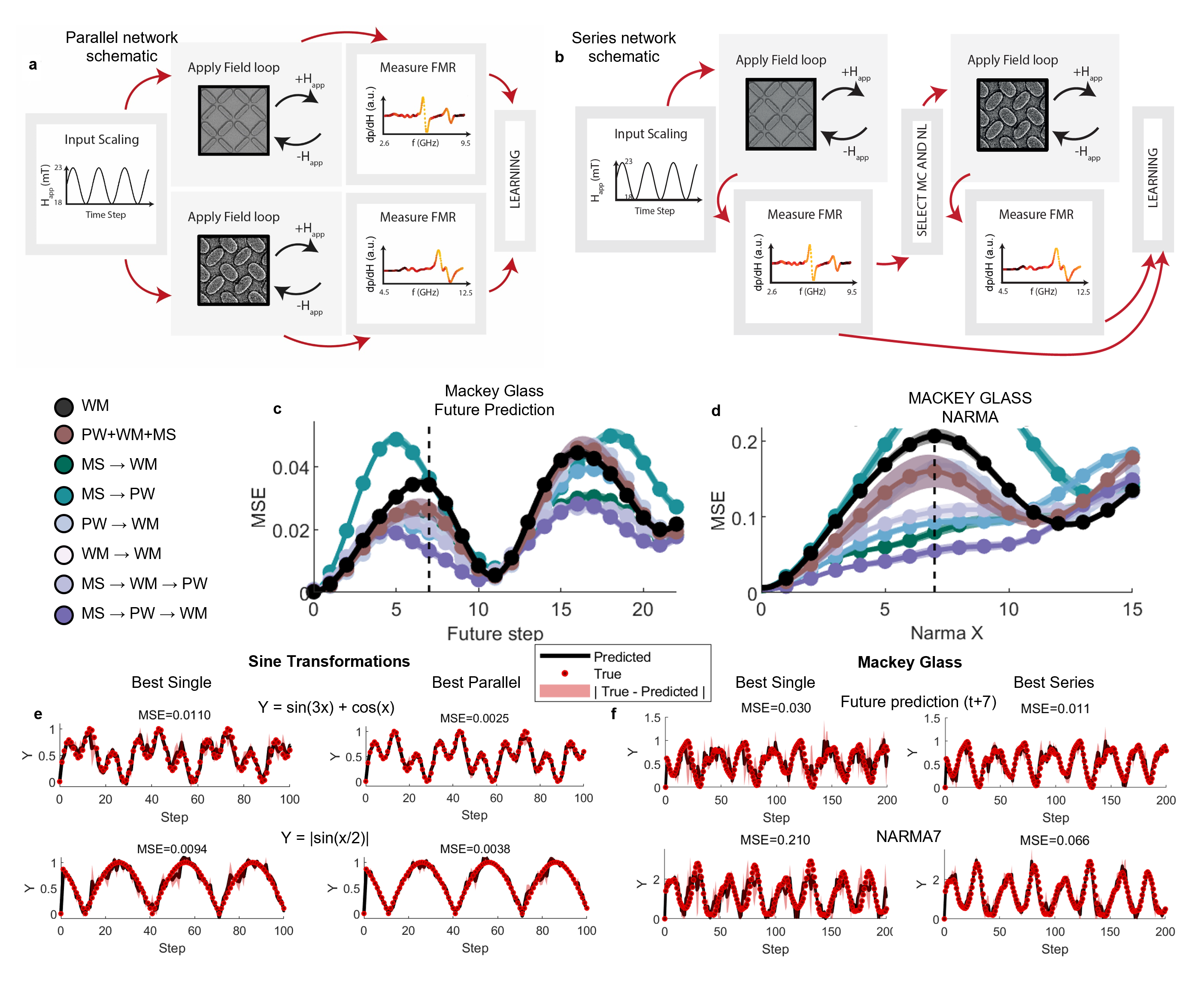}
    \captionof{figure}{\textbf{Interconnecting physical reservoirs.}
    a) Schematic of the parallel network architecture. Data is input to two separate reservoirs in parallel and their FMR spectra are combined for offline training. 
    b) Schematic of a series network architecture. Specific output-channels from the first reservoir are input into the second reservoir. Outputs of both reservoirs are combined for offline training.
    c) Mackey-Glass future prediction and d) NARMA-transformed Mackey-Glass for the best series network (WM, black), best parallel network (PW+WM+MS, brown) and each series network architecture. Series networks go from low memory-capacity to high memory-capacity reservoirs. Parallel networks only marginally reduce error.
    MSE profiles are flattened in series networks, with improvements up to 4 $\times$.
    e) Example predictions for sine-transformations for the best single and best parallel network.
    f) Example predictions are shown for t+7 and NARMA7 for the best single and series networks.
    }
   \label{Networks}
\end{figure*}

No individual physical system excels across all tasks, and any single system tends to underperform for harder tasks. This is a well-known symptom of single reservoir systems in software and hardware\cite{dambre2012information,inubushi2017reservoir,goldmann2020deep,love2023spatial,manneschi2021exploiting}. In software, multiple reservoirs with distinct responses have been combined into networks to harness the benefits of different dynamical behaviours\cite{dambre2012information,inubushi2017reservoir,goldmann2020deep,love2023spatial,manneschi2021exploiting}. In such networks, each reservoir can be viewed as a complex node\cite{d2023controlling} with high-dimensionality and distinct memory and nonlinearity scores.

Here, we construct parallel and series (often termed hierarchical or deep) networks of physical reservoirs, combining a synergistic suite of distinct nanomagnetic arrays with substantially enhanced performance. The increased output dimensionality gained from multiple physical arrays is beneficial for computation, but increases the likelihood of overfitting - a common challenge in machine-learning.  One way to avoid this is to increase the size of the training dataset, however, this luxury is often unavailable in real-world applications, especially in remote physical use-cases which must expensively acquire their own data using sensors. In response, we implement a feature selection scheme \cite{manneschi2021exploiting,manneschi2021sparce} to reduce the number of network outputs by discarding less useful channels, avoiding overfitting and providing robust, accurate performance (see Methods). 

Schematics of the parallel and series networks are shown in Supplementary Figure \ref{Networks} a) and b) respectively. In parallel networks, data is input into multiple arrays and the FMR response of each array measured. During offline training, the response of different arrays is concatenated to give the network output (i.e. two arrays with 300 frequency output channels produce a network with 600 outputs/parameters).

In series networks, data is input to the first array/reservoir (R1) and FMR response is measured. The 300-dimensional R1 FMR response now must be converted to a 1D field input for the second reservoir array R2. To accomplish this, we take the amplitude of a specific FMR frequency-channel and map it to an input field sequence for R2. The R1 output frequency-channel is selected via per-channel memory-capacity and nonlinearity evaluation. This analysis and scaling step is performed offline, with future prospects for on-chip implementations. The FMR spectra of R1 and R2 are then combined for offline training and prediction\cite{manneschi2021exploiting}. 

Parallel networks are capable of enhancing performance at tasks demanding high nonlinearity and low memory. Example transformations are shown in Supplementary Figure \ref{Networks} e) for sin(3x) + cos(x) and |sin(x/2)| where improvements up to 4.4$\times$ vs. the best single array observed when combining all three arrays in parallel (further transforms are discussed later).

Supplementary Figure \ref{Networks} c,d) shows series network MSE vs $t$ for future prediction (c) and NARMA-transformation (d) of Mackey-Glass time-series. The best single reservoir (WM, black) parallel network (PW + WM + MS, brown) are also shown. Parallel networks do not show significant enhancement due to the lack of memory-transfer in this architecture. Low memory-capacity (R1) to high memory-capacity (R2/R3) series network architectures significantly improve performance, up to 2.7 $\times$ and 4 $\times$ in three-layer deep networks for challenging t+7 and NARMA7 tasks respectively with predictions shown in Supplementary Figure \ref{Networks} i). As in the brain, reservoir ordering is critical\cite{manneschi2021exploiting}, with high memory-capacity to low memory-capacity network architectures showing weak improvements of $<$1.5 for all prediction tasks (supplementary note 7 discusses deep architecture ordering). Three-layer deep networks substantially outperform two layers, highlighting the benefits of more complex networks.

The best prediction is found for MS -> PW -> WM  (3 layers, memory-capacity = 6.7, nonlinearity = 0.75). While the 3-layer deep network has a memory-capacity improvement of 1.31 vs. the best single reservoir, MSE values are up to 3.7$\times$ lower than the single WM sample for future prediction tasks due to the expanded range of temporal responses/timescales in the deep network. This expanded range of temporal responses results in strong flattening of MSE periodicity vs. single and parallel reservoirs, especially evident in the NARMA-transform task (f). This enhanced temporal richness is not accurately reflected in a single memory-capacity/memory-capacity score. Memory-capacity and nonlinearity metrics are valuable guides for basic reservoir evaluation, but lack the granularity for accurately predicting performance.

\begin{figure}[h!]
    \centering
    \includegraphics[width=\textwidth]{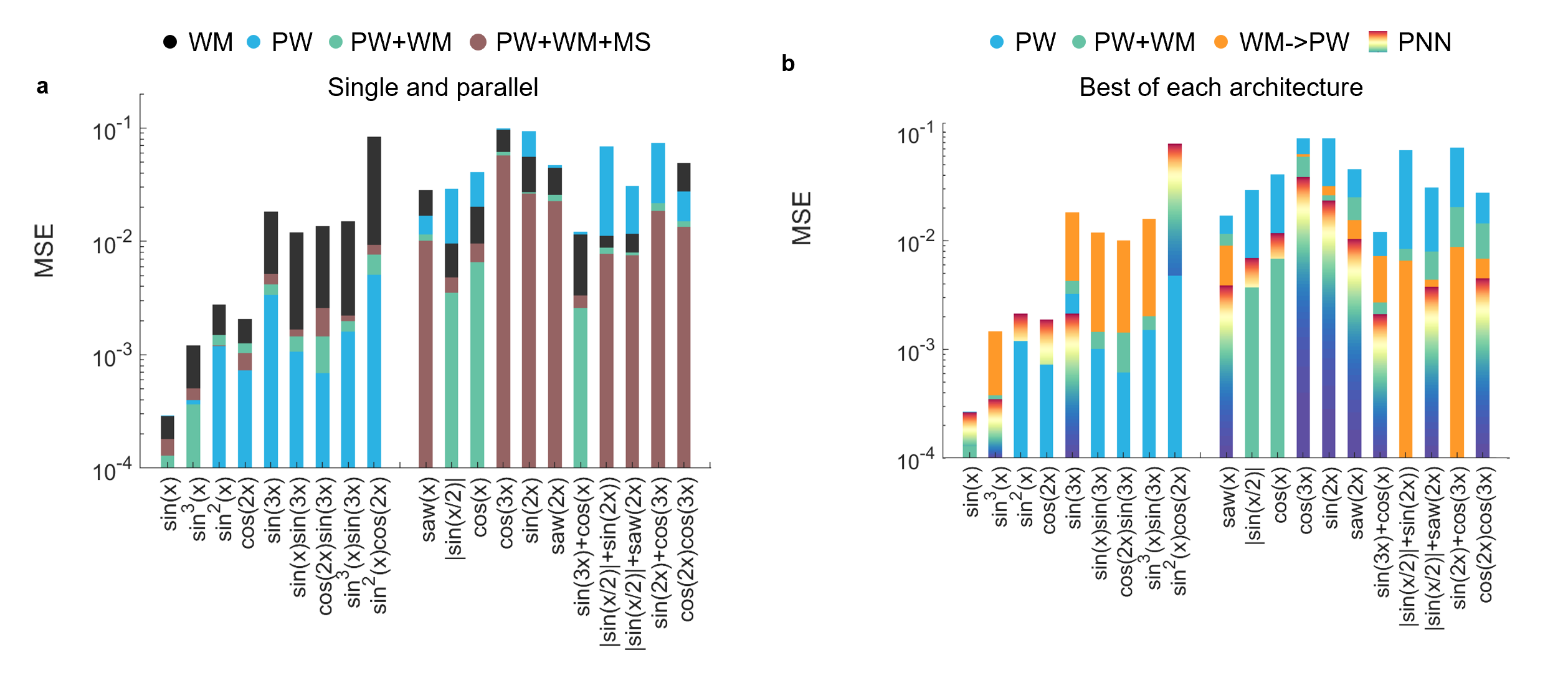}
    \caption{\textbf{Sine transformations with interconnected networks.} a) MSE for single and parallel configurations when transforming a sine input. Improvements are obtained across asymmetric (memory-capacity + nonlinearity) tasks, up to 4.4 $\times$ for sin(3x) + cos(x). Performance often worsens for symmetric (nonlinearity only) tasks. b)  MSE profiles the best single, parallel, series and PNN networks when transforming a sine input. Series improvements are only observed when connecting WM and PW for more complex asymmetric tasks. PNN outperforms other architectures for 9/20 tasks.}
    \label{Sine_transforms}
\end{figure}

Supplementary Figure \ref{Sine_transforms} a) shows MSEs of the best single and parallel networks for each signal-transformation task. Only the best 2- and 3-layer parallel architectures are shown. 
Lower MSE is observed for parallel architectures for asymmetric tasks requiring memory-capacity and nonlinearity. 
Parallel networks are able to harness memory-capacity and nonlinearity characteristics from different artificial spin reservoirs to reduce MSE for asymmetric transforms.
Lowest MSEs are observed when combining WM + PW, or all three artificial spin reservoirs with performance gains up to 4.4 $\times$ for more complex asymmetric tasks. For some tasks, adding MS to create a 3 layered parallel architecture worsens performance due to the limited transformation capabilities of that reservoir replacing more useful outputs from the other artificial spin reservoirs.

Interestingly, the best parallel architectures have lower nonlinearity than the single PW (0.65 vs. 0.75). The feature selection process optimises outputs across all tasks. Introducing high memory-capacity outputs from WM to improve asymmetric tasks comes at the cost of sacrificing high nonlinearity outputs from PW, reducing nonlinearity-only task performance. 

Supplementary Figure \ref{Sine_transforms} b) shows MSE values for single, parallel and series networks for signal-transformation. Series networks show enhanced performance for complex asymmetric tasks requiring both MC and NL (e.g. sin(2x) + cos(3x)). For simpler symmetric and asymmetric tasks, single and parallel networks dominate. 


\begin{figure}[h!]
    \centering
    \includegraphics[width=0.6\textwidth]{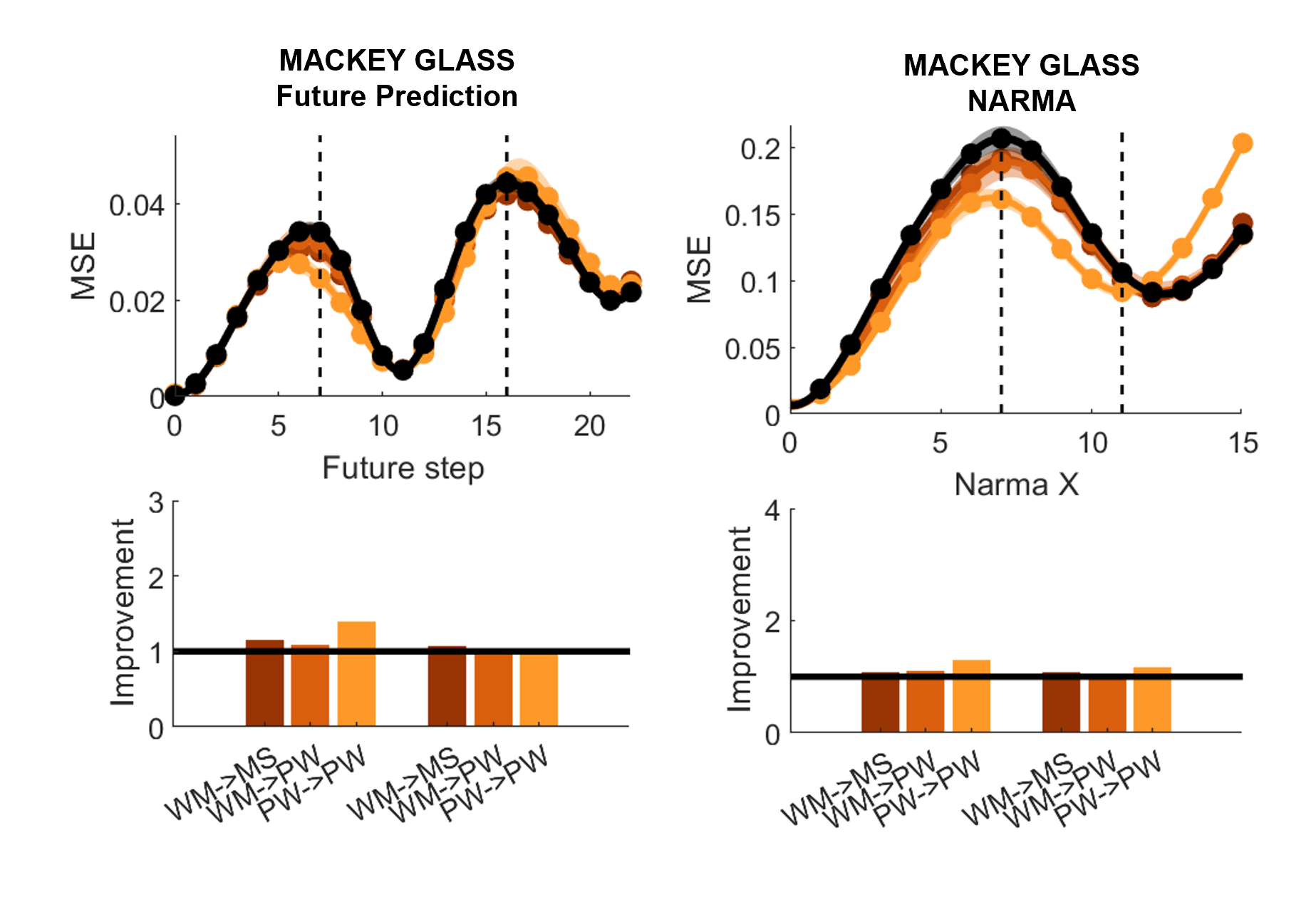}
    \caption{\textbf{Reservoir ordering effects on prediction performance}. Mackey-Glass future prediction and NARMA transformation in a deep network when going from a reservoir with high memory-capacity to a reservoir with low memory-capacity. Low improvements are seen across all tasks.
    }
    \label{fasttoslow}
\end{figure}

Supplementary Figure \ref{fasttoslow} shows the future prediction and NARMA transformation performance in a deep network when going from a reservoir with high memory-capacity to a reservoir to low memory-capacity. Low improvements are observed throughout. The initial reservoir response obscures short term information. The second reservoir simply mimics this response. As such, short term information is lost.

The ordering of artificial spin reservoir's plays an important role: if R1 has low memory-capacity, its response is captures short term behaviour. When this information is transferred, a high memory-capacity R2 receives information about short-term behaviour which it can retain for longer. As such, short-term memory is retained in R1 and long-term memory is retained in R2. For high-to-low memory-capacity, R1's output will be obscured by history-dependence, limiting the amount of information retained about short term behaviour. When this information is transferred, R2 simply mimics this information. As such, neither R1 or R2 retain a reasonable short-term memory, only long-term correlations are present, reducing the overall memory-capacity. 

Complex memory-capacity + nonlinearity signal-transformations require higher harmonic generation and responses shifted in phase w.r.t the input. By going from a high memory-capacity reservoir (WM) to a high nonlinearity reservoir (nonlinearity), both a linear and nonlinear representation of the input signal is present in the overall network (linear from high memory-capacity, nonlinear from high nonlinearity) producing a diverse set of lagged responses and higher harmonics.

\subsection*{Supplementary note 7 - Mackey-Glass attractor reconstruction}
\begin{figure*}[ht!]
    \includegraphics[width=\textwidth]{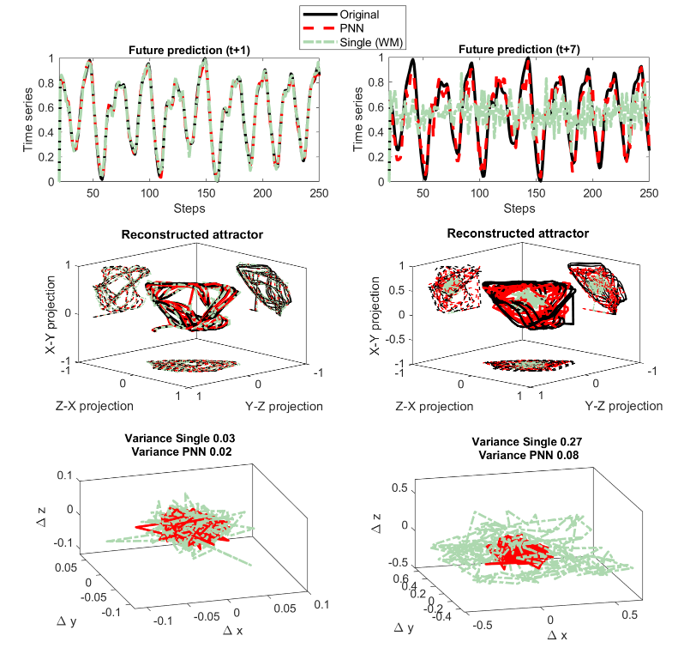}
    \captionof{figure}{\textbf{Mackey-Glass attractor reconstruction.}
    Reconstruction of the Mackey-Glass attractor using state space reconstruction for a prediction window of 1 and 7 time steps. Top panels show the original Mackey-Glass time series versus single and PNN networks. We visually see that at larger future prediction steps the combined reservoirs perform better. This can also be quantified in the reconstructed attractor (middle panels) via State Space Reconstruction, with a delay of 16 points and an embedding dimension 3, for the single and combined reservoirs. The difference between the combined and single attractor trajectories and the reconstructed attractor for the original time series is shown in the bottom right figure. We quantify the difference using the metric  $v=\sqrt{varx+vary+varz}/\sqrt{3}$. 
    }
   \label{MG_attract}
\end{figure*}

We use another methodology to understand the efficacy of our method in reconstructing the original time-series, even in the case of the prediction task. In Supplementary Figure \ref{MG_attract} we plot show two prediction tasks (t+1 and t+7) for the Mackey-Glass time series (top curves). We see that at low prediction windows, both a single reservoir and a network of reservoirs reconstruct the time series. We can analyse the efficacy of the task by looking at the reconstructed attractors using the State Space Reconstructor method \cite{ssr1,ssr2,ssr3}, embedding it in 3-dimensions with a delay of $T=16$. We compare the original reconstructed attractor (black curves) versus the single (green curves) and PNN models (red curves). These plots contain much more information than MSE  (center middle plots). As we can see, at low prediction windows (t+1) the reconstructed attractor is well approximated by both the RC models. However, at longer prediction windows (t+7) the PNN follows more closely the attractor. This can be quantified by looking at the multi-dimensional variance of the 3-dimensional time series obtained by subtracting the trajectory of the attractor from the original time-series versus the two reconstructed ones (bottom plots). We see error is much more spread at longer prediction windows for the single RC.

\subsection*{Supplementary note 8 - PNN in the underparameterised regime}

In Figure 2 of the main text, the featuer selection algorithm is allowed to freely chose between all outputs. The total number of outputs used for each architecture is: WM - 34, MS+PW+WM - 44, MS$\rightarrow$WM - 42, MS$\rightarrow$WM$\rightarrow$PW - 93 and PNN - 13567. The enriched readout dimensionality is a distinct advantage of the PNN in terms of computational performance. 

\begin{figure*}[ht!]
    \includegraphics[width=\textwidth]{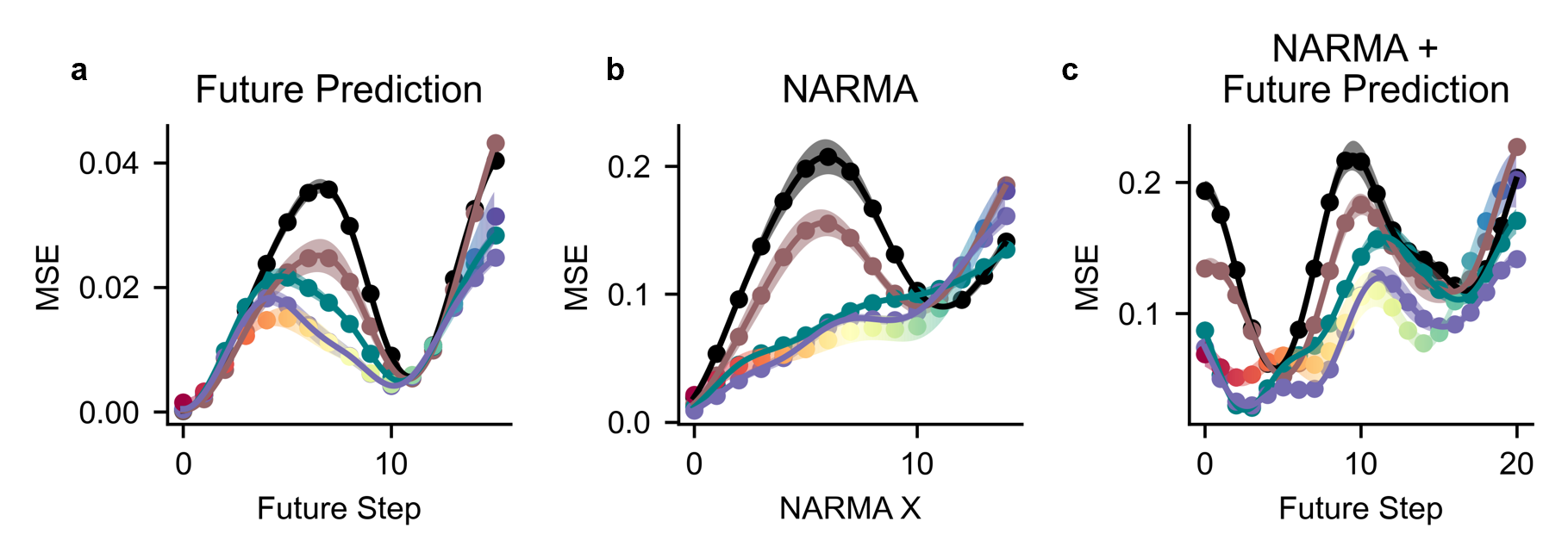}
    \captionof{figure}{\textbf{Underparameterised PNN performance.} MSE profiles for a) Mackey-Glass future prediction, b) NARMA transformation and c) future prediction of NARMA-7 processed Mackey-Glass for the best single, parallel, series and PNN when the PNN is forced to be underparameterised (i.e. the number of parameters is less than the size of the training set). PNN MSE is similar to the series networks as it is not able to harness all of the outputs.
    }
   \label{PNN_UP}
\end{figure*}

Supplementary Figure \ref{PNN_UP} shows the performance when the PNN is constrained to have a similar number of outputs the other networks. Here, 97 PNN are used during training. We find that performance is at a similar level to the three-series network. Performance is worse for some tasks. This is because the feature selection algorithm is stochastic and requires more iterations to explore the entire readout space. Additionally, the algorithm finds a set of outputs that performs well at all tasks. Given enough iterations, the performance of the PNN will at least match other architectures, as the PNN contains outputs from all other network architectures.

\subsection*{Supplementary note 9 - Software comparison}
We now compare the performance of the various network architectures to software models, specifically echo state networks (ESN) and multilayer perceptrons (MLP). For ESN, we vary the number of internal nodes (N\textsubscript{nodes}) and for each node we randomly initialise 50 networks as described in the Methods section \textit{ESN Comparison}. For MLPs, we vary the size of the hidden layers (N\textsubscript{hidden}) from 1-500 and number of previous inputs the MLP receives (T\textsubscript{seq}) from 1-10 (see Methods for MLP details).

\begin{figure}[ht!]
    \centering
    \includegraphics[width=\textwidth]{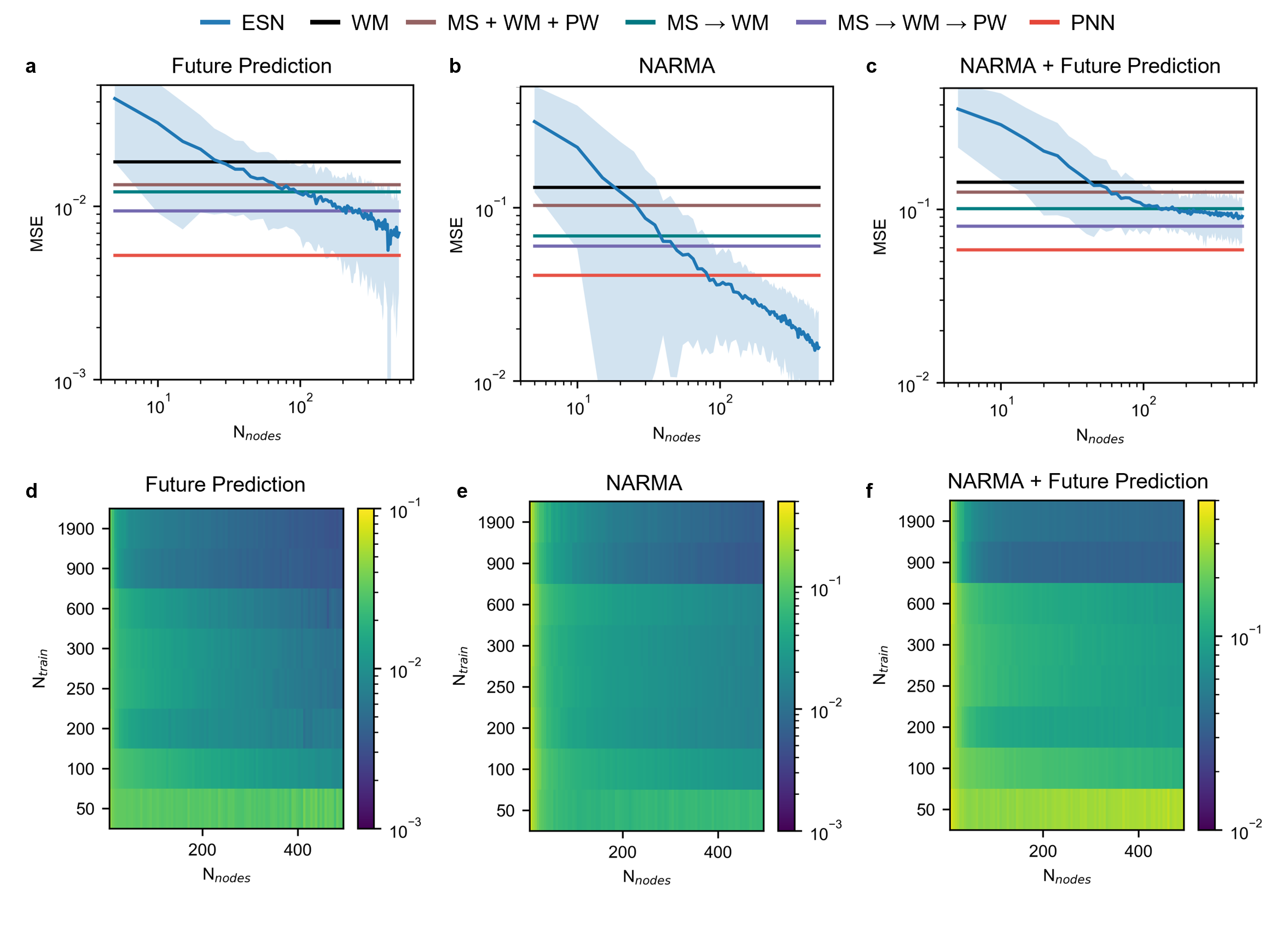}
    \caption{\textbf{Comparison to echo state networks.} ESN performance when varying the number of ESN nodes for a) Mackey-Glass future prediction, b) NARMA transform, c) NARMA transform plus future prediction as presented in the main text. In each plot, the blue line and shaded region represent the average MSE and error over 50 randomly initialise ESNs at a given number of nodes (N\textsubscript{nodes}). Solid flat lines represent the MSE from the physical networks presented in the main text. Here, the training data size (N\textsubscript{train}) = 200. MSE is an average over multiple tasks: t+0 to t+12 for future prediction (a), NARMA 0 - 12 for NARMA transforms (b) and t+0 to t+20 for NARMA plus future prediction (c). Panels d-f) show heatmaps of MSE for varying N\textsubscript{train} and N\textsubscript{nodes}.}
    \label{ESN_comparison}
\end{figure}

Supplementary Figure \ref{ESN_comparison} shows the performance of ESNs with varying number of nodes (N\textsubscript{nodes}) for a) Mackey-Glass future prediction, b) NARMA transform, c) NARMA transform plus future prediction as presented in the main text. Panels d-f) show the ESN performance when varying N\textsubscript{train} and N\textsubscript{nodes} for each task. In a-c), the blue line represents the average error of 50 randomly initialised ESNs for a given number of internal nodes with shaded area representing $\pm 2\theta$. Coloured lines show the performance of the various networks explored in the main text. In a-c), N\textsubscript{train} = 200. MSE represents the average performance over multiple tasks:  t+0 to t+12 for future prediction (a), NARMA 0 - 12 for NARMA transforms (b) and t+0 to t+20 for NARMA plus future prediction (c). For ESNs, as N\textsubscript{nodes} and N\textsubscript{train} increases, MSE reduces as expected. For each task, we find that single arrays are matched by ESNs with 20 - 40 nodes. As the physical network complexity increases, the corresponding ESN size to match performance also increases. The ESN size required to match the PNN performance is 100 for the NARMA task and $>$500 for prediction tasks. 

\begin{figure}[ht!]
    \centering
    \includegraphics[width=\textwidth]{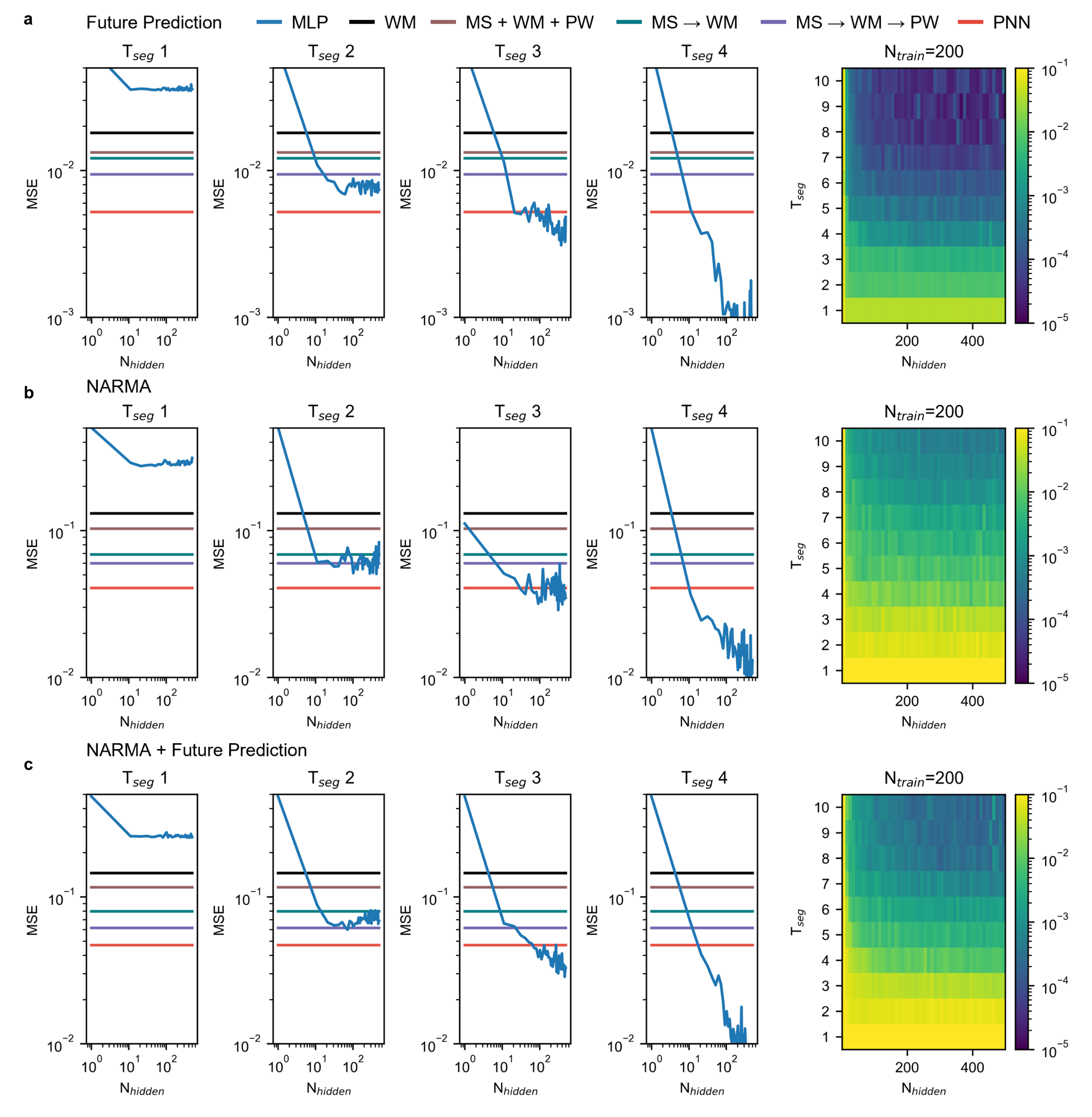}
    \caption{\textbf{Comparison to mulitlayer perceptrons (MLP).} MLP performance when varying the size the hidden layers (N\textsubscript{hidden}) and the number of previous inputs given to the model during training (T\textsubscript{seg}) for a) Mackey-Glass future prediction, b) NARMA transform, c) NARMA transform plus future prediction as presented in the main text. In each plot the blue line represents the MSE of the MLP for a given number of hidden nodes. Solid flat lines represent the MSE from the physical networks presented in the main text. Here, the training data size (N\textsubscript{train}) = 200. MSE is an average over multiple tasks: t+0 to t+12 for future prediction (a), NARMA 0 - 12 for NARMA transforms (b) and t+0 to t+20 for NARMA plus future prediction (c).}
    \label{MLP_comparison}
\end{figure}

We now compare the physical network performance to MLPs. MLPs are static, i.e. they hold no information about previous inputs and will fail at predictive tasks. As such, we vary the number of previous inputs given to the MLP during each time step. Supplementary Figure \ref{MLP_comparison} shows the performance of MLPs when varying the size of the hidden layers (N\textsubscript{hidden}) and the number of previous inputs provided to the MLP (T\textsubscript{seg}) for a) Mackey-Glass future prediction, b) NARMA transform, c) NARMA transform plus future prediction as presented in the main text. As with the ESN, MSE represents the average performance over multiple tasks:  t+0 to t+12 for future prediction (a), NARMA 0 - 12 for NARMA transforms (b) and t+0 to t+20 for NARMA plus future prediction (c). The MLP MSE reduces as N\textsubscript{hidden} and T\textsubscript{seg} increases as expected. For T\textsubscript{seg} = 1, the physical networks outperform all trialled MLPs due to the lack of past information in the MLP. We find that series networks are well matched to MLPs with T\textsubscript{seg} = 2 and N\textsubscript{hidden} $\sim$ 10 and the PNN is matched to an MLP with T\textsubscript{seg} = 3 and N\textsubscript{hidden} $\sim$ 10-50 for all tasks. Whilst the resulting MLP is simple, the network interconnections must be trained using gradient descent. This is fairly inexpensive for small networks, but as network and task complexity increases, so does the expense of training MLP weights.

We now implement the same networking methodology described in the main text with software echo-state networks.

\begin{figure}[ht!]
    \centering
    \includegraphics[width=\textwidth]{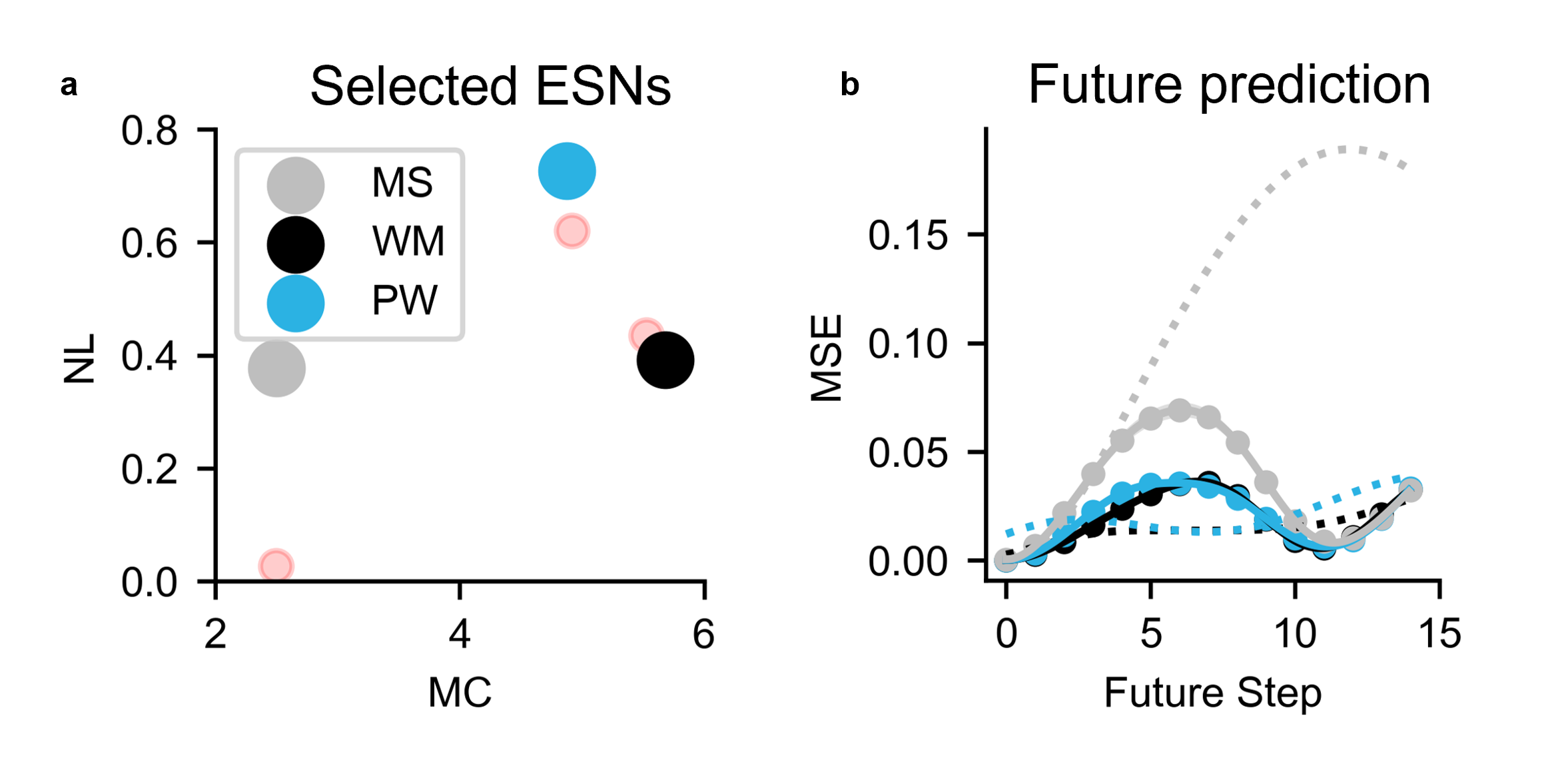}
    \caption{\textbf{Metrics and MSE of echo-state networks.} a) Memory-capacity (MC) and non-linearity (NL) of the nanomagnetic arrays and three ESNs (red markers) chosen to have similar metrics. b) Performance when predicting future values of the Mackey-Glass equation for nanomagnetic arrays (solid line) and ESNs with similar metrics (dotted lines).}
    \label{Selected ESNs}
\end{figure}

We begin by initializing ESNs with 200 nodes (to ensure a similar number of output channels in comparison to nanomagnetic arrays) and select three which display similar memory-capacity and non-linearity to the three nanomagnetic arrays. Supplementary Figure \ref{Selected ESNs} shows the memory-capacity and non-linearity of the three nanomagnetic arrays alongside three ESNs with similar properties (red circles). The ESNs were found via randomly varying the hyperparameters of the ESN and selecting ESNs which had the closest characteristics.

We find that, despite the ESNs and nanomagnetic arrays displaying similar metrics, the performance shows different characteristics. For the two high memory-capacity ESNs, the MSE rises slightly from future step 1 – 3, and then becomes flat, whereas the hardware reservoirs display a periodic profile in the MSE. We find that it is challenging to find an ESN that displays the same characteristics as the hardware reservoirs. This is not surprising, as nanomagnetic reservoirs and software ESNs are governed by different underlying dynamics. In nanomagnetic arrays, coupling arises from the dipolar field which decays spatially and hence coupling only occurs locally, whereas in ESNs, coupling is randomly assigned across the network. In ESNs, the coupling strengths between nodes are initialised and remain fixed, whereas in nanomagnetic arrays, the coupling between neighbouring elements depends on the state of those elements (e.g. uniform magnetisation or vortex). Furthermore, in ESNs, only certain nodes are coupled to the input, which enhances memory-capacity as certain nodes require a number of network updates before receiving information about a particular input. On the other hand, in nanomagnetic arrays, each node is subject to the input data when it is first applied. Finally, experimental nanomagnetic systems are subject to noise, whereas ESNs are noise free. It is possible that, with extensive optimisation, an ESN could be initialised that has more similar characteristics to nanomagnetic arrays, however, achieving this is beyond the scope of this report.

We note that when initialising ESNs, the majority of them had memory-capacity far beyond the nanomagnetic arrays (close to 8). 

We now interconnect these arrays following the same methodology used to connect different nanomagnetic arrays. We begin by analysing the channel specific time correlation and non-linearity of the three ESNs displayed in Supplementary Figure \ref{ESN per channel}.

\begin{figure}[ht!]
    \centering
    \includegraphics[width=\textwidth]{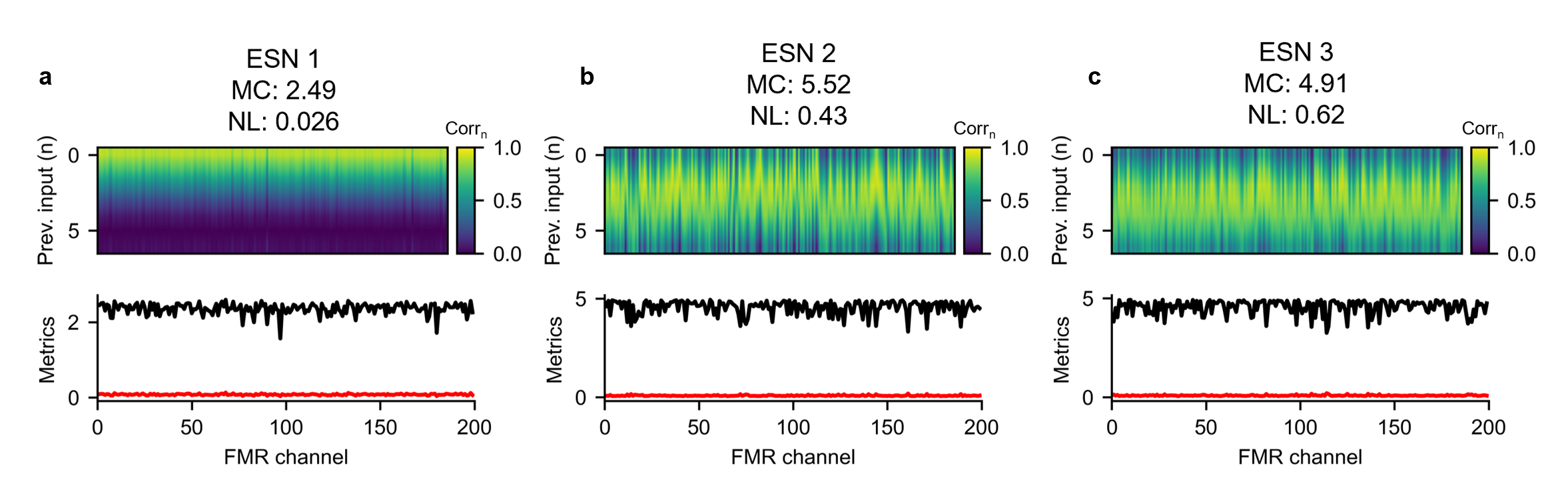}
    \caption{\textbf{ESN metrics}. Per-channel correlation and non-linearity for the three selected ESNs are shown in a-c.}
    \label{ESN per channel}
\end{figure}

When comparing these results to the nanomagnetic reservoirs (SI Supplementary Figure \ref{perchannel}) we again see fundamental differences in the dynamics of ESNs vs nanomagnetic hardware. For the low memory-capacity ESN (ESN 1), we find that all channels are only correlated with short-term previous inputs. Conversely, for the higher memory-capacity ESNs (ESN 2$\&$3), there are a broad range of correlations. Some channels are correlated with short term previous inputs, whereas some are correlated with previous inputs many time-steps ago. In contrast, in nanomagnetic arrays, all of channels are most strongly correlated with short-term inputs and only a handful are correlated with long-term previous inputs. This is a symptom of the differences between internode couplings in ESN when compared to nanomagnetic arrays. The random, sparse connections in ESN allow many different dynamic timescales whereas the spatially constricted, local couplings in nanomagnetic systems prevent this behaviour. 

For each ESN, we now select six channels: three with the highest memory capacity and three with the highest non-linearity, giving a total of 18 input time-series to feed into the next layer. Each ESN is the subject to all input sequences to produce 54 2-series networks. We then create 3-series networks by selecting the 2-series architectures which: 1) begin with the low memory-capacity reservoir (ESN 1) and 2) which display the lowest MSE. We then perform the same per-channel metric analysis, generating 9 input sequences which are passed to the final ESN to form a 3-series network. Finally, we combine the responses from all single, 2-series and 3-series networks using the same methodology as the PNN. We evaluate the performance of the ESN networks using the same feature selection methodology described in the Methods section of the manuscript.

\begin{figure}[ht!]
    \centering
    \includegraphics[width=\textwidth]{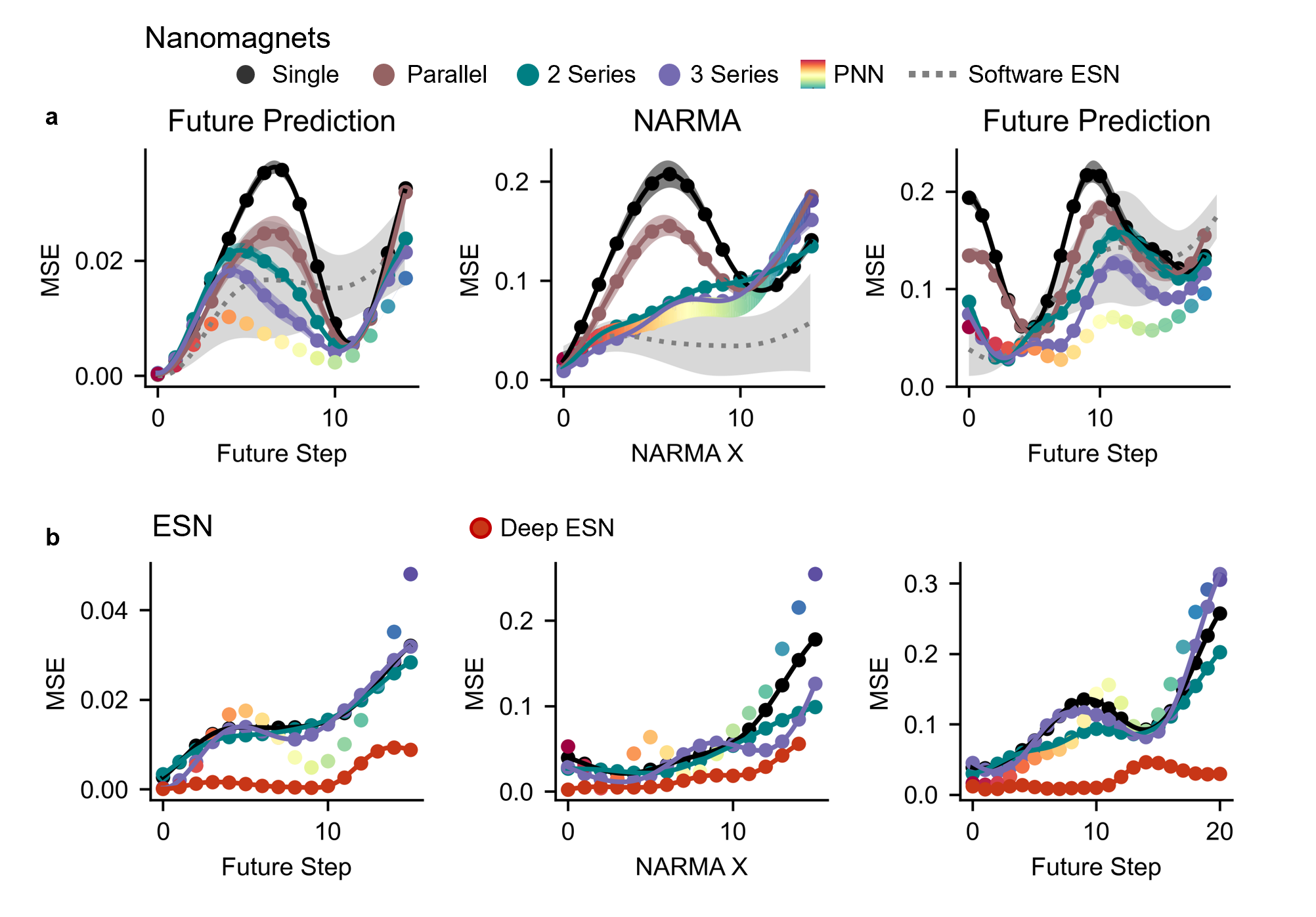}
    \caption{\textbf{Comparison between nanomagnetic arrays and software ESNs when interconnecting networks.} a) shows the experimental arrays from the main manuscript. b) shows the ESN networks.}
    \label{ESNvssystem}
\end{figure}

Supplementary Figure \ref{ESNvssystem} compares the performance of single, 2-series, 3-series and PNN architectures for the three prediction tasks presented in the main text. The top row shows the nanomagnet response previously included in the manuscript and the bottom row shows the ESN network results. The ESN results displayed are taken from the architecture with the lowest MSE over all future / NARMA steps for a given task. When interconnecting ESNs, we find that whilst improvements are observed for some tasks, they are less pronounced than for the physical system. Series networks only show improvements for future prediction t+1 and t+2, NARMA 10 and above, and NARMA + Future prediction t+4 to t+13, whereas series networks improve all predictions for the nanomagnetic arrays. Software analogues of the PNN architecture show improvements for more tasks (notable future prediction t+7 – t+12), but also have worse performance for other tasks. We believe this is because single ESNs already display a rich set of memory time-scales, and hence interconnecting arrays does not enrich the overall network to the same extent. These results give us an indication that interconnecting physical systems can give MSE improvements which produce a similar level of performance to software counterparts. However, we stress here that the initial ESNs selected are not optimised for performance. If we were to optimise ESNs for performance, then it is likely that the software network would outperform nanomagnetic hardware.

In addition, we test deep echo state networks, where multiple interconnections between different ESN layers are made (i.e. multiple nodes from one ESN connect to multiple nodes in another) in a similar manner to previous work of some of our authors\cite{manneschi2021exploiting}. We connect the three ESNs with similar characteristics to the physical nanomagnetic reservoirs into a 3 x 3 network, whereby each ESN is present in each layer. Layer interconnections are initialised randomly and we average over 10 random trials. The prediction results for the three tasks are shown in Supplementary Figure \ref{ESNvssystem} (red curve, ‘Deep ESN’). As expected, the performance dramatically improves compared to our proposed method as there is an increase in the amount of information being transferred between different layers. Currently, such interconnections are not possible with the hardware approach as each input is only able to accept one input per time step. If a physical system is used which can take multiple inputs at each time step (e.g. memristors), we can connect it in this way and achieve a computationally powerful network with just a handful of nodes.

\newpage
\subsection*{Supplementary note 10 - Further overparameterisation details}

We now provide further details regarding the overparameterisation regime, specifically, how train length, network sub components and task affect the ability to reach a beneficial overparamterised regime. 

\begin{figure}[ht!]
    \centering
    \includegraphics[width=\textwidth]{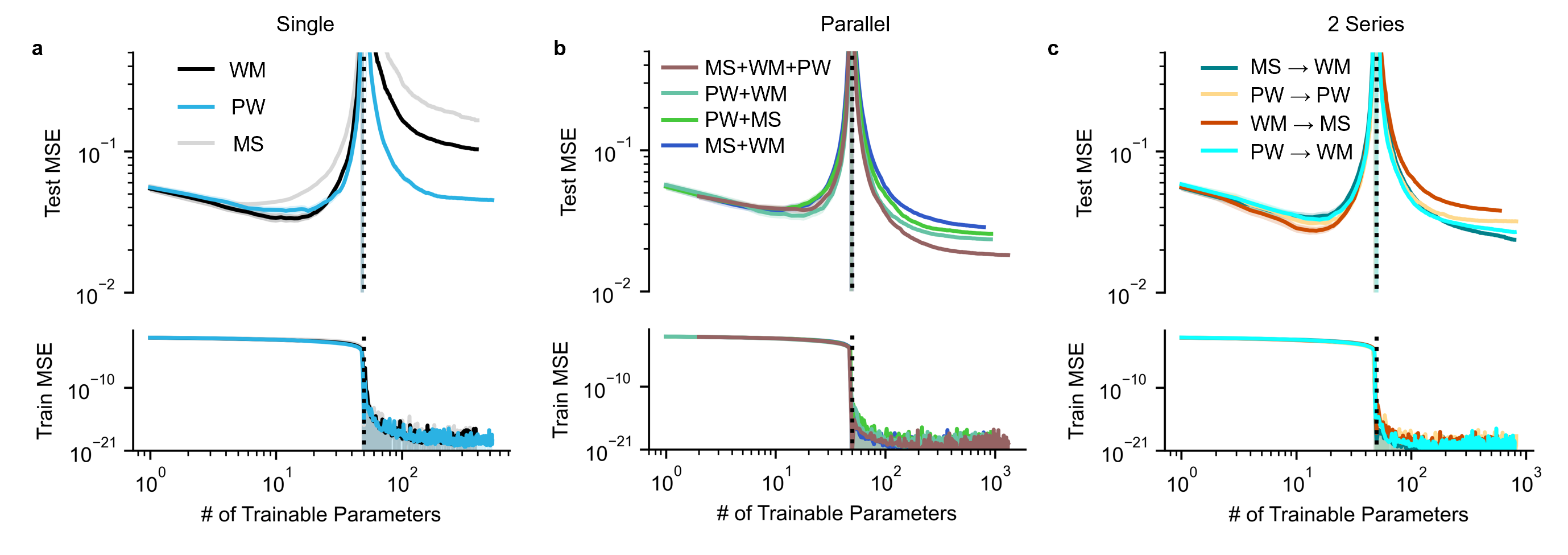}
    \caption{\textbf{Overparameterisation with all PNN architectures.} MSE dependence on the number of trainable parameters for a) Single networks, b) Parallel networks and c) 2-series networks}
    \label{OP_morenets}
\end{figure}

We begin with a discussion of how the network subcomponents and architecture affect the effective dimensionality of the output and therefore ability to reach a beneficial overparameterised regime. Supplementary Figure \ref{OP_morenets} shows the dependence of the number of trainable parameters (i.e. number of output channels used during regression) on the MSE for a) single, b) parallel and c) 2-series networks. Here, MSE is an average over 6 prediction tasks (see methods for details) encompassing tasks that all networks perform well and those which simpler networks struggle. Beginning with the single arrays, we see that no single array benefits from being in an overparameterised regime as the minimum MSE in this region is higher than the minimum MSE in the underparameterised regime. However, we see that the underlying array affects the ratio of MSE\textsubscript{OP} to MSE\textsubscript{UP}. PW shows a far greater reduction in the overparameterised regime compared to WM and MS. Combining these results with the memory capacity and non-linearity calculations in and the spectral evolution during computation in Figure 1 of the main text, we see that having a diverse spectral output is key to achieving a beneficial overparameterised regime. PW has structural diversity, producing a large variety of magnetic states which are detected at different resonant frequencies. MS is a highly linear reservoir, with relatively simple dynamics and magnetic elements which all occupy the same frequency space. WM is a linear sample with strong memory capacity, with magnetic elements largely occupying the same frequency space, but behaving in a complex manner. For an equivalent frequency range and resolution, PW fills out the frequency space, producing a diverse set of spectral outputs. When in the overparameterised regime, the regression can harness this diverse response to produce a lower MSE in comparison with the spectrally limited WM and MS arrays. 

Moving to parallel networks, we see that all combinations of parallel networks benefit from an overparameterised regime. This is expected when considering the effective dimensionality of the parallel network outputs. Each of the three networks has a distinct set of internal dynamics and output spectra, as such, combining them in parallel enhances the spectral diversity, enabling networks to benefit from an overparameterised regime. Comparing the parallel networks with different architectures, we see that the improvements gained depend on which networks are included. Combining the two low non-linearity reservoirs (MS+WM) shows a weaker improvement than combining creating a 2 parallel network with the non-linear PW sample as expected. 
For series networks, the level of improvements depend on the arrangement of nodes as seen in the underparameterised regime. WM->MS shows worse overparameterised performance than MS$\rightarrow$WM despite the theoretical internal dynamics being the same for each network. As we are assessing performance on a prediction task, both the separated node dynamics and combined dynamics of the system plays a role. MS$\rightarrow$WM is more capable of transferring previous information and increasing memory capacity, leading to better predictions and a lower MSE. As such, both the range of dynamics of each node, and there arrangement play a role. Interestingly, WM$\rightarrow$MS displays better improvements than when combining MS and WM in parallel. Here, the internal dynamics are the same and the difference arises from the input that the second layer receives. By feeding the output of one array to the input of the next, an additional set of diverse dynamics arises from the input sequence. 

\begin{figure}[ht!]
    \centering
    \includegraphics[width=\textwidth]{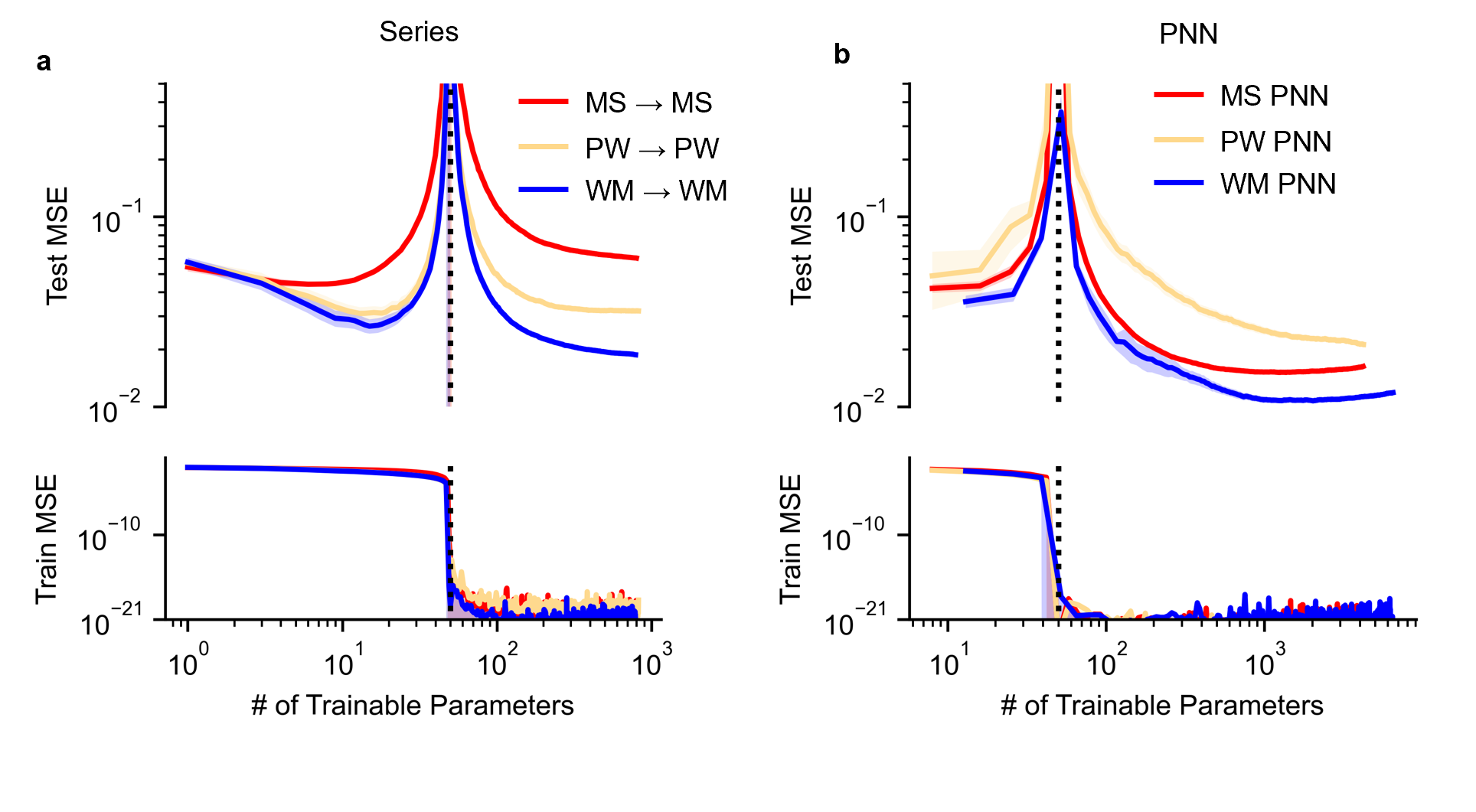}
    \caption{\textbf{Single array overparameterisation.} MSE dependence on the number of trainable parameters for network with one physical system. a) MSE curve for 2-series networks comprising one physical system for each reservoir explored in this work. b) MSE curve for physical neural networks comprising one physical system. For 2 series networks, only WM$\rightarrow$WM reaches a beneficial overparameterised regime. For PNNs, all systems can reach a beneficial overparameterised regime.}
    \label{OP_single}
\end{figure}

 Supplementary Figure \ref{OP_single} a) shows the double descent curves for three series networks MS$\rightarrow$MS, WM$\rightarrow$WM and PW$\rightarrow$PW. As mentioned, MS is a low non-linearity, low memory capacity reservoir. The readout diversity for this array is low. When creating a 2 series network with MS in both layers MSE OP is higher than MSE UP as the readout diversity is not sufficient to overcome overfitting. Creating series networks with systems that have more complex behaviour and higher readout diversity can lead to a beneficial overparameterised regime. As such, by designing an appropriate system, one can reach overparameterisation with just a single 2 series network. Note that if we combined two copies of a node in parallel, the results would be equivalent to a single system as the two sets of outputs would be identical. Figure X b shows the double descent curve when creating a PNN out of a single system. Here, we collate all responses from each sample from the 1st and 2nd layer of the PNN, such that a network will only contain one system as its node, but will receive a variety of different inputs. In each case, the system is able to reach overparameterisation. The implications of this are that even with a low non-linearity, low memory capacity system, one can reach a beneficial overparameterised regime by feeding multiple inputs to that system to access different dynamics. These inputs can be produced from the output of the first layer, or by masking the original input signal. Note than we can not compare the final MSE’s of the networks in this plot as the inputs for a specific array were not necessarily optimised to produce a high-memory response. For example, here, PW PNN has higher MSE than MS PNN. This is due to many of the PW arrays receiving highly non-linear inputs from the first layer in order to assess how output and input metrics are related in Figure 2 e) of the main text. 

\begin{figure}[ht!]
    \centering
    \includegraphics[width=\textwidth]{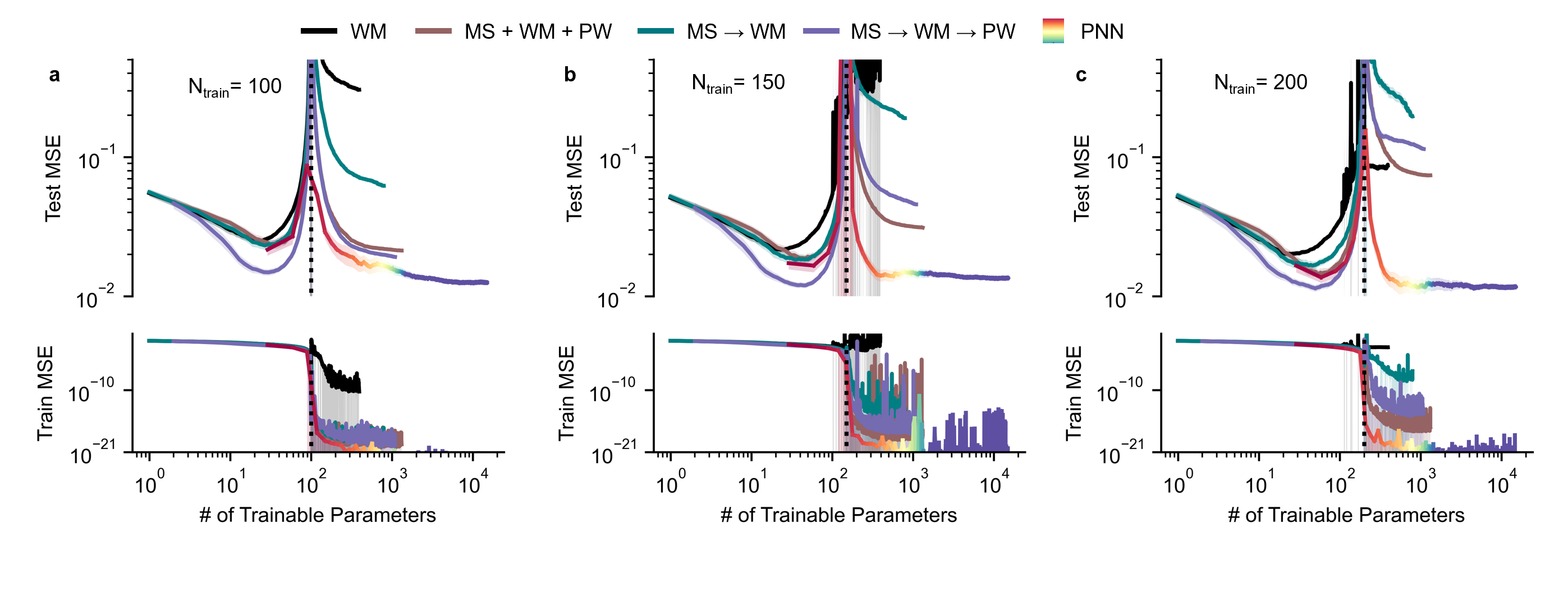}
    \caption{\textbf{MSE dependence on the number of trainable parameters when varying the size of the training set.} a,b,c) are training lengths of 100, 150 and 200 respectively. For larger training dataset sizes, single, parallel and series networks do not reach a beneficial overparameterised regime due to their limited effective readout dimensionality. PNN can reach a beneficial overparameterised regime in all cases.}
    \label{OP_varytrain}
\end{figure}

Supplementary Figure \ref{OP_varytrain} shows the double descent curves for selected architectures when varying the train length from 100 – 200. As N\textsubscript{train} increases, the MSE in the OP regime inreases for single, series and parallel networks as the effective dimensionality of the readout is too small to overcome overfitting. The PNN strongly benefits from an overparameterised regime in all cases.

\begin{figure}[ht!]
    \centering
    \includegraphics[width=\textwidth]{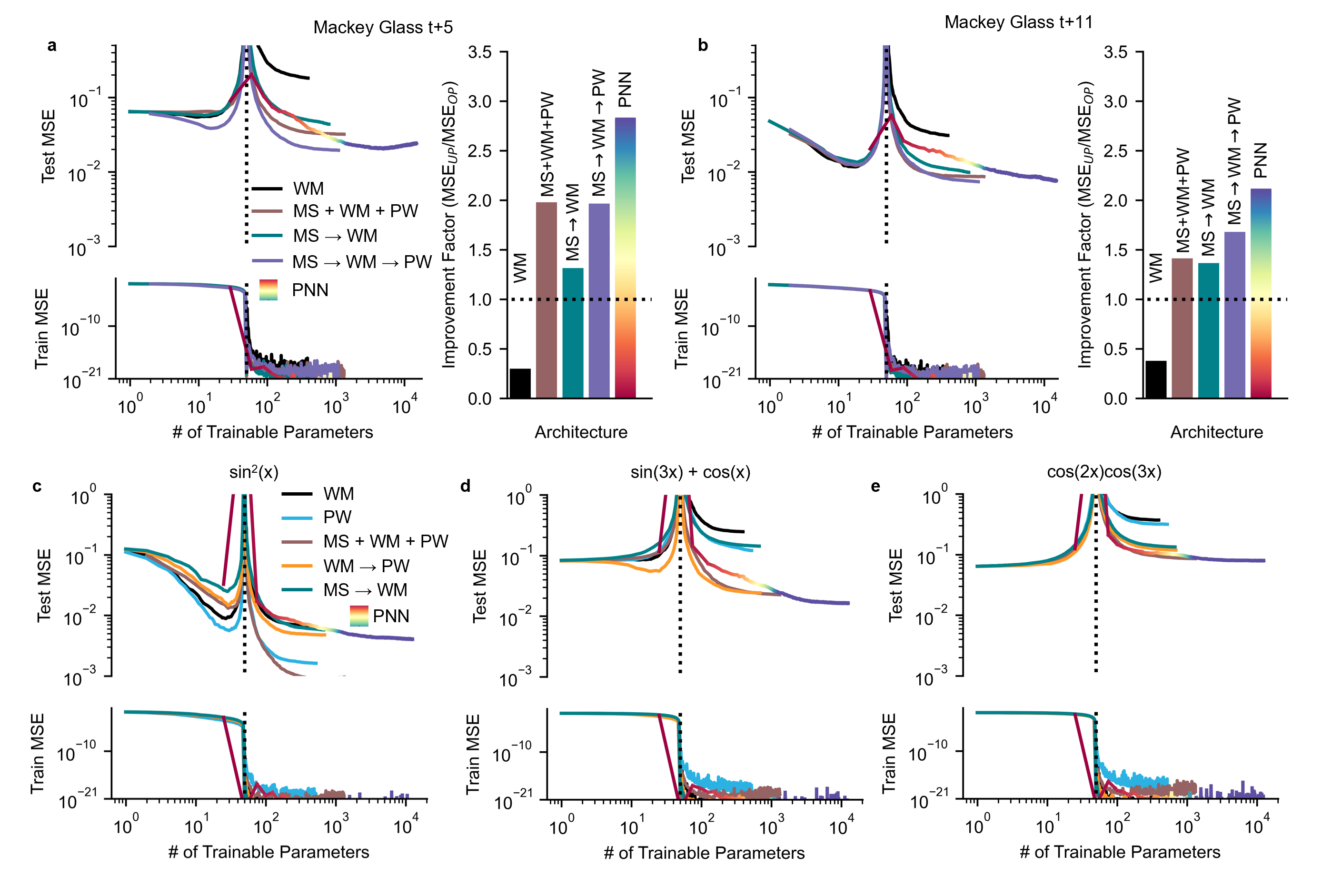}
    \caption{\textbf{Effect of task on overparameterisation.} MSE dependance on number of parameters when predicting a) t+5 and b) t+11 of the Mackey-Glass equation as well as transforming a sinusoidal input to c) sin$^2$(x), d) sin(3x) + cos (x) and e) cos(2x)cos(3x). The benefits of overparameterisation are task dependent.}
    \label{OP_tasks}
\end{figure}

Finally, we discuss how task affects overparmeterisation. Supplementary Figure \ref{OP_tasks} shows the MSE dependence on the number of parameters when predicting a) t+5 and b) t+11 of the Mackey-Glass equation. A similar trend is observed in both cases. This is true for all future predictions evaluated (not shown). Here, whilst the task changes, the requirements of the network do not. Both tasks require strong memory capacity to achieve good performance as seen in Figure 2 of the main text.

By changing the requirements of the task, we start to see greater differences in the benefits of overparameterisation. Supplementary Figure \ref{OP_tasks} shows double descent curves transforming a sinusoidal input to c) sin$^2$(x), d) sin(3x) + cos (x) and e) cos(2x)cos(3x). Here the tasks require different levels of non-linearity and memory. 

\subsection*{Supplementary note 11 - Overparameterisation with gradient descent}
\begin{figure}[ht!]
    \centering
    \includegraphics[width=\textwidth]{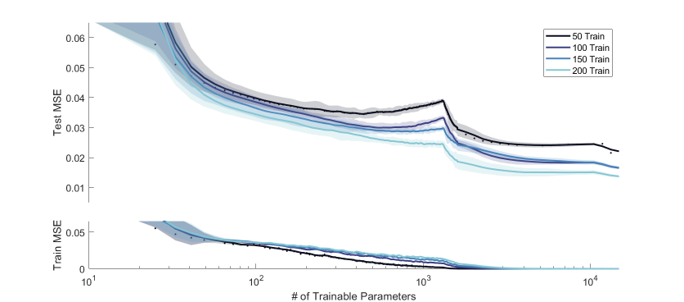}
    \caption{\textbf{Overparameterisation with gradient descent.} Train and test MSE for the PNN when increasing the number of outputs and the length of the training set. Here, gradient descent for weight training. Here network weights are added by layer order, whereby outputs are added from the first PNN layer, then second layer, then final layer.}
    \label{GradDec}
\end{figure}

Supplementary Figure \ref{GradDec} shows the train and test MSE for the PNN architecture when increasing the number of parameters. Here, gradient descent is used to optimise network weights. Here, network outputs are added in the order that they appear in the PNN network i.e. outputs are added from the first layer, then the second layer, then the final layer. In the main text, outputs are sampled randomly from all layers. We observe the same qualitative trends as with linear regression, whereby the train and test MSE reduce in the underparameterised regime. Test MSE increases in the overfitting regime and then reduces in the overparameterised regime. The number of outputs required to reach an overparameterised state is higher than in the main text due the way network weights are added. This shows that the observed results are robust to the training method used, and the method of sampling outputs.

\subsection*{Supplementary note 12 - Further meta-learning predictions}
\begin{figure}[ht!]
    \centering
    \includegraphics[width=0.85\textwidth]{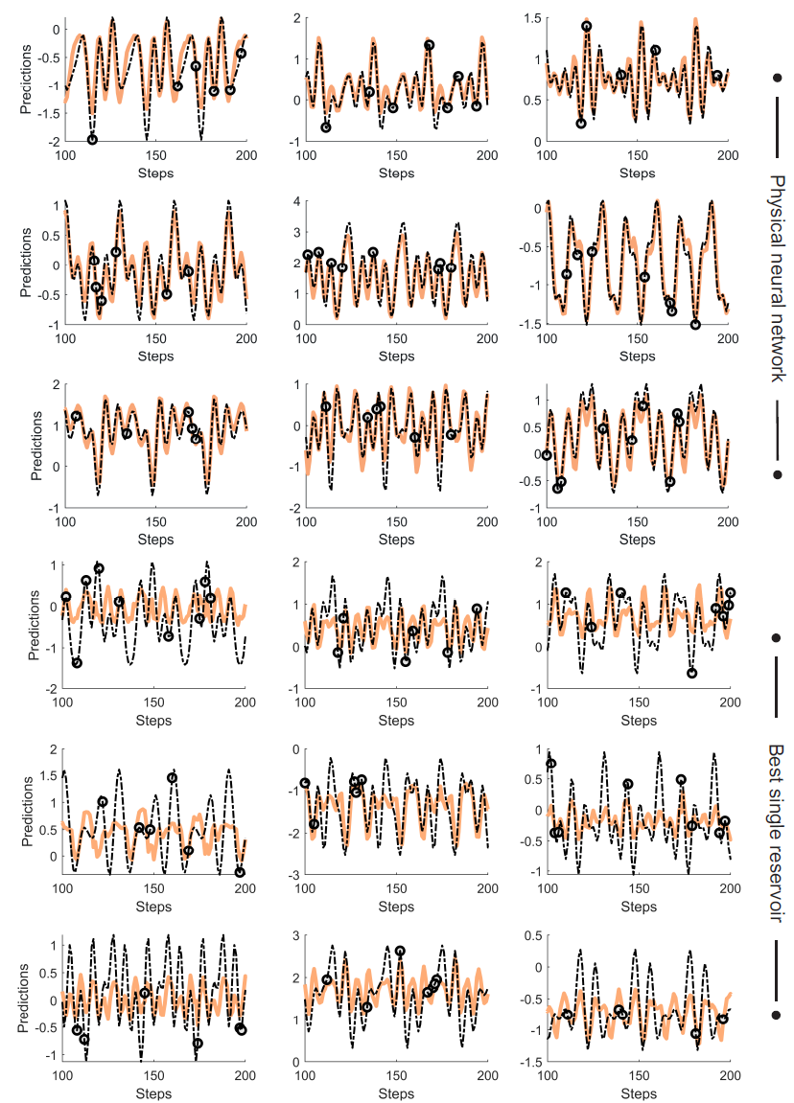}
    \caption{\textbf{Few-shot learning examples.} Further examples of frequency decomposition task with the PNN and comparison with the best single reservoir (WM).}
    \label{Meta_supp}
\end{figure}

Supplementary Figure \ref{Meta_supp} shows further examples of meta-learning predictions for a variety of targets. Both the PNN and the best single reservoir (WM) are shown. The PNN is able to accurately reproduce a wide variety of targets whereas the single reservoir fails at all tasks. As mentioned in supplementary note 4, the symmetry of the signal with respect to the input signal determines whether a signal transformation task requires non-linearity only, or memory and non-linearity. The  sin$^2$(x) transformation is symmetric, requiring only non-linearity. In this case, we see tht all networks considered reach beneficial overparameterisation. The task is relatively simple, and effectively dimensionality required to perform well is low. Here, we see that PW and MS+WM+PW outperform the series and PNN networks. As mentioned in supplementary note 4, this is because series and PNN networks are designed to increase history-dependent response, which hinders performance for symmetric sine transforms. Both sin(3x) + cos (x) and cos(2x)cos(3x) are asymmetric with respect to the input, requiring the same input value to be transformed to different outputs across the waveform and therefore require both memory capacity and non-linearity. For sin(3x) + cos (x),  only PNN, parallel and WM$\rightarrow$PW networks reach beneficial overparameterisation. For cos(2x)cos(3x), no networks reach beneficial overparameterisation. As such, care must be taken to match the network architecture to the task to achieve the best performance in the overparameterised regime. 

\subsection*{Supplementary note 13 - Performance comparison to conventional hardware}

We now benchmark the resources required to run echo-state networks (the closest software analogue to physical reservoirs) and compare this to the power consumption, speed, and energy consumption of both our current computing architecture and proposed future devices.

In software, an ESN is initialised with random interconnections. Applying input and updating the ESN takes the form of a set of matrix calculations as:

\begin{align}
    \mathbf{x}(t+\delta t)=(1-\alpha)\mathbf{x}(t)+\alpha f(\mathbf{W}_{in}\mathbf{s}(t)+\mathbf{W}_{esn}\mathbf{x}(t))
\end{align}
Where $\mathbf{x}(t)$ is the ESN state, $\mathbf{W}_{in}$ and $\mathbf{W}_{esn}$ are matrices representing the randomly initialised input and internode weights, $\alpha$ is a scaling factor and \textit{f} is a non-linear activation function (here tanh). This is repeated for the entire input sequence, with states saved at every input. Following this, regression is performed on the saved states to give a computational output. 

For nanomagnetic arrays, a state update corresponds to application of an input (global magnetic field, Oersted field or spin-orbit torque). The array updates itself naturally through the physics. The state is then readout via ferromagnetic spectroscopy and saved via analogue to digital conversion. After this, regression is then performed via conventional CMOS, as it is in the software reservoir. 
As such, any differences in performance arise from the input / update process, the readout, and analogue to digital conversion.

We begin by analysing the resources of software reservoirs. A common way of benchmarking the resources of software algorithms is to analyse the number of floating-point operations (FLOPs). We can dissect the reservoir update formula to analyse the number of FLOPs in relation to the number of nodes in an ESN (N). For a reservoir with N nodes, $\mathbf{x}(t)$ is a matrix of size [N,1], $\mathbf{W}_{in}$ is a matrix with size [N,1], $\mathbf{W}_{esn}$ is a matrix with size [N, N], $\alpha$ is a float, \textit{f} is a non-linear activation (here tanh). We note than the number of FLOPs for addition of two matrices of size [m,n] is \textit{mn} and the multiplication of two matrices of size [m,n], [n,p] is \textit{nm(2p-1)}. From this we get the theoretical number of FLOPs for each subcomponent of the update formula:

\begin{itemize}
    \item (1-$\alpha$) – 1 FLOP
    \item (1- $\alpha$)$\mathbf{x}(t)$ – multiplication of a float and [N,1] matrix = N FLOPs
    \item $\mathbf{W}_{in}$$\mathbf{s}(t)$ – multiplication of [N,1] and [1,1] matrix = N FLOPs
    \item $\mathbf{W}_{esn}$$\mathbf{x}(t)$ – multiplication of [N,N] and [N,1] matrix = N(2N-1) FLOPs
    \item $\mathbf{W}_{in}$$\mathbf{s}(t)$ + $\mathbf{W}_{esn}$$\mathbf{x}(t)$ – addition of two [N,1] matrices = N FLOPs
    \item \textit{f} – activation function of a [N,1] matrix. tanh takes ~20 FLOPs per value. = 20N FLOPs
    \item Final addition of two [N,1] matrices = N FLOPs

\end{itemize}

In total, to update a reservoir we have N + 2N-1 + N(2N-1) + N + 20N + N = 2N$^{2}$ + 24N - 3 FLOPs. In practice, this value is higher due to the initialisation of a reservoir, and the optimisation of reservoir properties (e.g. running multiple random iterations to find an optimal configuration). We must also take reservoir sparsity into account (echo-state networks do not have all-to-all connections, instead they are sparse networks where only $\sim$20$\%$ of connections are non-zero) giving a final equation of FLOPS = 2(sN)$^{2}$ + 24sN-3 FLOPs where s is the sparsity factor (between 0 and 1).

From this, we can benchmark the time and energy required to perform a reservoir update on a variety of different hardware. We compare the FLOPS (FLOPs per second) and power consumption of GeForce RTX 4070\cite{TechPowerUp_2024}, Intel Core i7-13700H\cite{CPU}, Raspberry Pi-4\cite{RASP}, and Zynq SoC Z-7020 FPGA\cite{FPGA} in Supplementary Table \ref{Conventional}. FLOPS/Watt is calculated as FLOPS / Thermal Design Power.

\begin{table}[h!]
\begin{tabular}{l|l|l|l}
\textbf{System}                                                            & \textbf{FLOPS} & \textbf{Thermal Design Power (W)} & \textbf{FLOPS / Watt} \\ \hline
GeForce RTX 4070                                                           & 4.30E+13       & 200                               & 2.15E+11               \\
\begin{tabular}[c]{@{}l@{}}Intel   Core i7-\\      13700H CPU\end{tabular} & 5.38E+11       & 45                                & 1.19E+10               \\
Raspberry Pi-4                                                             & 1.35E+10       & 6.6                               & 2.05E+09               \\
Zynq SoC Z-7020                                                            & 1.80E+11       & 2.5                               & 7.20E+10              
\end{tabular}
\caption{Performance of three conventional computing hardware options}
\label{Conventional}
\end{table}

We can use this information to calculate the time and energy required to update a reservoir using: Time = FLOPs\textsubscript{update} / FLOPS and Energy = FLOPs\textsubscript{update} / FLOPS/Watt where FLOPs\textsubscript{update} is the number of FLOPs required to update an ESN of a particular size. Supplementary Table \ref{Conventional_energy} compares the time and energy required to update a 30 node and 500 node ESN for each of the above hardware with a sparsity of 0.2.

\begin{table}[h!]
\begin{tabular}{l|l|l|l|l}
\textbf{System}            & \textbf{Time (30 nodes) (s)} & \textbf{Energy (30 nodes) (J)} & \textbf{Time (500 nodes) (s)} & \textbf{Energy  (500 nodes) (J)} \\ \hline
GeForce RTX 4070           & 4.95E-12                     & 9.91E-10                       & 5.21E-10                      & 1.04E-07                         \\
Intel   Core i7-13700H CPU & 3.96E-10                     & 1.78E-08                       & 4.17E-08                      & 1.87E-06                         \\
Raspberry Pi-4             & 1.58E-08                     & 1.04E-07                       & 1.66E-06                      & 1.09E-05                         \\
Zynq SoC Z-7020            & 1.18E-09                     & 2.96E-09                       & 1.24E-07                      & 3.11E-07                        
\end{tabular}
\caption{Energy and time when updating an echo-state network with 30 and 500 nodes using conventional hardware.}
\label{Conventional_energy}
\end{table}

We can compare this to the time, energy and power required to input and readout data in a nanomagnetic array. Our existing system uses a NanOsc CryoFMR PPMS probe to record spectra, and hence uses a superconducting magnet to apply fields. Operating this system carries a vast energy cost. Instead, we can compare this to a similar experimental set-up that uses an electromagnet. The electromagnet supplies a magnetic field using ~ 1 A and 5 V for 100 ms, resulting in a power of 5W and energy consumption of 0.5 J. For readout, we use a microwave source with power of 17 dBm (0.05 W), which takes $\sim$20s to record the entire spectra. This gives an energy of 1 J. Finally, the output is fed into a lock-in amplifier, which runs at 60 W, requiring 1.2 kJ, by far the dominating factor. As such, the total power, time, and area of the existing array is 65 W, 20.1s and ~1 m$^{2}$.

We now calculate the power time and energy of a projected device. Below we separate out each of the components described in the manuscript. We assume Oersted field pulses of 1 ns, spin orbit torque (SOT) switching pulse of 0.25 ns, RF power of 0 dBm (1 mW) for sequential measurement, RF noise source power of 150 mW for parallel measurement and measurement time per channel of 14 ns (steady state precession requires ~100 oscillations which gives $\sim$14 ns for a frequency of 7 GHz resonance\cite{ross2023multilayer}, for 200 channels this gives 2800 ns). Magnetic array area is based on 5 x 5 um projected device.

We compare Oersted field and spin-orbit torque input methods. To calculate the current required to produce a Oersted field from a current carrying microstrip, we use the method of Ref. \cite{kiermaier2012electrical}. The Oersted field generated in a current carrying wire is defined by Ampere's law. We assume a wire with a 10 $\mu m$ width and infinite length, split into finite elements of size . A differential current $dI_{z}$ at position ($x_j$, $y_j$),  produces a differential magnetic field $dH_{x,y}$ described as follows:

\begin{equation}
   dH_{x,y} = \frac{dI_{z}}{2\pi\left((x-x_j)^2+(y-y_j)^2)\right)} \left( \frac{-(y-y_j)}{(x-x_j)} \right)
\end{equation}

\begin{figure}[ht!]
    \centering
    \includegraphics[width=0.85\textwidth]{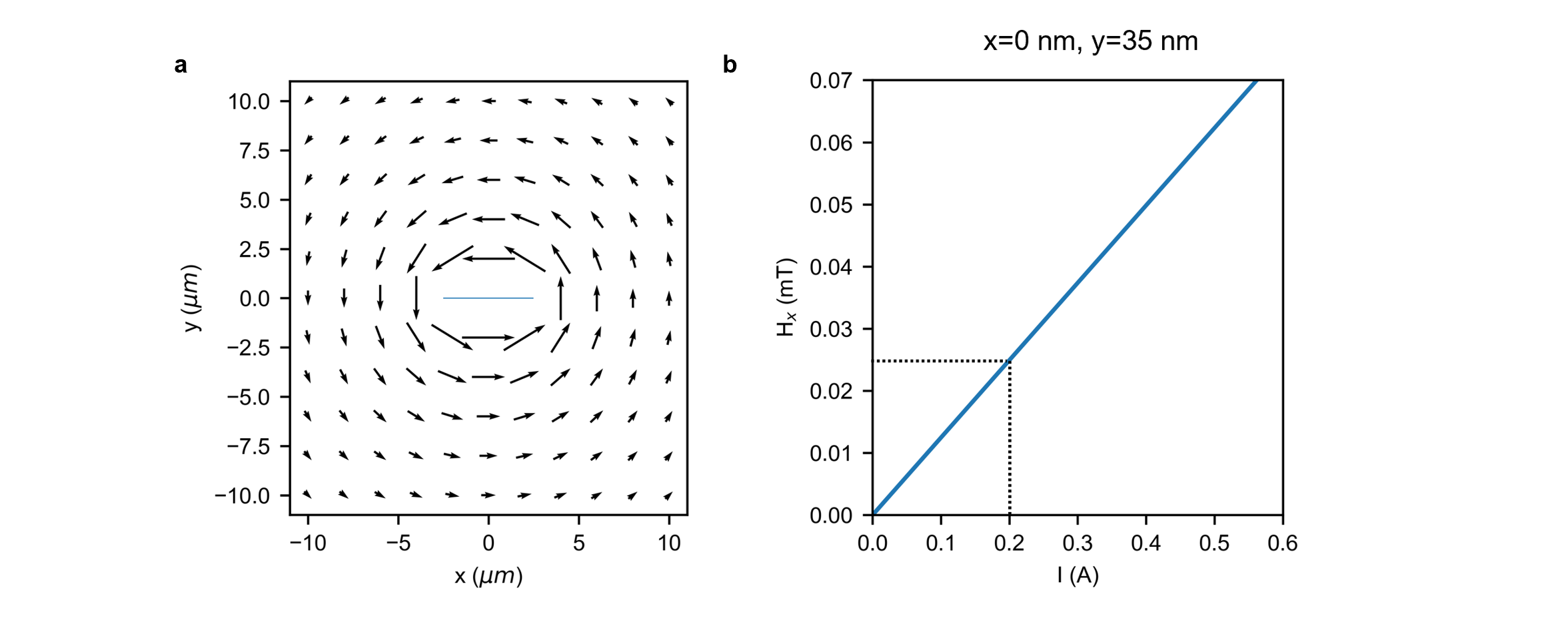}
    \caption{\textbf{Oersted field calculation.} a) Oersted field profile from an infinite wire with width of 5 $\mu m$ and thickness of 50 nm. The blue line represents the stripline. b) x-component of magnetic field at x = 0 nm and y = 35 nm as a function of current. Dotted lines represent the current needed to produce a 25 mT field.}
    \label{Energy}
\end{figure}

By numerically integrating over all positions in the wire, the magnetic field at $(x,y)$ is obtained. Supplementary Figure \ref{Energy} a) shows the magnetic field profile of a wire with width 5 $\mu m$, thickness 50 nm and infinite length. Supplementary Figure \ref{Energy} b) shows the x-component of the magnetic field at x = 0 nm, y = 35 nm. To generate a 25 mT field, a current of 200 mA (8$\times 10^{11} A m^{-2}$) is required.

Assuming the strip-line is made of Cu with a resistivity $\rho$ = 16.8 $n\Omega m$, the resistance of a Cu strip with dimensions of 5 $\mu m$ $\times$ 5 $\mu m$ $\times$ 50 nm is given by R =  $\rho l / A$ = 0.336 $\Omega$. The power consumption can be calculated as P = $I^2 R$ = 13.4 mW.

To calculate the Ta power consumption for spin torque switching we follow the same methodology. For a current density of 10$^{11}$ A m$^{-2}$ and Ta sheet dimensions of 5 $\mu m$ $\times$ 5 $\mu m$ $\times$ 5 nm, we have a current of 2.5 mA. Assuming a Ta resistivity of 131 $n\Omega m$ gives a resistance R = 26.2 $\Omega$ and P = 0.164 mW.

The torque generated on a ferromagnet through spin-obit effects is inversely proportional to the thickness of the ferromagnetic material\cite{manchon2019current}. For thicker elements (20 nm) considering torques alone, we can expect at least a 10 X increase in the current density required to switch. The coercive field of the nanomagnets also increases as thickness increases (we estimate up to 4 X based on micromagnetic simulations), as such we can expect between 10 – 40 X increase. This equates currents in the range of 25 – 100 mA and powers in the range of 16.8 – 269 mW range. Therefore, it may be preferable to use the Oersted field switching scheme for thicker nanomagnets.

\begin{figure}[ht!]
    \centering
    \includegraphics[width=0.85\textwidth]{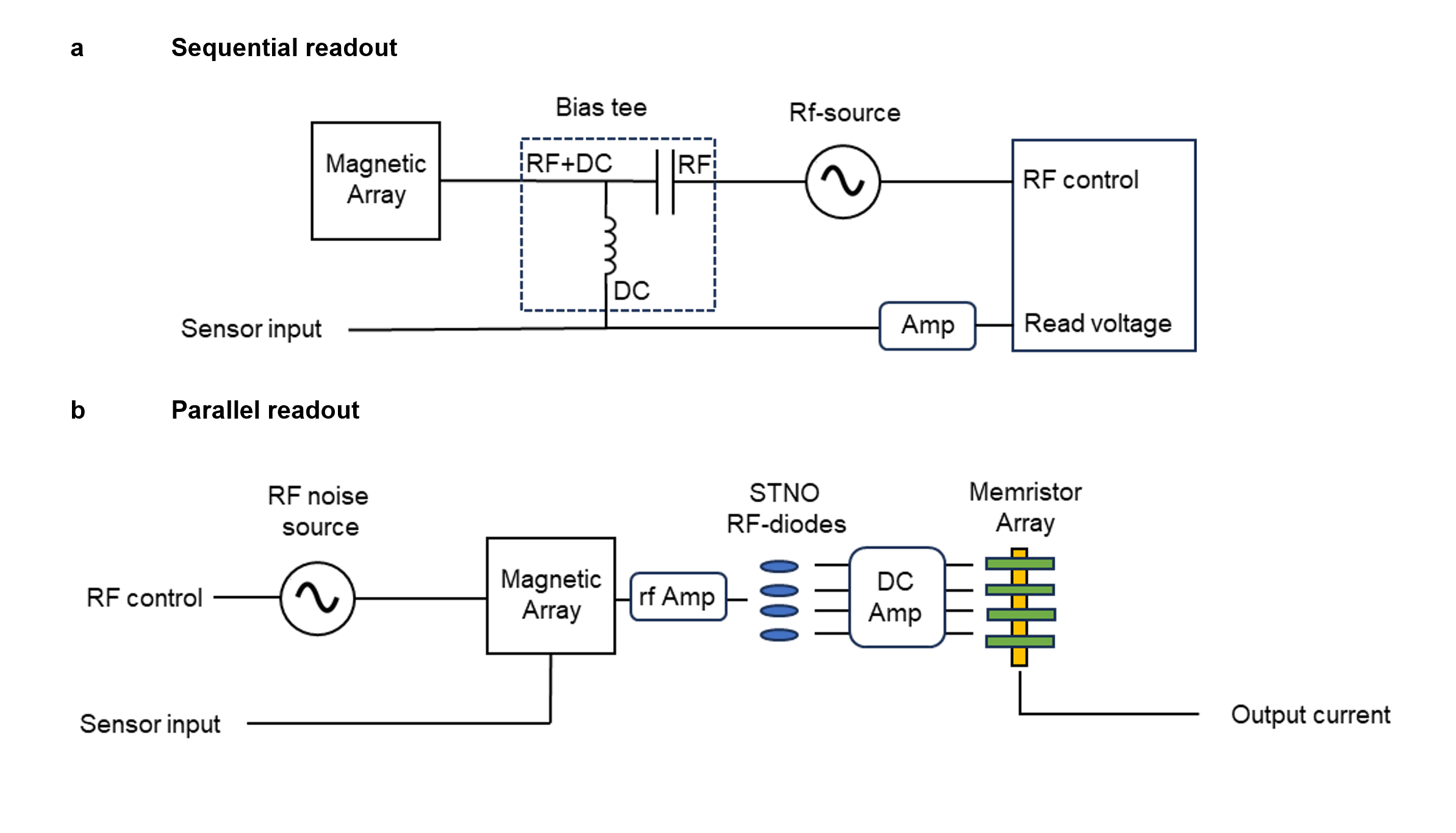}
    \caption{\textbf{Device schematic.} a) Schematic of sequential spin-torque ferromagnetic resonance readout device. Sensory input is provided through a bias tee. Readout utilises spin-torque ferromagnetic resonance whereby a voltage is generated from spin-wave precession. b) Schematic of the parallel readout scheme where an RF noise source excites all modes in parallel. Signal is then passed to spin torque nano-oscillator (STNO) RF-diodes and then to a memristor array for weight multiplication.}
    \label{Schematics}
\end{figure}

Supplementary Table \ref{Components Energy} shows the power, energy, time and area of the various components required to make a projected device. From this, we can calculate the power and energy of different device architectures. Supplementary Figure \ref{Schematics} shows two possible device architectures.
\begin{table}[]
\begin{tabular}{l|l|l|l|l|l}
\textbf{System}                            & \textbf{Power (W)} & \textbf{Time (s)} & \textbf{Energy (J)} & \textbf{Dimensions (L x W)} & \textbf{Area (mm\textasciicircum{}2)} \\ \hline
Oersted field (pulsed)                     & 1.34E-02           & 1.00E-09          & 1.34E-11            & 5e-3 x 5e-3 mm              & 0.000025                              \\
SOT switching 2nm (pulsed)                 & 1.64E-04           & 2.50E-10          & 4.1E-14            & 5e-3 x 5e-3 mm              & 0.000025                              \\
SOT switching 20nm (pulsed)                & 3.87E-01           & 2.50E-10          & 9.68E-11            & 5e-3 x 5e-3 mm              & 0.000025                              \\
RF source (0 dBm, sequential 200 channels) & 1.00E-03           & 2.00E-06          & 2.00E-09            & 4 x 4 mm                    & 16                                    \\
RF source (noise source, parallel)         & 1.50E-01           & 1.00E-08          & 1.50E-09            & 20 x 20 mm                  & 400                                  
\end{tabular}
\caption{Power, time, energy and area of the various components required to operate projected nanomagnetic arrays.}
\label{Components Energy}
\end{table}

We begin with a scheme which sequentially measures each frequency channel illustrated (Supplementary Figure \ref{Schematics} a)). In this scheme, a sensor input in the form of a current is applied to the device via a bias tee. Following this, a broadband RF source is swept and the DC voltage output at each frequency is recorded. The power energy and time when operating this scheme is presented below in Supplementary Table \ref{Sequential_Energy} for the two input schemes (ignoring DC amplification for now).

\begin{table}[h!]
\begin{tabular}{l|l|l|l}
\textbf{Input}      & \textbf{Power (W)} & \textbf{Time (s)} & \textbf{Energy (J)} \\ \hline
Oersted field       & 1.44E-02           & 2.80E-06          & 2.80E-09            \\
SOT switching (2nm) & 1.17E-03           & 2.80E-06          & 2.80E-09           
\end{tabular}
\caption{Power, time, and energy for the sequential readout scheme for Oersted field and spin-orbit torque (SOT) switching.}
\label{Sequential_Energy}
\end{table}

In this scheme, the time and energy for one update is dominated by the RF readout. Total power is dependent on the input scheme. This scheme requires temporary storage of each channel in a separate memory cache which would add to the power, time, and energy. A simple, but unoptimized, way of achieving this would be to connect the hardware to a low-power microcontroller such as an Arduino Uno operating at ~100 mW.

Alternatively, we can consider the parallel readout scheme illustrated in Supplementary Figure \ref{Schematics} b). In this scheme, a sensory input is applied to the device to switch the state. An rf-noise source simultaneously excites all frequencies, and the resulting signal is sent to a series of spin-torque nano oscillator RF diodes tuned to different frequencies, serving as RF diodes. The spin-torque nano oscillators convert the RF signal into a set of DC voltages which are then passed, to a memristor array for weight multiplication. For input and readout, Supplementary Table \ref{Parallel_Energy} shows the powers, energies, and update times for the two different input schemes.

\begin{table}[h!]
\begin{tabular}{l|l|l|l}
\textbf{Input}      & \textbf{Power (W)} & \textbf{Time (s)} & \textbf{Energy (J)} \\ \hline
Oersted field       & 1.63E-01           & 1.50E-08          & 2.11E-09            \\
SOT switching (2nm) & 1.50E-01           & 1.43E-08          & 2.10E-09           
\end{tabular}
\caption{Power, time, and energy for the parallel readout scheme for Oersted field and spin-orbit torque (SOT) switching.}
\label{Parallel_Energy}
\end{table}

Here, the time per update has reduced by two orders of magnitude compared to the previous scheme as the readout has been parallelised. On the contrary, the power has now increased by one order of magnitude and is dominated by the RF source. The energy per update in this scheme has reduced slightly. 

We must also consider the power, energy, time and area required from any amplifiers in the circuit. In the sequential readout scheme (Supplementary Figure \ref{Architectures} a)) an amplifier is necessary to convert the low amplitude output voltage to a compatible read voltage for the microcontroller (we assume the sensory input is appropriately scaled to be compatible with input). The output voltage from the arrays at a power of 0 dBm is in the range of 1 – \SI{10}{\micro\volt} which must be scaled to 0 – 5 V range. The required amplification to convert a \SI{10}{\micro\volt} signal to 5 V is 5 / 10e-6 = 550,000. Low power instrumental amplifiers operating at $\sim$ \SI{10}{\micro\watt} powers have been demonstrated with the necessary gains\cite{kim2016dynamic} and chip sizes around 4 mm$^2$. Time and energy for this amplification would have minimal impact on the previously calculated values.

For the parallel readout scheme (Supplementary Figure \ref{Schematics} b)), we require a series of DC amplifiers in between the spin torque nano oscillators and the memristor array. We can use the above DC amplifiers, requiring a total power of \textit{N} $\times$ \SI{10}{\micro\watt}, where \textit{N} is the number of channels measured (2 mW for 200 channels). This process requires stable precession in the spin torque nano oscillator which takes 14 ns, giving an energy of 2.8e-11 J. As such, the time per update would double to $\sim$ 30 ns, whereas the energy would stay approximately the same.

In both the conventional hardware and nanomagnetic hardware, analogue to digital conversion is required. For conventional hardware, data gathered from a sensor will undergo conversion, with processing performed digitally. For nanomagnetic hardware, processing takes place in the analogue domain with readout being converted. For the sequential scheme, each frequency value will need to be converted whereas in the parallel scheme, only the readout value requires conversion. Therefore, the energy of the sequential scheme will increase as a result. Converters with high sample rate (~1 GS/s) and operating powers or 11 mW are commonly available \cite{zahrai2018review}.

We can summarise the findings by comparing the four CMOS processors to the sequential and parallel schemes in Supplementary Figure \ref{Performance_comp}.

\begin{figure}[ht!]
    \centering
    \includegraphics[width=\textwidth]{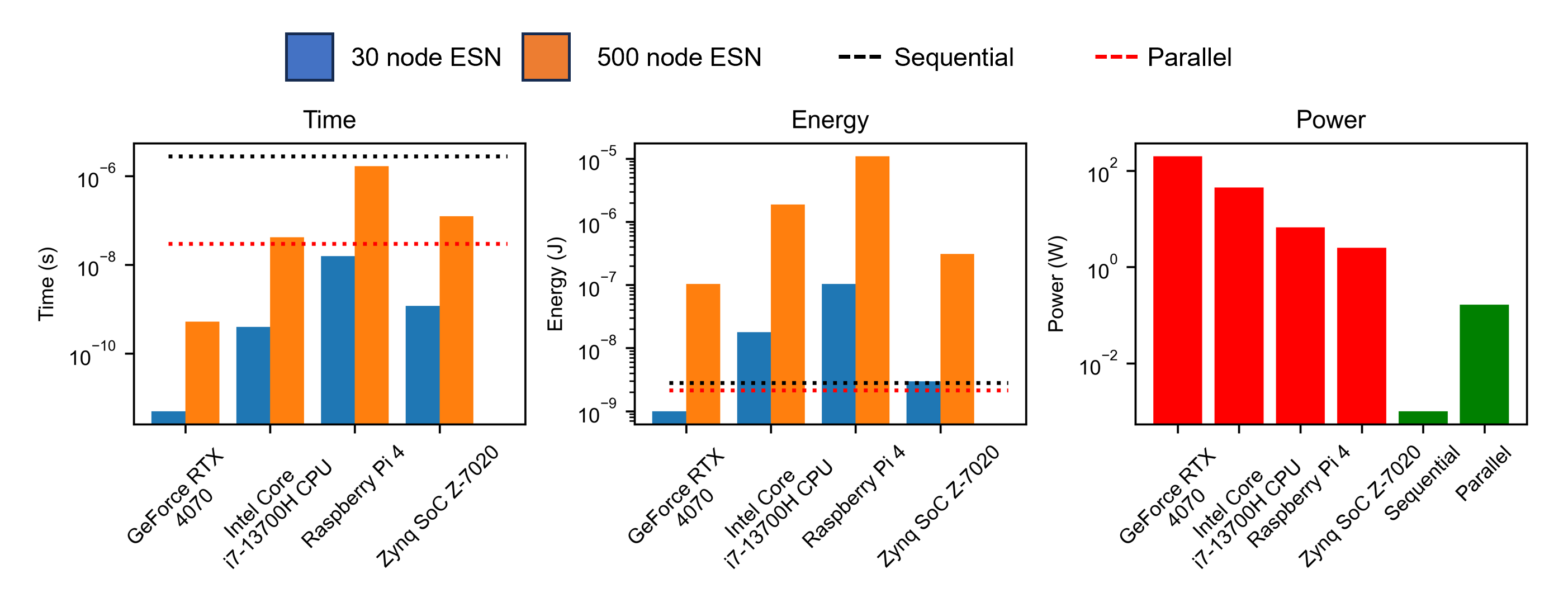}
    \caption{\textbf{Performance comparison.} Comparison of the power, time and energy of updating an ESN vs our projected nanomagnetic devices}
    \label{Performance_comp}
\end{figure}

We find that the time for a single update is slower than conventional hardware for small ESNs but more comparable for large ESNs. As nanomagnetic systems are further optimised to function as well as large ESNs, then the nanomagnetic schemes will become advantageous. Our projected devices are lower energy than conventional hardware (except GPUs) even for small ESN sizes. In terms of power, our schemes are $\sim$16– 2,100 X lower compared to Zynq FPGA hardware. The conventional hardware powers here are quoted as base powers, as opposed to the power required to update the ESN. The conventional hardware can be made to operate at lower powers, but at an additional time cost.

So far, we have considered single arrays. When creating a PNN using conventional hardware, the energy and time costs scale linearly with the number of echo-state networks, as adding an additional network requires the same number of FLOPs to input and update. The total power remains the same. 

\begin{figure}[ht!]
    \centering
    \includegraphics[width=0.85\textwidth]{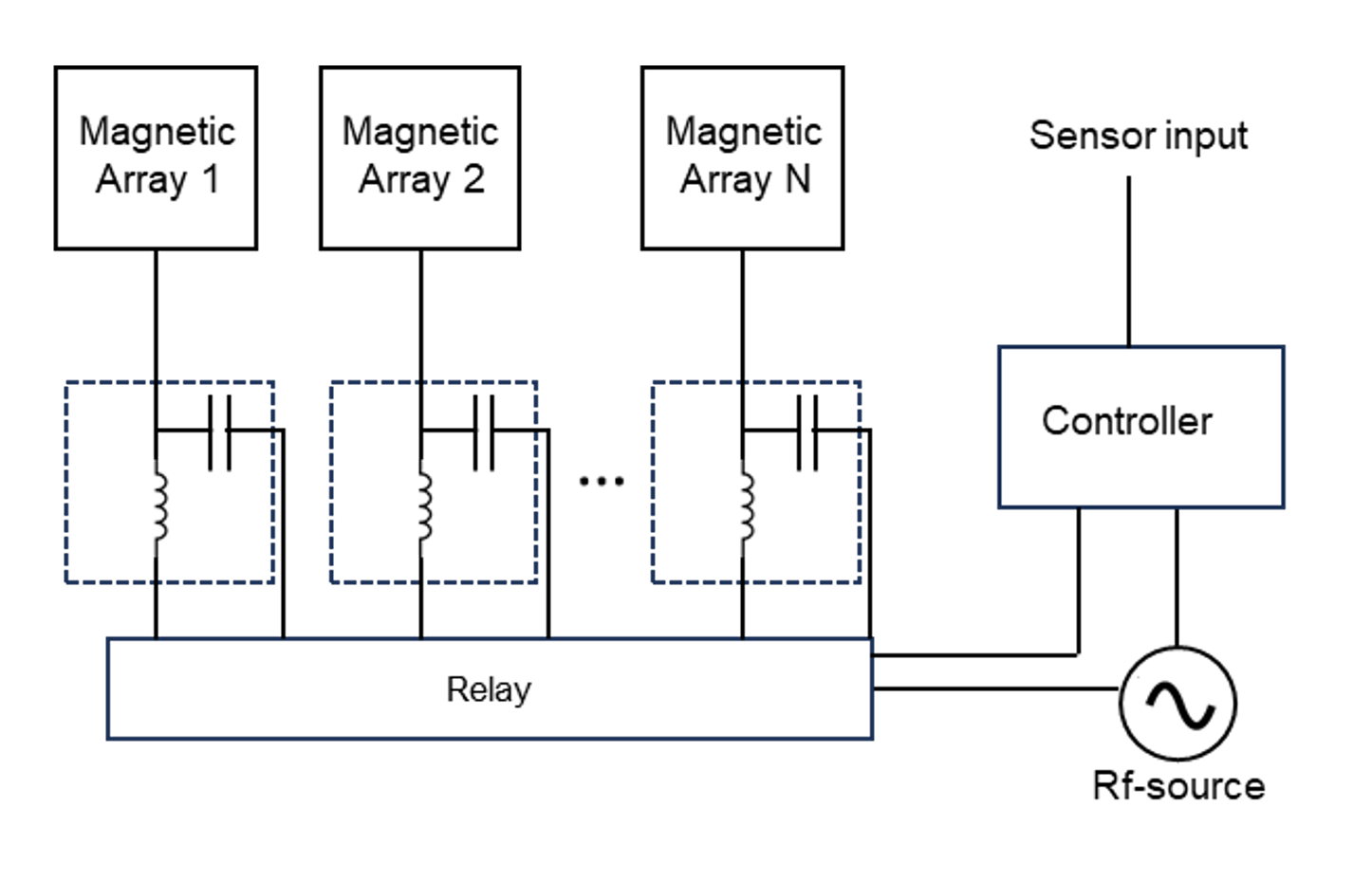}
    \caption{\textbf{PNN device schematic.} Example schematic of a hardware PNN with sequential RF readout method.}
    \label{PNN_Schematic}
\end{figure}

For the nanomagnetic arrays, adding additional reservoirs to the sequential scheme would require a series of relay switches to control where the input signal is sent at any given time (Supplementary Figure \ref{PNN_Schematic}). This could, in principle, be achieved with transistors or MEMS components. The RF signals can also be sent to certain locations via rf-mems components. These components require ~0.1 mW to operate and hence do not greatly affect the overall power consumption of the proposed devices.

In this case, the time and energy would by a factor of N larger, where N is the number of magnetic arrays. Here, the area of the device remains dominated rf-source. For the parallel readout scheme, each array would need to be sequentially measured to receive input from the previous array in the network. As such, we again find that the time and energy would increase by a factor of N for this system.

In summary, in both software and hardware options, the energy and time scales in the same manner with increasing PNN size, therefore the previous comparison of performance and energy with a single array is sufficient for comparison. Whilst the focus here is on nanomagnetic reservoirs, our PNN scheme can be used with other low-power / low-energy neuromorphic technologies. One example is low power memristor arrays which run at $\sim$22 µW \cite{zhong2022memristor}. If one were to interconnect these dynamic memristors into PNNs, one can benefit from the computational advantages of the networked approach, at a fraction of the power consumption of our proposed devices.  

\subsection*{Supplementary note 14 - Architecture details}
\begin{table}[h!]
\begin{tabular}{l|l|l}
Network                                         & Data \# on repo. & Figure             \\ \hline
MS                                              & 0                & 1i, 2f,g, 3a-d     \\
WM                                              & 7                & 1i, 2f,g, 3a-d, 4f \\
PW                                              & 14               & 1i, 2f,g, 3a-d     \\
Parallel (MS + PW + WM)                         & 0,7,14 (Pall)    & 1i, 2f,g, 3a-d     \\
2 series (MS $\rightarrow$ WM)                  & 0, 36 (S34)      & 1i, 2f,g, 3a-d     \\
3 series (MS $\rightarrow$ WM $\rightarrow$ PW) & 0, 36, 47 (S45)  & 1i, 2f,g, 3a-e     \\
PW$\rightarrow${}WM                             & 14, 17 (S15)     & 2f                 \\
PW$\rightarrow${}PW                             & 14, 25 (S23)     & 2f                 \\
WM$\rightarrow${}MS                             & 7, 29 (S27)      & 2f                 \\
MS$\rightarrow${}PW$\rightarrow${}WM            & 0, 18, 41 (S39)  & 2f                
\end{tabular}
\caption{Table of architectures used in the main figures of the manuscript. The network architecture and sub-reservoirs are displayed in the `Network' column. The data numbers and architecture names from the GitHub repository are displayed under `Data \# on repo.'. The figures which contain each network are displayed under `Figure'.}
\label{Architectures}
\end{table}

Supplementary Table \ref{Architectures} displays the network, data labels and figures which contain this network. Data numbers and network names match those within the `Data' folder of the GitHub repository.